%% file: DraftFermions.tex
\documentclass[11pt]{article}
\pdfoutput=1

\usepackage{a4wide}
\usepackage{Packages}

\usepackage{Definitions}

\usepackage{comment}
\usepackage{setspace}

\hypersetup{pdfauthor={R.Dijkgraaf, D.Orlando, S.Reffert},pdftitle={Dimer Models, Free Fermions and Super Quantum Mechanics}}

\author{
  \begin{minipage}{.97\linewidth}
    \vspace{1cm}
    \begin{center}
        \textbf{Robbert Dijkgraaf}${}^{1,2}$, \textbf{Domenico
          Orlando}${}^3$ and \textbf{Susanne Reffert}${}^1$
    \end{center}
    \vspace{1cm}
    \hspace{2cm}\begin{minipage}{.7\linewidth}
     {\it \begin{footnotesize}
        \begin{itemize}
        \item[${}^1$] Institute for Theoretical Physics, University of Amsterdam,\\
          Valckenierstraat 65, 1018XE Amsterdam, The Netherlands.
        \item[${}^2$] KdV Institute for Mathematics, University of Amsterdam,\\
          Plantage Muidergracht 24, 1018 TV Amsterdam, The Netherlands.
        \item[${}^3$] Universit\`a di Milano-Bicocca and INFN, Sezione di
          Milano-Bicocca,  \\
          P.zza della Scienza, 3, I-20126 Milano, Italy
        \end{itemize}
      \end{footnotesize}}
    \end{minipage}
    \vspace{1cm}
  \end{minipage}
}

\date{}

\title{\vspace{1.5cm}
  \begin{huge}
    \textbf{Dimer Models, Free Fermions and}\\[.35em]\textbf{
     Super Quantum Mechanics}
  \end{huge}
}

\begin{document}

\begin{titlepage}
  \maketitle
  \thispagestyle{empty}

  \vspace{-14cm}
  \begin{flushright}
    ITFA-2007-18
  \end{flushright}

  \vspace{14cm}

  \begin{center}
    \textsc{Abstract}\\
  \end{center}
    
  This note relates topics in statistical mechanics, graph theory and
  combinatorics, lattice quantum field theory, super quantum mechanics
  and string theory. We give a precise relation between the dimer
  model on a graph embedded on a torus and the massless free Majorana
  fermion living on the same lattice. A loop expansion of the fermion
  determinant is performed, where the loops turn out to be
  compositions of two perfect matchings. These loop states are
  sorted into co--chain groups using categorification techniques
  similar to the ones used for categorifying knot polynomials.  The
  Euler characteristic of the resulting co--chain complex recovers the
  Newton polynomial of the dimer model. We
  re--interpret this system as supersymmetric quantum mechanics, where
  configurations with vanishing net winding number form the ground states. Finally, we
  make use of the quiver gauge theory -- dimer model correspondence to
  obtain an interpretation of the loops in terms of the physics of
  $D$--branes probing a toric Calabi--Yau singularity.

\end{titlepage}

\onehalfspace

\tableofcontents

\newpage

\input{Introduction}

\input{Dimer}

\input{Fermions_D}

\input{Loops_D}

\input{Categorification}

\input{SQM_D}

\input{Gauge}

\input{Results}

\input{Conclusions}

\subsection*{Acknowledgements}

We would like to thank Luis \'Alvarez-Gaum\'e, David Cimasoni, Davide Forcella,
Emanuel Scheidegger, and Alberto Zaffaroni for enlightening
discussions. Furthermore, we would like to thank David Cimasonifor detailed comments on the manuscript.

D.O. and S.R. would like to thank \textsc{CERN} for hospitality, where
part of this work was carried out.

The research of R.D. was supported by a NWO Spinoza grant and the FOM program "String Theory and Quantum Gravity."
D.O. is supported in part by \textsc{INFN} and \textsc{MIUR} under contract 2005-024045 and by 
the European Community's Human Potential Program \textsc{MRTN}-\textsc{CT}-2004-005104.
S.R. is supported by the EC's Marie Curie Research Training Network under the contract \textsc{MRTN}-\textsc{CT}-2004-512194 "Superstrings".

\appendix

\renewcommand{\pb}[1]{{\parbox{25mm}{\centering \vspace{2mm} #1 \vspace{2mm}}}}
\input{Appendix}

\newpage
\bibliography{DimerReferences}

\end{document}

%% file: Introduction.tex
\section{Introduction}
\label{sec:intro}

The dimer model is concerned with the statistical mechanics of close
packed dimer arrangements on a bipartite graph.  The physical system
which can be thought of as a real world representation of the dimer
model is the adsorption of diatomic molecules on a crystal surface. 

In the 1960s, the question of how many perfect matchings exist on a plane
graph was solved independently by Kasteleyn~\cite{Kasteleyn1,
Kasteleyn2}, and Temperley and Fisher~\cite{Temperley, Fisher}: the total number is given by the Pfaffian of a signed, weighted adjacency matrix of the graph (the Kasteleyn matrix).
Much of
the original interest in the dimer model arose because it provides a simple
and elegant solution for the 2--dimensional Ising model~\cite{Hurst}.

The problem of enumerating perfect matchings is of course a classical problem in graph theory and combinatorics, see \emph{e.g.}~\cite{Kenyon2}. It can also be phrased in terms of domino tilings~\cite{Kenyon1}.
During the last years, the interest in the dimer model was revived
thanks its manifold connections to other branches of
mathematics and physics. The dimer model is related to
\begin{itemize}
\item configurations of a melting crystal corner and the
topological string A--model \cite{Okounkov:2003sp, Iqbal:2003ds},
\item real algebraic geometry~\cite{okounkov1,okounkov2}, 
\item BPS black holes from $D$--branes wrapping collapsed cycles~\cite{Heckman:2006sk}.
\end{itemize}
Furthermore, a correspondence between the dimer model and quiver
gauge theories arising from $D3$--branes probing a singular toric
surface was discovered and worked out in great detail
\cite{Hanany:2005ve, Franco:2005rj, Franco:2005sm, Hanany:2005ss,
Hanany:2006nm}. An explanation of this
correspondence via mirror symmetry was given in~\cite{Feng:2005gw}.

\bigskip
That the dimer model must be related to the free massless fermion on
the same lattice has been known for a long time already. Its relation
to the Ising model, which in turn corresponds to a free fermion model
clearly indicates this.  Moreover, Kasteleyn also showed that the
partition function of the dimer model living on a graph embeddable on
a genus $g$ Riemann surface is given by a linear combination of
$2^{2g}$ Pfaffians. This is, of course, reminiscent of the number of
spin structures on a Riemann surface. In fact, a one-to-one
correspondence between equivalence classes of Kasteleyn orientations
and spin structures was proved in~\cite{cimasoni-2006}. 

In the following, we will make the connection between the dimer model and the
free fermion precise by identifying the Dirac operator with the Kasteleyn matrix and relating the expansion of the determinant
appearing in the free fermion path integral to loop configurations
composed of two perfect matchings. This picture of free fermions as a
loop gas already appeared in a different context, for example in
\cite{Fermionloops}. 

The perfect matchings form a
basis for the space of fermion loop configurations.  If the loop gas lives on a surface of non--trivial homotopy, non--contractible loops of non--zero winding
number can occur, and the Pfaffian of the Kasteleyn matrix becomes a
polynomial. Drawing inspiration from the categorification programme of
Khovanov~\cite{Khovanov1, Khovanov2, Khovanov3}, we 
interpret this polynomial as the Euler characteristic of a co--chain
complex. 
The loop configurations of the free fermion model can
be naturally classified by the 
weight of the constituent matchings of the loop state and used to generate Abelian groups. %
For the Euler
characteristic to coincide with the Newton polynomial, a differential
operator must be constructed accordingly.  In our fermion loop gas,
all loop states are paired since they come in two opposite
orientations, except for states consisting only of double line perfect
matchings. This is reminiscent of the supersymmetric ground states in
supersymmetric quantum mechanics. Indeed, it is possible to
re--interpret the fermion loop gas as \textsc{sqm} and to construct a differential
operator which can be identified with the $Q$ operator. Like this, the
Newton polynomial becomes the generalized Witten index of the system.
Since the fermion loop states are bilinear in the perfect matchings, they can also be interpreted as maps from one matching to another. 

Thanks to the correspondence between the dimer model and quiver gauge theory, 
we know that the perfect matchings, \emph{i.e.} our zero modes, parametrize the toric surface being probed by the $D$--branes. 
In the following, we make use of this correspondence to give the loops states an interpretation in terms of the quiver gauge theory. Loops of vanishing net winding can be related to sequences of double Seiberg dualities on the associated acyclic quivers. In the special case of the hexagon graph, they correspond to certain BPS black hole configurations which are parametrized by melting crystal configurations~\cite{Heckman:2006sk}. 

\bigskip The plan of this paper is as follows.  In Section
\ref{sec:preliminaries}, we review the necessary background and
definitions of the dimer model.  In Section \ref{sec:dirac},
we relate the Kasteleyn matrix to the discretized Dirac operator
of the free fermion. In Section~\ref{sec:scaling-limit-theta}, we go to the limit of vanishing lattice
spacing in which we recover the partition function of the Dirac
fermion, which in the light of the interpretation of the dimer model
as a free fermion model is no longer surprising. In Section~\ref{sec:loop-expans-ferm}, we
expand the determinant arising in the free fermion path integral into
cyclic permutations, which correspond to closed loops on the graph. Furthermore, we give an interpretation of the loops as fermionic states in Subsection~\ref{sec:loops-as-fermionic}.

In Section \ref{sec:categorification}, the basic idea of
categorification is explained, and in Section \ref{sec:sqm}, the
fermion loop system is re--interpreted as supersymmetric quantum
mechanics. The simpler case of a square graph on the cylinder
is worked out first, together with a simple explicit example in Section~\ref{sec:example:-one-square}.
In Section~\ref{sec:loops-as-operators}, we interpret the loops as operators mapping from one perfect matching to another. Finally, Section~\ref{sec:sqm_torus} gives the full construction on the torus.

In Section \ref{sec:geometry}, the basics of quiver gauge theories and the
correspondence to the dimer model is reviewed. The interpretation of
the loop states in terms of the physics of $D$--branes probing a singular toric CY is attempted in Section \ref{sec:results}. In Section~\ref{sec:acyclic}, the relation between perfect matchings and corresponding acyclic quiver graphs is clarified. In Section~\ref{sec:doubleseiberg}, we show that sequences of double Seiberg dualities correspond to cyclic permutations of the nodes of the acyclic quiver and to isomorphisms of the full gauge quiver. With these results, we can interpret the loops with vanishing overall winding in Section~\ref{sec:0-0-loops}. The loops with non--trivial winding are discussed in Section~\ref{sec:winding_loops}.

The results of this paper are summarized in a dictionary given in
Section~\ref{sec:conclusions}, where we close with some concluding
remarks.


%% file: Dimer.tex
\section{Dimer Model Preliminaries}
\label{sec:preliminaries}

Take a plane \emph{bipartite} graph $\gra$, \emph{i.e.} one in which all
vertices can be colored black and white, such that each black vertex
is only connected by links to white vertices and vice versa. Let $M$
be a subset of the set $E$ of edges of $\gra$. $M$ is called a
matching, if its elements are links and no two of them are
adjacent. If every vertex of $\gra$ is saturated under $M$, the
matching is called \emph{perfect}. Such a link which joins a black and
a white vertex is also called a \emph{dimer}. The dimer model describes the
statistical mechanics of a system of random perfect matchings.  In the
simplest case, we ask for the number of close packed dimer
configurations, \emph{i.e.} the number of perfect matchings.

Let us label the vertices of the underlying graph $\gra$ on which the
dimer model lives consecutively.  The topology of $\gra$ can be encoded
in its adjacency matrix $A$, where $A_{x,y}=1$ if the vertices $x$ and
$y$ are joined by an edge, and $A_{x,y}=0$ otherwise. The number of
perfect matchings is given by the number of ways in which the vertices
can be partitioned into adjacent pairs.

The \emph{Hafnian} of a symmetric matrix $A$, introduced by Caianiello
\cite{Caianiello}, is defined as
\begin{equation}
  \label{eq:hafnian}
  \Hf(A)=\sum_{\pi \in \mathcal{S}}A_{\pi(1),\pi(2)}A_{\pi(3),\pi(4)} \dots A_{\pi(n-1),\pi(n)},
\end{equation}
where $\mathcal{S}$ is the set of all permutations $\pi$ of $\set{1, \dots 
 n}$ satisfying $\pi(1)<\pi(3)< \dots <\pi(n-1)$ and $\pi(2i-1)<\pi(2i)$
 for $1\leq i \leq n$, $n$ even.  If we take $A$ to be the adjacency
 matrix of ${\gra}$, $\Hf(A)$ counts the perfect matchings, since
 terms in (\ref{eq:hafnian}) which contain non--adjacent pairs are
 zero.  The \emph{permanent} of an $n\times n$ matrix $A$ is, like the
 determinant, given by the sum over all permutations, except that the
 sign of the permutations is not taken into account:
\begin{equation}
  \label{eq:permanent}
  \perm(A)=\sum_{\pi\in S_n}\prod_{i=1}^n A_{i,\pi(i)}.
\end{equation}
While the determinant can be evaluated in polynomial time via Gaussian
elimination, computing the permanent is $\#P$--complete
\cite{Valiant}.  The Hafnian can be regarded as a generalization of
the permanent. In fact~\cite{Kuperberg1},
\begin{equation}
  \Hf(A)=\perm(B)\ \  \text{for $A$ of the form}\ \ A= \left( \begin{array}{c|c}
      0&B\\ \hline
      B^t&0
    \end{array} \right).
\end{equation}
This identity accounts for the fact that sometimes, the number of
perfect matchings is given as the permanent of the adjacency matrix in
the literature.  Like the permanent, also the Hafnian is difficult to
evaluate and does not satisfy any useful identities.

The counterparts of the Hafnian and permanent are the \emph{Pfaffian} and determinant, which take into account the sign of the permutations:
\begin{align}
  \label{eq:pfaffian}
  \Pf(A)&= \sum_{\pi \in \mathcal{S}}\sign(\pi)\,A_{\pi(1),\pi(2)}A_{\pi(3),\pi(4)} \dots A_{\pi(n-1),\pi(n)},\\
  \label{eq:det}
  \det(A)&=\sum_{\pi\in S_n}\sign(\pi)\prod_{i=1}^n A_{i,\pi(i)},
\end{align}
where the permutations contributing to the Pfaffian are subject to the same conditions as for the Hafnian.
Here, we have
\begin{equation}
  \Pf(A')=(-1)^{\binom{n}{2}}\det(B)\ \  \text{for $A'$ of the form}\ \ A'=\left(\begin{array}{c|c}
      0 & B \\ \hline
      \! \! -B^t & 0 \end{array} \right).
\end{equation}
Moreover, for every anti--symmetric matrix of even order, the relation
\begin{equation}
  \label{eq:rel}
  \det A= \left( \Pf A \right)^2
\end{equation}
holds.  Already in 1913, Polya suggested to change some signs in a
matrix $A$, such that the determinant of the new matrix $\tilde A$
would be the permanent of $A$~\cite{polya}.  Kasteleyn
\cite{Kasteleyn1, Kasteleyn2} actually does exactly this: he
introduces an orientation on \gra, which leads to a signed adjacency
matrix $K$, now called the \emph{Kasteleyn matrix}. The Pfaffian of $K$
gives the number of perfect matchings.

A \emph{Kasteleyn orientation} fulfills the following
condition: the product of all edge weights around a face must equal $-1$
if the number of edges around the face is $0 \mod 4$.  If the number
of edges equals $2 \mod 4$, the product must equal $1$~\cite{Kenyon2}. One
can choose an orientation by consistently assigning arrows to
the edges of the graph, as originally suggested by Kasteleyn
\cite{Kasteleyn1, Kasteleyn2}. For our purposes, it turns out to be
more convenient to allow the roots of 1 as (complex) edge weights.

Already Caianiello~\cite{Caianiello} remarked that the expectation
value of any product of free Fermi fields is a Pfaffian, while the
expectation value of any product of free Bose fields is a Hafnian. In
this sense Kasteleyn's method of attaching signs to the edges of a graph \gra\
could be seen as a fermionization of a bosonic system. Given the
correspondence of the dimer model to the free fermion on \gra, one
could look for its bosonic equivalent, but this goes beyond the scope
of the present work.

\begin{figure}[h!]
  \begin{center}
    \includegraphics[width=50mm]{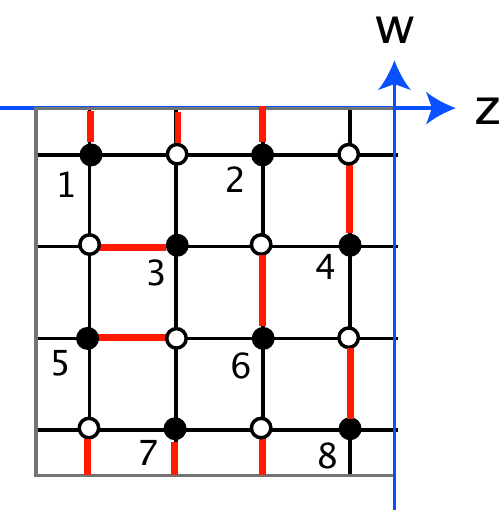}
    \caption{Example of a square graph on the torus}
    \label{fig:example}
  \end{center}
\end{figure}

In the following, we will consider graphs embedded on a cylinder or a
torus. The treatment can be straightforwardly generalized to any genus
$g$ Riemann surface.  In essence, all of the above remains true. The
only change is that on the torus (cylinder), we have two (one)
non--trivial cycles, which we will denote by $z$ and $w$. In the case
of the plane graph, the edge weights originated solely from the
Kasteleyn orientation. More generally, the Kuperberg flatness condition must be respected~\cite{Kuperberg2}. We choose a positive direction on the dimers,
say $\bullet\to\circ$. Now we assign the weight $z$ ($w$) to each edge
which crosses the cycle $z$ ($w$) in positive direction and the weight
$1/z$ ($1/w$) to each edge which crosses it in negative
direction. While the Pfaffian of the Kasteleyn matrix yielded a number
in the case of the plane graph, it now becomes a polynomial in $z$ and
$w$, the so--called characteristic polynomial or \emph{Newton
polynomial} of the graph. The coefficient of each monomial $z^pw^q$
gives the number of matchings with \emph{weight} $(z,w)=(p,q)$.
These are matchings with the number of dimers crossing $z$ in positive
direction minus the number of dimers crossing $z$ in negative
direction equal to $p$ (analogous for $q$). In the literature,
what we call the weight is usually referred to as the slope of a height
function defined on the composition of two matchings. The height
function is defined as follows. Choose a reference matching $\mathrm{PM}_0$. To
find the slope of a matching $\mathrm{PM}$, compose it with the reference
matching, $\mathrm{PM}-\mathrm{PM}_0$, where the minus serves to change the orientation of
$\mathrm{PM}_0$ to $\circ\to\bullet$. This results in closed loops (composition
cycles) and double line dimers. The rule is that when an edge in $\mathrm{PM}$
belonging to a closed loop is crossed such that the black node is to
its left (right), the height changes by $+1$ $(-1)$. If an edge
belonging to $\mathrm{PM}_0$ is crossed, the signs are reversed. This height
function is defined up to the choice of the reference matching
$\mathrm{PM}_0$. Crossing the boundary of the fundamental region of the torus, this function
can jump. If the height function jumps by $p$ units crossing $z$, it
is associated to the power $z^p$ in the Newton polynomial of the graph
(and equivalently for $w$). Choosing a different reference matching
results in a common prefactor of $z^{p_0}w^{q_0}$ for all
monomials. Our method of assigning weights to a matching corresponds
to choosing a reference matching of weight $(0,0)$ that does not
intersect the $z$ or $w$ cycle. The matching shown in Figure
\ref{fig:example} has weight $(1,0)$, where $1=2-1$.
\\
The characteristic, or Newton polynomial, of the dimer model on
the torus takes the form
\begin{equation}
\label{eq:torus-partition}
  P(z,w)=\sum_{n_z, n_w} N_{n_z,n_w}\,(-1)^{n_z+n_w+n_zn_w}z^{n_z}w^{n_w},
\end{equation}
where the $N_{n_z,n_w}$ count the number of matchings that have weight
(height change) $(n_z,n_w)$.
Furthermore, the partition function which counts the total number of matchings for the square graph on the torus is given by
\begin{equation}\label{eq:totalnumber}
Z=\frac{1}{2}\left(-P(1,1)+P(1,-1)+P(-1,1)+P(-1,-1)\right),
\end{equation}
where the first term is always zero. $P(z,w)$ evaluated in $z,w=\pm1$ corresponds to the contributions of the four different spin structures.


%% file: Fermions_D.tex

\section{The dimer model as a free fermion}
\label{sec:fermions-dimers}

As already mentioned in the introduction, many clues point towards the
equivalence of the dimer model and the free fermion living on the
same lattice. On one hand, we know that the dimer model corresponds to
the two--dimensional Ising model. In the continuum limit, the critical
Ising model corresponds in turn to the free Majorana fermion. Another clue is
that the partition function of the dimer model on a genus $g$ Riemann
surface is a linear combination of $2^{2g}$ Pfaffians, one for each of
the different boundary conditions. The one--to--one correspondence
between the $2^{2g}$ equivalence classes of Kasteleyn orientations and
the spin structures was proved in~\cite{cimasoni-2006}. 

Note that our construction given in the following is different from the transfer matrix approach used
in~\cite{lieb:2339,Sutherland:1968,alet:041124}.


\subsection{The Dirac operator and the Kasteleyn matrix}
\label{sec:dirac}

In this section, we show that the dimer model on a square lattice can be
naturally mapped to the dynamics of a free massless fermion on the
same lattice. 
More precisely, we will see how introducing a Kasteleyn orientation on
the graph can be thought of as the projection of the $\del$ operator
on $\gamma $ matrices, see also~\cite{KenyonDirac, Kenyon2}.

\begin{figure}
  \centering
  \includegraphics{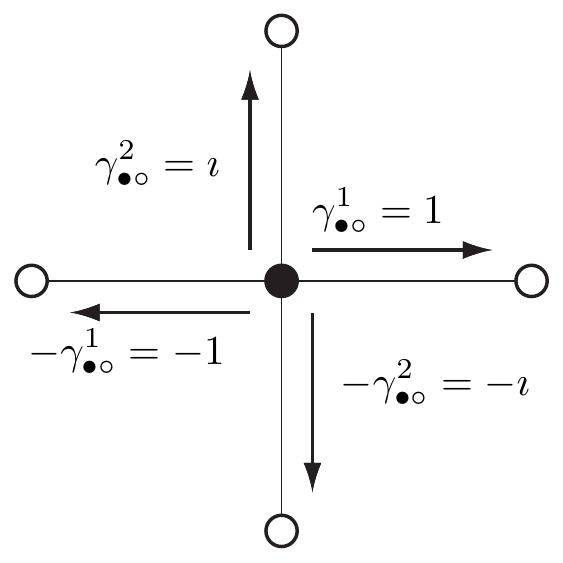}
  \caption{Dirac operator}
  \label{fig:Dirac-Kasteleyn}
\end{figure}

Let us consider the theory for a free massless fermion on a bipartite
square graph\footnote{ A different definition for the Dirac operator,
  such as the one given in~\cite{Rabin:1981qj,Becher:1982ud} would be
  needed to deal with a general random graph. This goes beyond the
  scope of the present note.}. At this stage, the construction is
independent of the boundary conditions, \emph{i.e.} we can think of
the model as living on the infinite plane. The graph being bipartite,
it is natural to use staggered fermions~\cite{PhysRevD.16.3031} and
associate a two--component real spinor to each pair of nodes:
\begin{equation}\label{eq:twocomponent}
  \Psi =
  \begin{pmatrix}
    \chi_{\bullet} \\ \chi_{\circ}
  \end{pmatrix} .
\end{equation}
The naive discretization of the free massless fermion action becomes then
\begin{equation}
  S = \int \di x \: \bar \Psi (x) \slashed{\del} \Psi (x) = \sum_{ \braket{x,  y}} \chi (x)\, \slashed{\del}(x,y)\, \chi(y)  ,
\end{equation}
where the sum runs over all neighbouring vertices $ x $ and $y$
on the lattice. The discretized Dirac operator
can be cast into the form
\begin{equation}
\label{eq:Lattice-Dirac}
  \slashed{\del}(x,y) = \gamma^\mu \frac{ x_\mu - y_\mu}{\abs{x-y}} \, ,  
\end{equation}
where $\gamma^\mu$ are two matrices satisfying the Clifford algebra
$\set{ \gamma^\mu, \gamma^\nu} = 2\, \delta^{\mu \nu}$ in two
Euclidean dimensions. On the square lattice, each node has
four neighbours, separated by unit vectors, so the action reads
\begin{equation}
  S = \sum_{x \in E (\gra)} \sum_{k=1}^4 \chi(x)\, \gamma^\mu e\ud{k}{\mu}\, \chi (x + e^k) \, ,
\end{equation}
where $e^k$ is the vector $e^k= ( \cos \frac{2k \pi}{4},\, \sin \frac{2k
  \pi}{4})$.  If $x$ is a $\bullet $ site, $x \pm  e^k$ is a
$\circ$ site and vice versa. This means that we can split the action
into two pieces:
\begin{equation}
  \label{eq:Dirac-action}
  S = \left(
    \begin{array}{c|c}
      \vec{\chi}_\bullet & \vec{\chi}_\circ 
    \end{array} \right) \left(
    \begin{array}{c|c}
      0 & \slashed{\del}_{\bullet \circ} \\ \hline 
      \slashed{\del}_{\circ \bullet} & 0 
    \end{array} \right) \left(
    \begin{array}{c}
      \vec{\chi}_\bullet \\ \hline
      \vec{\chi}_\circ 
    \end{array} \right) =  \vec{\chi}_\bullet \cdot \slashed{\del}_{\bullet \circ} \cdot \vec{\chi}_\circ + \vec{\chi}_\circ \cdot \slashed{\del}_{\circ \bullet} \cdot \vec{\chi}_\bullet \, ,
\end{equation}
where $\vec{\chi}_\bullet$ is the vector of Grassmanian variables
living on the $\bullet $ nodes.  Let us consider the first term. The
operator $\slashed{\del}_{\bullet \circ}$ is non--vanishing if and only
if $\bullet $ and $\circ$ are neighbours, and takes values
proportional to the components of the $\gamma $ matrices. More
precisely, if we think of $\slashed{\del}_{\bullet \circ}$ as a
matrix, each line will contain four non--vanishing entries $\set{
  \gamma_{\bullet \circ}^1, \gamma_{\bullet \circ}^2, -
  \gamma_{\bullet \circ}^1, - \gamma_{\bullet \circ}^2}$. Choosing the
representation
\begin{align}
  \gamma^1 =
  \begin{pmatrix}
    0 & 1 \\ 1 & 0
  \end{pmatrix} \, ,&&
  \gamma^2 =
  \begin{pmatrix}
    0 & \imath \\ - \imath & 0
  \end{pmatrix} \, ,
\end{align}
these components read $\set{1, \imath, -1, - \imath}$. These are the
weights of the four links around each $\bullet$ node. Since their
product equals to $\left( - 1 \right)$ and on this graph the product of weights around a node is the same as the product around a plaquette, this is precisely a Kasteleyn
orientation (see Fig.~\ref{fig:Dirac-Kasteleyn})\footnote{This argument
  can be easily generalized to any regular tessellation of the
  hyperbolic plane (leading to higher--genus graphs). One can show that
  for an $n$-gon, the product of the weights around a node is $\left(
    -1 \right)$ if $n = 0 \mod 4$ and $\left(+1 \right)$ if $n = 2
  \mod 4$. For the hexagonal lattice (genus one), the
  weights around a $\bullet$ site read $\set{ 1, e^{2 \pi \imath /3},
    e^{-2\pi \imath /3} }$ and their product is $\left( + 1
  \right)$. It is worth to remark that for higher genus surfaces, there is no unique global choice of complex structure.}

The same construction can of course be repeated for the second term in
Eq.~\eqref{eq:Dirac-action}. In this case, we get for each line of the
operator $\slashed{\del}_{\circ \bullet}$ the four non--vanishing elements
$\{ \gamma_{\circ \bullet}^1, \gamma_{\circ \bullet}^2,$ $ -
  \gamma_{\circ \bullet}^1, - \gamma_{\circ \bullet}^2\}$ which, in the
same representation for the $\gamma$ matrices that we used above, read
$\set{ 1, - \imath, - 1, \imath} $, yielding again a Kasteleyn
orientation.

Adjacency matrices with equivalent Kasteleyn orientations have the same determinant (up to an overall sign), therefore we have
\begin{equation}
  \det ( \slashed{\del}_{\bullet \circ} ) = \det ( \slashed{\del}_{\circ \bullet}) = \Pf ( K )\, ,
\end{equation}
where $K$ is a Kasteleyn matrix for the graph.
Writing the partition function for the free massless fermion gives
\begin{equation}
  \label{eq:fermion-path-integral}
  Z_{\text{fermion}} = \int \di \psi \: e^{-S[\psi]} = \sqrt{ \det (\slashed{\del})} = \sqrt{ \det (\slashed{\del}_{\bullet \circ} )  \det ( \slashed{\del}_{\circ \bullet}) } = \Pf (K) \, ,
\end{equation}
which is precisely the number of perfect matchings on the lattice for a given boundary condition.

Explicit expressions for both the dimer model on a square graph and
the free fermion theory are known. 
Take a  triangular lattice describing a torus with
periodicity
\begin{align}
  z = z + i N + j M e^{\imath \theta} \, ,&& i,j \in \setZ,\ z\in\setC,  
\end{align}
\emph{i.e.} with modular parameter $\tau = \tfrac{M}{N} e^{\imath
  \theta}$.  In~\cite{Nash:1996kn}, it is shown that the partition
function for a vortex with winding $(a,b)$ on this lattice,
which can be mapped to a free fermion, is given by the infinite product
\begin{multline}
  Z_{\text{vortex}}  \oao{a}{b} (M,N, \theta) =  \prod_{k=-M/2}^{M/2} \prod_{l=-N/2}^{N/2} 2 \abs{ \alpha \cos \left( 2 \pi \frac{k}{M} + \pi \frac{a}{M} \right)  + \beta  \cos \left( 2 \pi \frac{l}{N} + \pi  \frac{b}{N} \right) + \right. \\
  \left. + \gamma \cos \left( 2 \pi \left( \frac{k}{M} - \frac{l}{N} \right) + \pi \left( \frac{a}{M} - \frac{b}{N} \right) \right) - \sigma}, 
\end{multline}
where
\begin{align}
  \alpha = \beta = \frac{1 - \cos \theta}{\sin^2 \theta}, && \gamma = \frac{\cos \theta}{\sin^2 \theta}, && \sigma = \frac{2 - \cos \theta}{\sin^2 \theta} . 
\end{align}
Specializing this formula for a square torus with $\theta = \pi /2 $ such that
$\alpha = \beta = 1, \gamma = 0, \ \sigma =2 $, we find:
\begin{equation}
  Z_{\text{vortex}}  \oao{a}{b} (M,N, \tfrac{\pi}{2}) =  \prod_{k=-M/2}^{M/2} \prod_{l=-N/2}^{N/2} 2 \abs{ \cos \left( \frac{2k \pi}{M} + a \frac{\pi}{M} \right) + \cos \left( \frac{2 l \pi}{N} + b \frac{\pi}{N} \right) - 2},
\end{equation}
which is precisely the partition function of the dimer model.

\subsection{The scaling limit}
\label{sec:scaling-limit-theta}

Having shown that the dimer model on a graph is equivalent to a free
Majorana fermion theory, we expect the torus partition function to be
modular invariant in the thermodynamical limit, and to consist of
theta functions.

Since this is a universal property, we specialize to a
particular lattice, for example the $M \times N$ square graph. The
partition function for this model was discussed in
\cite{Kasteleyn2,Kenyon1} and takes the form
\begin{equation}
  Z_{MN} = \tfrac{1}{2} \sum_{a,b=0}^1 \left( -1 \right)^{a+b+ab} Z_{MN} \oao{a}{b} \, ,
\end{equation}
where
\begin{equation}
\label{eq:partition-square-mn}
  Z_{MN} \oao{a}{b} = \prod_{k=-M/2}^{M/2} \prod_{l=-N/2}^{N/2} 2 \abs{ \cos \left( \frac{2k \pi}{M} + a \frac{\pi}{M} \right)  +  \cos \left( \frac{2 l \pi}{N} + b \frac{\pi}{N} \right) - 2 } \, .
\end{equation}
If we regard the $M \times N$ lattice as a discretization of a square
torus with modular parameter $\tau = \imath t$ into $M \times N$
squares, the scaling limit is obtained by taking
\begin{align}
  M, N \to \infty && \frac{M}{N} = t , \quad \text{$t\,$ fixed}.  
\end{align}
This corresponds to sending the lattice size $\frac{1}{M}$ to zero.

As we have previously pointed out, this is precisely the same
expression found for a soliton on a triangular lattice~\cite{Nash:1995ba}. After a
careful analysis, the authors prove that in the continuum limit, this
reproduces the Ray--Singer result for the $\bar \del$--torsion of the
torus~\cite{Ray:1973sb} (see also~\cite{Ferdinand1,Ferdinand2}):
\begin{equation}
  Z_{MN} \oao{a}{b} \xrightarrow[\genfrac{}{}{0pt}{}{M,N \to
    \infty}{M/N = - \imath \tau }]{} \left[
    \frac{\vartheta \oao{a}{b} (\tau)}{\eta(\tau)} \right]^2 \, .
\end{equation}

Instead of reproducing the proof,
we will limit ourselves here to a heuristic argument. In the infrared
limit, the cosine can be approximated by a parabola and the functions
above behave as follows:
\begin{multline}
  Z_{MN} \oao{a}{b} = \prod_{k=-M/2}^{M/2} \prod_{l=-N/2}^{N/2} 2
  \abs{\, \cos \left( \frac{2k \pi}{M} + a \frac{\pi}{M} \right) +
    \cos \left( \frac{2 l \pi}{N} + b \frac{\pi}{N} \right) - 2\, }
  \sim \\
  \sim \prod_{k=-M/2}^{M/2} \prod_{l=-N/2}^{N/2} \frac{8 \pi^2}{M^2}
  \abs{ \left(k+ \frac{a}{2} \right)^2 + \left(l+ \frac{b}{2}
    \right)^2 t^2} \, .
\end{multline}
In the $\zeta$--function regularization, one recognizes the usual product
expansion for the theta functions. More precisely, introducing $\tau =
\imath t$ we find %
%
\begin{gather}
  Z_{MN} \oao{a}{b} \xrightarrow[\genfrac{}{}{0pt}{}{M,N \to
    \infty}{M/N = - \imath \tau }]{} \left[ \prod_{k \in \setZ}
    \prod_{l \in \setZ} \abs{ k + \frac{a}{2} + \left( l + \frac{b}{2}
      \right) \tau} \right]^2= \left[
      \frac{\vartheta \oao{a}{b} (\tau)}{\eta(\tau)} \right]^2 \, ,  \\
    Z_{MN} = \tfrac{1}{2} \sum_{a,b=0}^1 \left( -1 \right)^{a+b+ab}
    Z_{MN} \oao{a}{b} \xrightarrow[\genfrac{}{}{0pt}{}{m,n \to
      \infty}{m/n = - \imath \tau }]{} \tfrac{1}{2} \sum_{a,b=0}^1
    \left( -1 \right)^{a+b+ab} \left[ \frac{\vartheta \oao{a}{b}
        (\tau)}{\eta(\tau)} \right]^2 = Z_{\text{Dirac}} (\tau).
\end{gather}
The final result is that in the scaling limit, the partition function
is the one of a Dirac fermion, or equivalently, for two Majorana
fermions. It is not surprising that  we encounter a fermion doubling effect~\cite{Nielsen:1981hk, Nielsen:1980rz} since we have started with a periodic
dispersion relation.


%% file: Loops_D.tex
\subsection{The loop expansion of the fermion determinant}
\label{sec:loop-expans-ferm}

After having shown the equivalence of the discretized Dirac operator
and the Kasteleyn orientations and the appearance of the Dirac fermion
partition function in the scaling limit, we will now directly relate
the dimer model to the free fermion via a diagrammatic expansion of
the fermion determinant.

Consider a Grassmanian field $\chi$ living on the nodes $V(\gra)$ of a
graph $\gra $ described by a quadratic action:
\begin{equation}
  S[\Psi] = \sum_{x,y \in V(\gra)} \chi_x A_{x,y} \chi_y  . 
\end{equation}
As we have seen in Eq.~(\ref{eq:twocomponent}), we can associate a
two--component Majorana fermion $\Psi ={\chi_\bullet \choose
  \chi_\circ}$ to each black--white pair in a bipartite graph. The
path integral for such an action is easily computed and is given by
the Pfaffian of the matrix $A$:
\begin{equation}
  Z_{\text{fermion}} = \int \left[  \prod_{x \in V(\gra )} \di \chi_x \right] e^{\frac{1}{2} \vec{\chi} A \vec{\chi}} = \Pf (A),
\end{equation}
where $\vec{\chi}$ is the vector $\vec{\chi} = \set{ \chi_1, \chi_2,
  \dots, \chi_{\abs{V(\gra)}} }$. In fact, expanding the exponential
(and using the fact that $\chi_x$ are Grassmanian variables), we find:
\begin{equation}
  Z_{\text{fermion}} = \int \left[ \prod_{x \in V(\gra )} \di \chi_x \right]
  \prod_{x<y} \left( 1 - \chi_x A_{x,y} \chi_y \right).
\end{equation}
All the factors in the product commute, therefore we can order them by
increasing $x$. Expanding, the terms that survive the integration are
those which contain each variable exactly once, hence they are products
of the $\abs{V(\gra)}/2$ matrix elements $A_{x,y}$. There is one such
product for each of the permutations of the nodes. Taking into account
the signs, we recover the expression for the Pfaffian in
Eq.~\eqref{eq:pfaffian}.

In the following, we give a diagrammatic expansion for the
partition function in terms of loops on the graph. Instead of the
Pfaffian, we can thanks to relation \eqref{eq:rel} expand
the determinant under the square root. By definition, the determinant
is a sum over all possible permutations $\pi$ of the lattice points,
see Eq.~\eqref{eq:det}.
The signature $\sign(\pi)$ of a permutation is easily found by
expressing $\pi$ as the composition of cyclic permutations:
\begin{equation}
  \pi = \pi_s \circ \pi_{s-1} \circ \dots \circ \pi_l \circ \dots \circ \pi_1  ,
\end{equation}
where $\pi_l$ is a cyclic permutation of $l_r$ elements. We define
\begin{equation}
  \sign (\pi_l) = \left( -  1\right)^{l_r + 1}  
\end{equation}
and
\begin{equation}
  \sign (\pi) = \prod_{l=1}^s \sign (\pi_l)  .
\end{equation}

Each cyclic permutation $\pi_l$ can be represented by a closed loop in
the graph by joining the vertex $x$ by a directed line to the point
$\pi_l(x)$, see \emph{e.g.}~\cite{Fermionloops}. Each vertex in $V(\gra)$ can be:
\begin{itemize}
\item isolated if the permutation does not act on it;
\item joined to another one by a double line if the permutation
  interchanges the two;
\item joined to two others as part of a cyclic chain.  
\end{itemize}
In particular, a permutation of two neighbouring points is a dimer
(with a double line). 

\begin{figure}
  \centering
  \includegraphics[width=.25\textwidth]{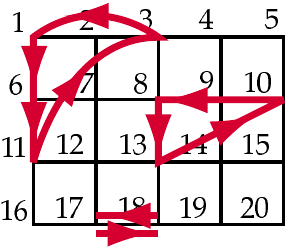}
  \caption{The diagrammatic representation for the permutation $\pi =
    \pi_3 \circ \pi_2 \circ \pi_1$ with $\pi_1 : (17, 18) \mapsto
    (18,17)$, $\pi_2 : (8,9,10,13) \mapsto (9,10,13,8) $ and $\pi_3 :
    (1,3,11,6) \mapsto (3,11,6,1)$.}
  \label{fig:example-loop}
\end{figure}

A general permutation $\pi$ is represented by a graph obtained as the
disjoint union of such loops, and the determinant is written as the
weighted sum of all possible such graphs.  Note that two points can be
joined by a line even if they are not neighbouring on the
lattice. If we do not impose further
constraints on the form of the matrix $A$, we have a sum over all the
$\abs{V(\gra)} !$ possible permutations and the general term
\emph{will not} be a subgraph of $\gra$ (see
Fig.~\ref{fig:example-loop}).

\bigskip

A given graph only contributes to the determinant if the product of
the $A_{x \pi(x)}$ along its edges is non-vanishing. This means that
imposing constraints on the form of $A$ reduces the number of terms
needed.
\begin{itemize}
\item $A$ is a \emph{local} operator if $A_{x,y} = 0 $ when $x $ and
  $y$ are not neighbouring. In this case, the only loops contributing
  to the determinant are those constructed as unions of edges in
  $E(\gra)$, and we can limit ourselves to summing over subgraphs of
  $\gra$.
\item If $A$ does not allow self--interactions, \emph{i.e.}
  $A_{x,x} = 0, \ \forall x \in V(\gra)$, we can consider only
  permutations such that $\pi(x) \neq x,\ \forall x \in
  V(\gra)$. Diagrammatically, this means that we only consider
  subgraphs without isolated points.
\end{itemize}
\begin{figure}
  \centering
  \includegraphics[width=.25\textwidth]{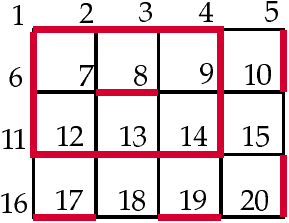}
  \caption{The diagrammatic representation for a permutation allowed
    by the action for a free massless fermion. Orientations are not represented.}
  \label{fig:local-loop}
\end{figure}
The Dirac operator on the graph in Eq.~\eqref{eq:Lattice-Dirac} enjoys
both properties.  In other words, when representing the path integral
for massless free fermions diagrammatically, we consider all subgraphs
containing all the vertices, composed of closed loops or dimers, and
without any isolated points. An example of such a configuration is
depicted in Fig.~\ref{fig:local-loop}.

\bigskip 

\begin{figure}
  \centering
  \begin{tabular}[c]{ccccc}
    \includegraphics[width=.25\linewidth]{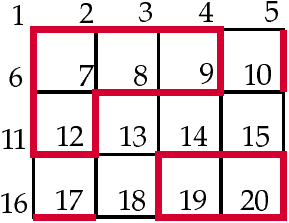} &
    \begin{minipage}{1em}
      \vspace{-1cm}$=$ \vspace{1cm}
    \end{minipage}
    & \includegraphics[width=.25\linewidth]{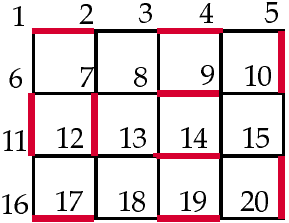} & \begin{minipage}{1em}
      \vspace{-1cm}$\otimes$ \vspace{1cm}
    \end{minipage} & \includegraphics[width=.25\linewidth]{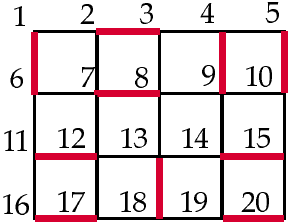} \\
    $\ket{\Psi}$ && $\kket{m_1}$ && $\kket{m_2}$
  \end{tabular}
  \caption{Decomposition of a subgraph $\ket{\Psi}$ containing two
    loops as the combination $\ket{\Psi} = \kket{m_1} \otimes
    \kket{m_2}$. Orientations are not represented.}
  \label{fig:loop-composition}
\end{figure}

On a bipartite graph, each of these contributing loop configurations
can be seen as the composition of two perfect matchings.  This
decomposition is indeed unique.  Let $\kket{m_i}$ and $\kket{m_j}$ be
two perfect matchings. We define the combination $\ket{\Psi} =
\kket{m_i} \otimes \kket{m_j}$ as the collection of all dimers
belonging to either of the matchings, taken with the usual $\bullet
\to \circ$ orientation if the edge is in $\kket{m_i}$ and with the
opposite one if it is in $\kket{m_j}$. Any subgraph $\ket{\Psi}$
contributing to the determinant can always be written in this way. In
fact:
\begin{itemize}
\item if $\ket{\Psi}$ does not contain any loops, its edges are in
  one--to--one correspondence with those of a perfect matching
  $\kket{m}$ and we define $\ket{\Psi} \doteq \ket{m}= \kket{m} \otimes \kket{m}$;
\item if $\ket{\Psi}$ contains one loop, this loop consists of an even
  number of edges since the graph is bipartite. Therefore we can
  decompose it into two disjoint sets of edges that join all the
  vertices. It follows that we can always find two matchings
  $\kket{m_i}$ and $\kket{m_j}$ that only differ by a shift by one
  dimer on the loop and are equal on all the isolated dimers in
  $\ket{\Psi}$. Note that the state $\ket{\tilde\Psi}$ containing the
  same loop with reversed orientation corresponds simply to
  $\ket{\tilde\Psi} = \kket{m_j} \otimes \kket{m_i}$;
\item if $\ket{\Psi}$ contains $n$ loops (which we regard here as unoriented), it can be constructed as the
  composition of any of the perfect matchings containing the disjoint
  sets of dimers making up each loop. It is easy to convince oneself
  that there are $2^{n}$ such combinations, corresponding to the
  $2^n$ possible orientations of the loops (see
  Fig.~\ref{fig:loop-composition}).
\end{itemize}

Note that the composition of two perfect matchings always generates an
allowed configuration. In fact, in each of the matchings, every point
in $V(\gra)$ is touched by an edge and in the resulting subgraph, each
point is either reached by two distinct lines (and then is part of a
loop) or by a double one (and is part of a dimer).

If on a graph $\gra$, there are $k$ perfect
matchings, the determinant is written as the sum over $k^2$ subgraphs.


  
\subsection{Loops as fermionic states}
\label{sec:loops-as-fermionic}

As we have seen in the last section, the partition function for the free fermion can be expanded as a sum
over loop configurations on the lattice. Here we show how these
configurations can be understood in terms of fermionic states.

Consider a square graph on the torus with $N$ horizontal and $M$
vertical cells, each identified by a vector $\mathbf{z} = \left( z,
  w \right) = z\, \mathbf{e}_1 + w\, \mathbf{e}_2$. Let $h
(\mathbf{z})$ be the height function defined on a loop
configuration. The function can be labelled by two integers $p,q$,
corresponding to the following periodicity conditions:
\begin{equation}
  h_{p,q} ( \mathbf{z} + n_1 N\, \mathbf{e}_1  + n_2 M\, \mathbf{e}_2  ) = h_{p,q} (\mathbf{z}) + n_1 p + n_2 q \, ,\hspace{1em} \forall n_i \in \setZ \, .
\end{equation}
The integers $(p,q)$ are clearly the
winding numbers for the loop configuration.\footnote{One can define a
consistent wavefunction for a $\setZ_k$ parafermion as
\begin{equation}
  \psi^{(k)}_{p,q} ( \mathbf{z} ) = e^{\frac{2 \pi \imath}{k} h (\mathbf{z})}  \, .
\end{equation}
Here, we concentrate on the $k=2$ fermionic case, but the formalism can
be extended easily.}

We would like to consider one of the two coordinates (say $z$) as
Euclidean time and hence see the system as the description of a
fermion in $1+1$ Euclidean dimensions. To each plaquette we associate
the operators $a (\mathbf{z})$, $a^\dagger (\mathbf{z})$,
$b(\mathbf{z})$, $b^\dagger (\mathbf{z})$ living on the left edge,
with the following interpretations. The operator
$a^\dagger(\mathbf{z})$ creates a particle at point $w$ for all
times $z^\prime \ge z$ and raises the height function by $1$ on
the right of point $\mathbf{z}$, while the operator $b^\dagger
(\mathbf{z})$ creates an antiparticle at point $\mathbf{z}$ and lowers
the height function by $1$ on the right. We impose the usual
anticommutation relations
\begin{align}
\label{eq:anticomm-ab}
  \set{a (\mathbf{z}), a^\dagger (\mathbf{z}^\prime )} =
  \delta_{\mathbf{z},\mathbf{z}^\prime} \, , && \set{b (\mathbf{z}),
    b^\dagger (\mathbf{z}^\prime )} =
  \delta_{\mathbf{z},\mathbf{z}^\prime} \, ,
\end{align}
all the other anticommutators vanishing. In particular, we recover the
constraint $\left( a^\dagger \right)^2 = \left( b^\dagger \right)^2 =
0 $, which can be translated to the fact that the height function
cannot jump by more than one unit when passing to an adjacent
plaquette.  As usual, one can construct a highest--weight
representation for the algebra by introducing a state $\ket{0}$ which
is annihilated by all the $a$ and $b$
\begin{equation}
  a (\mathbf{z} ) \ket{0} = b (\mathbf{z}) \ket{0} = 0,  
\end{equation}
and defining a general state as the result of the action of a sequence
of $a^\dagger $ and $b^\dagger$ on $\ket{0}$:
\begin{equation}
  \ket{\Psi} = b^\dagger (\mathbf{z}_1 ) \dots b^\dagger(\mathbf{z}_{N_b}) a^\dagger (\mathbf{z}_{N_b+1}) \dots a^\dagger (\mathbf{z}_{N_b+N_a}) \ket{0} \, .
\end{equation}
In this operator formalism, the height function is written as
\begin{equation}
  h ( z, w ) = \sum_{\zeta = 0}^{z} a^\dagger (\zeta, w) a ( \zeta, w) - b^\dagger (\zeta, w) b ( \zeta, w) = \sum_{\zeta = 0}^{z} N_a (\zeta, w) - N_b ( \zeta, w) \, ,
\end{equation}
where we introduced the number operators $N_a (\mathbf{z}) =
a^\dagger(\mathbf{z}) a(\mathbf{z})$.\footnote{In a typical Fermi
  system, $h(z,w)$ is the total charge at point $w$ in
  space and time $z$.}

Even though one can reproduce any height function by just using the
$a$ and $b$ operators, the relations in Eq.~\eqref{eq:anticomm-ab} are
not yet enough to describe all the states. This is because we have
arbitrarily chosen a time direction. One can show that the
anticommutation constraints should be supplemented by a set of
non--local constraints in order to ensure the consistency of the
height function. Yet it is more convenient to introduce two other sets
of operators $c, c^\dagger$ and $d, d^\dagger$ which when acting on a
point $(z, w)$ create a particle (respectively an antiparticle) at
fixed time $z$ for each $w^\prime \ge w$. They also satisfy
anticommutation relations of the type
\begin{align}
  \set{c (\mathbf{z}), c^\dagger (\mathbf{z}^\prime )} =
  \delta_{\mathbf{z},\mathbf{z}^\prime} \, , && \set{d (\mathbf{z}),
    d^\dagger (\mathbf{z}^\prime )} =
  \delta_{\mathbf{z},\mathbf{z}^\prime} \, ,
\end{align}
and anticommute with all the $a$s and $b$s. In this way, we
again obtain $\left( c^\dagger \right)^2 = \left( d^\dagger \right)^2
= 0$, which means that the height function cannot jump by more than one unit
when passing from a plaquette to the one on top or below. Redefining the vacuum
$\ket{0}$ as the state annihilated by $a (\mathbf{z})$,
$b(\mathbf{z})$, $c (\mathbf{z})$ and $d(\mathbf{z})$, we finally find
that a general state is written as
\begin{multline}
  \ket{\Psi} = d^\dagger (\mathbf{z}_1 ) \dots d^\dagger
  (\mathbf{z}_{N_d}) c^\dagger (\mathbf{z}_{N_d+1}) \dots c^\dagger
  (\mathbf{z}_{N_d+N_c}) b^\dagger (\mathbf{z}_{N_d+N_c+1}) \dots
  b^\dagger (\mathbf{z}_{N_d+N_c+N_b}) \\
  a^\dagger (\mathbf{z}_{N_d+N_c+N_b+1}) \dots a^\dagger
  (\mathbf{z}_{N_d+N_c+N_b+N_a}) \ket{0} \, .
\end{multline}

\begin{figure}
  \centering
  \includegraphics[width=.3\textwidth]{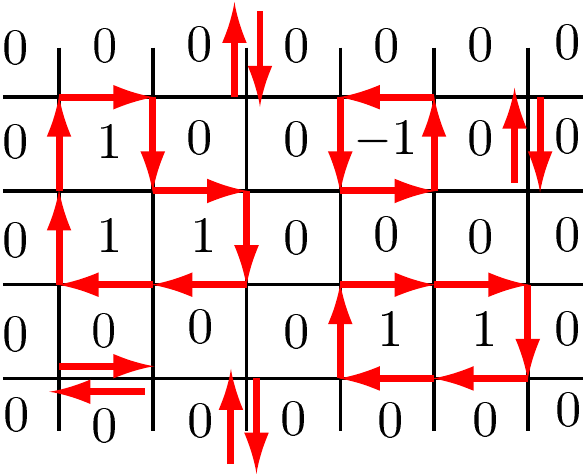}
  \caption{Height function for a loop state. }
  \label{fig:height-function-loop}
\end{figure}

Some consistency conditions must be satisfied. In particular, it should
be verified that the actions of $b^\dagger(z + 1, w)\,
a^\dagger(z,w)$ and of $d^\dagger(z,w + 1 )\,
c^\dagger(z,w)$ change the height function in the same way, namely
they raise $h(z,w)$ by one unit (see Figure
\ref{fig:height-function-loop}). It is easy to see that this is
equivalent to requiring that on any allowed state $\ket{\Psi}$:
\begin{equation}
  \label{eq:height-equality}
  h (z, w ) \ket{\Psi} = \left[ \sum_{\zeta = 0}^{z} N_a ( \zeta, w) - N_b ( \zeta, w) \right] \ket{\Psi} =   \left[ \sum_{\omega = 0}^{w} N_c ( z, \omega) - N_d ( z, \omega) \right] \ket{\Psi} \, ,
\end{equation}
where by allowed we mean that it contributes to the Berezin integral.\\
Note that $h (\mathbf{z}) \ket{0} = 0$, and since the operator
$a^\dagger (\mathbf{z})\, b^\dagger (\mathbf{z})$ anticommutes with $h
(\mathbf{z})$, one can easily see that an allowed state $\ket{\Psi}$
corresponds to a perfect matching if and only if it is annihilated by
$h(\mathbf{z})$:
\begin{equation}
  h (\mathbf{z}) \ket{PM} = 0 \, .  
\end{equation}

Other interesting operators can be defined:
\begin{subequations}
  \label{eq:global-height}
  \begin{align}
    H_z \ket{\Psi}&= \left[ h (N, \bar w) - h (0, \bar w)
    \right] \ket{\Psi} = \left[\sum_{\zeta = 1}^{N} N_a ( \zeta,
      \bar w) - N_b ( \zeta, \bar w) \right] \ket{\Psi} \, , \\
    H_w \ket{\Psi}&= \left[h (\bar z, M) - h (\bar z, 0)\right]
    \ket{\Psi} = \left[\sum_{\omega = 1}^{M} N_c ( \bar z, \omega) -
      N_d ( \bar z, \omega) \right] \ket{\Psi}\, ,
  \end{align}
\end{subequations}
where $\bar w$ (resp. $\bar z$) is an arbitrary value. Note that
both operators are global since they are defined for the whole graph
and do not depend on the choice of $\bar z$ or $\bar w$. We can consider $h
(\mathbf{z})$ as their local counterpart. $H_z,\,H_w$ count the winding
numbers of a state $\ket{\Psi}$ in the $z$ and $w $ directions:
\begin{subequations}
  \begin{align}
    H_z \ket{\Psi_{p,q}} &= p \ket{\Psi_{p,q}} \, , \\
    H_w \ket{\Psi_{p,q}} &= q \ket{\Psi_{p,q}} \, .
  \end{align}
\end{subequations}
In particular, states with vanishing winding numbers will be
annihilated by both $H_z$ and $H_w$. Another operator is the one that
interchanges each $a^\dagger $ with a $b^\dagger$, each $c^\dagger$
with a $d^\dagger$ and viceversa:
\begin{equation}
  \label{eq:Q1-creator}
  Q_1 \ket{\Psi} = \prod_{\mathbf{z}} \left[ a^\dagger (\mathbf{z} ) b (\mathbf{z} )  + b^\dagger (\mathbf{z} ) a (\mathbf{z} ) \right] \left[ c^\dagger (\mathbf{z} ) d (\mathbf{z} )  + d^\dagger (\mathbf{z} ) c (\mathbf{z} ) \right] \ket{\Psi} \, .
\end{equation}
One can easily verify that $\set{Q_1, h (\mathbf{z})} = 0$. It follows
that $Q_1$ leaves the perfect matchings invariant and pairs all the
other states such that $Q_1 \ket{\Psi_{p,q} } =
\ket{\tilde{\Psi}_{p,q}} = \ket{\Psi_{-p,-q} }$.

\begin{figure}
  \centering
  \includegraphics[width=.3\textwidth]{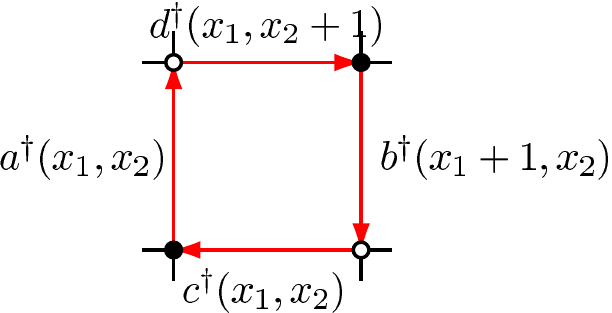}
  \caption{Diagrammatic representation for the four creation operators
    around a plaquette with coordinates $(z,w)$.}
  \label{fig:operators}
\end{figure}

A diagrammatic representation for a general state $\ket{\Psi} $ is
easily obtained as follows. For each operator $a^\dagger (\mathbf{z})$,
one adds a vertical upward arrow on the left edge of the
plaquette at point $\mathbf{z}$, for each $b^\dagger (\mathbf{z})$ a
downward arrow on the left edge of the plaquette, for each
$c^\dagger(\mathbf{z})$ a left pointing arrow on the bottom of the
plaquette and for each $d^\dagger(\mathbf{z})$ a right pointing arrow
on the bottom of the plaquette (see Fig.~\ref{fig:operators}).

In Sec.~\ref{sec:loop-expans-ferm}, we found the loop states to be
combinations of two perfect matchings. We can indeed recover the same
result in terms of this operator formalism, see
Appendix~\ref{sec:perf-match-oper} for the details, where we also give
a necessary and sufficient condition for a state $\ket{\Psi}$ to be
allowed.

\bigskip

Let us summarize the contents of this rather technical section. The
terms contributing to the expansion of the path integral in
Eq.~\eqref{eq:fermion-path-integral} can be represented as resulting
from the action of a set of operators $a^\dagger, b^\dagger, c^\dagger
$ and $d^\dagger$, defined on the lattice, on a vacuum $\ket{0}$. This
set is redundant in the sense that the knowledge of the $a^\dagger $
and $b^\dagger$ operators, plus the consistency conditions in
Eq.~\eqref{eq:state-consistency} uniquely identifies the $c^\dagger$
and $d^\dagger$.

We interpret the two directions in the graph as a Euclidean time ($z$)
and a space coordinate ($w$). The operator $a^\dagger (z,w)$
creates a particle at time $z$ and position $w$ that propagates at
the same position for all times $z \le z^\prime \le N$. In the
same way, $b^\dagger (z, w)$ creates an antiparticle. The height
function $h(z,w)$ computes the local charge ($+1$ for particles
and $-1$ for antiparticles) at point $w$ and time $z$. The charge
gradient $H_w$ between the points $w = 0 $ and $w = M$ at fixed
time $\bar z$ is conserved during the time evolution. \emph{A loop
  configuration with winding numbers $\left( H_z, H_w \right)$
  represents the time evolution of a system with initial charge
  gradient $H_w$, whose net charge in any fixed point in space $\bar
  w$ jumps by $H_z$ between $z=0$ and $z=N$}. In particular, a
$\left(0 , H_w \right)$ loop describes a time evolution where no net
charge is created globally, while a perfect matching corresponds to a
system where no net charge is present at the initial time nor is
created locally.


%% file: Categorification.tex
\section{Categorification of the Newton polynomial}
\label{sec:categorification}

What follows is inspired by the categorification programme for knot
polynomials of Khovanov~\cite{Khovanov1, Khovanov2, Khovanov3}. The
basic idea of his work is that for a knot $K$, a doubly graded
homology theory $H_{i, j}(K)$ can be constructed, whose graded Euler
characteristic with respect to one of the gradings yields the knot
polynomial in question, \emph{e.g.} the Jones polynomial, or the
Alexander polynomial.  The homology groups are constructed through a
categorification process which starts with a state sum representation
of the invariant polynomial.  A group is constructed for each term in
the summation, and differential maps between these groups are
appropriately defined.  Similar categorifications were subsequently
performed for other invariants with state sums, \emph{e.g.} various
graph polynomials, such as the chromatic polynomial~\cite{Helme1,
  Helme2}, and the more general Tutte polynomial~\cite{Rong}.

But exactly what is meant by the term categorification?  When we
categorify, we replace set theoretic concepts by category theoretic
ones. Sets become categories, functions become functors, and equations
between functions become natural isomorphisms between functors
(fulfilling certain relations). The opposite process,
decategorification, is much more familiar to us. When we decategorify,
we are basically throwing away extra information by taking all isomorphic
objects to be equal. When we say that all vector spaces of the same
dimension are the same, we decategorify and are left with a set of
isomorphism classes.  Obviously, the reverse process of inventing the
missing information is much more difficult.  The category
$\mathrm{FinSet}$ of all finite sets for example is a categorification
of the set of natural numbers. For an easy to read introduction to the
concept of categorification, we refer the reader to~\cite{Baez}.

\bigskip

After this general digression, we return to our own system of
fermion loops. Its partition function, the Newton polynomial, is, once
more, a state sum. From the point of view of the dimer model, its
integer coefficients and polynomial nature are obvious. From the point
of view of the free fermion, this is much less so. Giving this
partition function the interpretation of a graded Euler characteristic
explains its integer nature and gives it a geometric interpretation,
while at the same time suggesting its relation to an index.

The task at hand is obvious: construct a bigraded co--chain complex from the
fermion loop states with a differential map preserving the grading, such
that the Euler characteristic of this co--chain complex yields the Newton
polynomial.

The fact that we can associate four different charges to each loop
state living on the torus (see Eq.~\eqref{eq:matching-charges}) gives
us a certain freedom in the choice of the gradings in the double
complex. Our emphasis is placed on the matchings, so we choose to sort
the loop states into co--chain groups according to the maximum winding
number of the two constituent matchings in one direction ($\max_i H_p
\kket{m_i}$) and to define an internal grading based on the maximum
weight in the other direction ($\max_i H_q \kket{m_i}$). This
seemingly arbitrary choice just amounts to picking a horizontal and a
vertical direction in the double complex. The Euler characteristic,
for example, is not sensitive to this choice.

When studying the loop states, there is an obvious observation to
make: because of the two possible orientations of the loops, all
states are paired with the exception of the double line perfect
matchings. This very much smells of supersymmetric Quantum mechanics
(\textsc{sqm}), where the supersymmetric ground states are the only ones which
need not be paired. Indeed, if one takes the total weight of a
configuration to be its energy, the perfect matchings, having zero
winding, become ground states.

The Witten index,
\begin{equation}
  \label{eq:wittenindex}
  \mathrm{dim}\mathcal{H}^B_{(0)}-\dim\mathcal{H}^F_{(0)}=\Tr(-1)^F,
\end{equation}
counts the number of bosonic minus fermionic ground states and is an
invariant of the system. This is, indeed, also what the Newton
polynomial is doing! Once we re--interpret our fermion loop gas as a
\textsc{sqm} system and construct a co--chain complex such that its boundary
operator can be identified with the $Q$ operator and states with
opposite orientations become superpartners, the partition function
becomes the generalized Witten index of the
system.

This will be done explicitly in the following sections.


%% file: SQM_D.tex
\subsection{The dimer model on the cylinder and supersymmetric QM}
\label{sec:sqm}

\begin{figure}
  \centering
  \includegraphics[width=.4\textwidth]{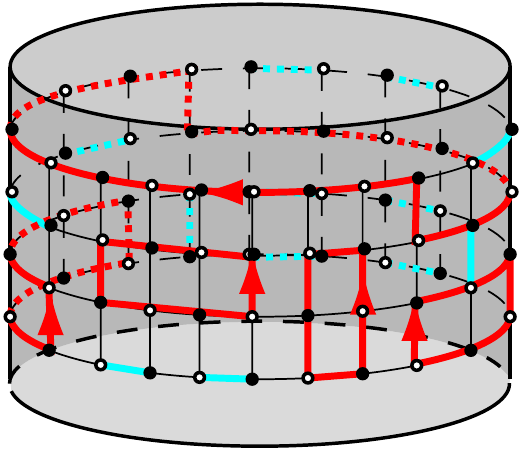}
  \caption{A loop configuration on the cylinder. In red: closed loops; in blue: double lines. This configuration has winding number $k-l = +1$}
  \label{fig:artistic-cylinder}
\end{figure}

In this section, we show how to construct a co--chain complex out of the
loop configurations of a free fermion on a lattice, and how this
system can be mapped to a \textsc{sqm} system in which the loop
configurations correspond to pure states.  For simplicity, we first
confine ourselves to a square graph embedded on a cylinder where only
one non--trivial cycle is present. A typical example of such a loop
configuration is represented in Figure~\ref{fig:artistic-cylinder}.

We have seen in Section~\ref{sec:loop-expans-ferm} that a loop
configuration can always be decomposed into a superposition of two
matchings. Let us introduce the following notation:
\begin{itemize}
\item a matching with weight $k$ is written as $\kket{k,a}$, where $a=1,
  \dots, N_k$ labels the matchings of the same weight;
\item we label a loop as $\kket{k,a; l,b}_\epsilon$ with $k \geq l$,
  $\epsilon = \pm 1$. This corresponds to one of the following
  combinations of matchings:
  \begin{equation}
    \begin{array}{ccll}
      \ket{k,a; l,b}_{+} &=& \kket{k,a} \otimes \kket{l,b} & \text{if $k>l$,} \\ 
      \ket{k,a; l,b}_{-} &=& \kket{l,b} \otimes \kket{k,a} & \text{if $k>l$,} \\ 
      \ket{k,a; k,b}_{+} &=& \kket{k,a} \otimes \kket{k,b} & \text{if $a>b$,} \\ 
      \ket{k,a; k,b}_{-} &=& \kket{k,b} \otimes \kket{k,a} & \text{if $a>b$,} \\ 
      \ket{k,a; k,a}_{+} &=& \kket{k,a} \otimes \kket{k,a} \, ,\\ 
      \ket{k,a; k,a}_{-} &=& 0 \, .
    \end{array}
  \end{equation}
  Opposite polarizations $\epsilon$ correspond to opposite orientations
  of the same loop configuration. States of the type $\ket{k,a;
    k,a}_+$ (which are in one--to--one correspondence with the perfect
  matchings) are singled out because they are not paired.
\end{itemize}

Since each loop configuration with a given orientation can be uniquely
decomposed into matchings, these quantum numbers can be assigned to a
given configuration by inspection. The weight of a matching is given
by the number of dimers crossing the boundary of the fundamental
region in positive direction minus the number of dimers crossing the
boundary of the fundamental region in negative direction (where we
choose $\bullet\to\circ$ as orientation).  $\epsilon \left( k-l
\right)$ is the total winding number of the loop configuration (which
coincides with the slope of the height function as defined in
Sec.~\ref{sec:preliminaries}). States of the type $\ket{k,a ; k,a}$
are in one--to--one correspondence with the perfect matchings. All other
states are paired (i.e. there are two polarizations). From now on, we
will omit to specify the degeneracy quantum numbers and write a loop
as $\ket{k ; l}_\epsilon$ (note the semicolon).

The quantum numbers can be seen as eigenvalues for the following operators:
\begin{align}
  H \ket{k;l}_\epsilon &= \left( k - l \right) \ket{k; l}_\epsilon \, ,\\
  K \ket{k;l}_\epsilon &= k \ket{k;l}_\epsilon \, ,\\
  \Pi \ket{k; l}_\epsilon &= \epsilon \ket{k;l}_\epsilon \, .
\end{align}
They commute on the above loop configurations by definition. It
follows that $H$ (which we want to identify with the energy) commutes
also with the operator
\begin{equation}
  F = 2 K - \frac{1}{2} \left( \Pi - 1 \right) \, ,
\end{equation}
which we identify with the fermion number operator.

In terms of the loops, the above operators have the following
interpretation. $H$ counts the (unsigned) total winding number of the
state. $K$ counts the maximum attainable total monodromy of a loop
configuration. The state $\ket{k; l}_\epsilon$ allows for maximally
$k$ overall (i.e. positively minus negatively oriented) homologically
non--trivial windings, of which $k-l$ are actually realized. If $H$ is
identified with the energy, $k$ gives the maximum possible
energy. Each loop with non--trivial winding which is added to the
configuration raises the energy of the state by one unit. Ground
states do not contain non--contractible loops. The matchings, in
particular, contain no loops at all. %
The eigenvalue of the polarization operator $\Pi$ is the sign of the
winding number. The fermion number operator $F$ is constructed such,
that positive polarization states $\ket{k; l}_+$ have even fermion
number $2k$, while the fermion number of the corresponding negative
polarization states $\ket{k; l}_-$ is raised by one, \emph{i.e.} is
$2k+1$. The operator $\left(-1\right)^F$ gives a $\IZ_2$ grading
according to the fermion parity. States with positive polarization
have even parity, while states negative polarization have odd parity.

\bigskip

To obtain the structure of a \textsc{sqm}, we need to construct two
conserved supercharges $Q$ and $Q^\dag$ which satisfy the following
algebra (see \emph{e.g.} Chapter 10 of~\cite{Hori}):
\begin{align}
  \acomm{Q, Q} &= 0 & \acomm{Q^\dag, Q^\dag} &= 0 \\
  \acomm{Q, Q^\dag} &= 2 H \\
  \comm{F, Q} &= Q & \comm{F, Q^\dag} &= - Q^\dag .
\end{align}
A possible choice is the following:
\begin{align}
\label{eq:Q-cylinder}
  \begin{cases}
    Q \ket{k; l}_+ = \sqrt{2 \left( k - l \right)} \ket{k; l}_- ,\\
    Q \ket{k; l}_- = 0 .
  \end{cases} &&
  \begin{cases}
    Q^\dag \ket{k; l}_+ = 0 ,\\
    Q^\dag \ket{k; l}_- = \sqrt{2 \left( k - l \right)} \ket{k; l}_+ .
  \end{cases}
\end{align}
It is also useful to define the operator $Q_1$ as in
Eq.~\eqref{eq:Q1-creator}:
\begin{equation}
  Q_1 = Q + Q^\dag.
\end{equation}
It reverses the polarization of a state,
\begin{equation}
  Q_1 \ket{k ; l}_\epsilon = \sqrt{2 \left(k -l \right)} \ket{k; l}_{-\epsilon} , 
\end{equation}
and squares to the energy:
\begin{equation}
  Q_1^2 = 2 H .  
\end{equation}

The space of states can be decomposed into subspaces according to the fermion number. In the following, we will concentrate our attention on $Q$.
\begin{subequations}
  \begin{align}
    C^p &= \Sp \set{\ket{k,a; l,b}_\epsilon | F \ket{k,a;
        l,b}_\epsilon = p
      \ket{k,a; l,b}_\epsilon } \, , \\
    C^{2k} &= \Sp \set{ \ket{k,a; l, b}_+ } \, ,\\
    C^{2k+1} &= \Sp \set{\ket{k,a ; l, b}_- } \, .
  \end{align}
\end{subequations}
We also define
\begin{subequations}
  \begin{align}
    C^+ &= \bigoplus_{k} C^{2k} \, , \\
    C^- &= \bigoplus_{k} C^{2k + 1} \, , \\
    C^* &= C^+ \oplus C^- = \bigoplus_{p} C^{p} \, .
  \end{align}
\end{subequations}
One verifies easily that
\begin{equation}
  Q : C^p \to C^{p+1}  
\end{equation}
(and conversely $Q^\dag : C^p \to C^{p-1}$). We can now construct the co--chain complex
\begin{equation}
\label{eq:SQM-complex}
  0 \hookrightarrow  C^0 \xrightarrow{Q} C^1 \xrightarrow{Q} \dots \xrightarrow{Q} C^N \xrightarrow{Q} 0 \, .  
\end{equation}

Let us now study the cohomology of such a complex. The $0$--energy states
$\ket{k,a; k,b}_\epsilon$ and the negative polarization states
$\ket{k,a;l,b}_-$ are $Q$--closed. On the other hand,
negative polarization states (with $k\neq l$) are also $Q$--exact, such that the
cohomology coincides with the set of supersymmetric ground states:
\begin{equation}
  \bigoplus_p H^p(Q) = \Sp \set{ \ket{k, a; l, b}_\epsilon| H \ket{k, a; l, b}_\epsilon = 0} = \Sp \set{\ket{k,a;k,b}_\epsilon} .
\end{equation}
Moreover, all the states with $a \neq b$ are paired (opposite parities). This means that when taking the Euler number, only the $\ket{k,a;k,a}$ states (perfect matchings) contribute:
\begin{equation}
  \chi = \Tr \left( - 1 \right)^F = \sum_p \left( - 1\right)^p \dim \left[ H^p (Q) \right] = \sum_p N_p .  
\end{equation}
As we remarked above, diagrammatically, ground states are
configurations with vanishing total winding number. The ones
contributing to the Euler number are only those that do not contain
any loops.

A different invariant can be constructed by taking the trace of the $\left(
  - 1\right)^F z^K$ operator (which is still a conserved quantity since
it is a function of conserved operators). It is easy to see that it is
given by
\begin{equation}
  \chi (z)= \Tr \left[ \left( - 1\right)^F z^K \right] = \sum_{k,\epsilon} \left( - 1\right)^p z^k \dim \left[ H^p (Q) \right] = \sum_k z^k N_k,
\end{equation}
where $p = 2k - \tfrac{1}{2}(\epsilon -1)$. This is the Poincar\'e
polynomial for the sequence in Eq.~(\ref{eq:SQM-complex}) and in
particular, $\chi(1) = \chi$. Moreover, $\chi(-z)$ is by construction
the partition function for the free Majorana fermion on the lattice.

It is worth stressing that even if, as it is often the case, the
zero--modes alone are enough to describe important features of the
physics of the system such as the partition function, we have complete
control over all the states in the theory. In Appendix~\ref{sec:generating-functions}, the generating functions capturing all loop states are given.

\subsection{Example: One square on the cylinder}
\label{sec:example:-one-square}

\begin{figure}[h!]
  \begin{center}
    \includegraphics[width=110mm]{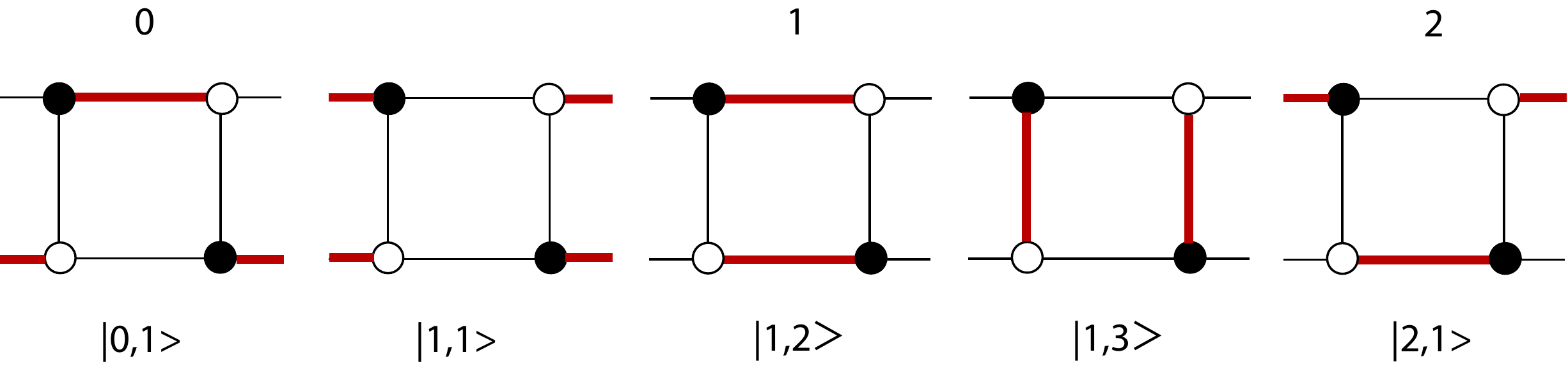}
    \caption{Matchings for the square on the cylinder}
    \label{fig:ex_matchings}
  \end{center}
\end{figure}

To illustrate the above, we now present the smallest possible example
on the cylinder, consisting only of one square.  The five possible
matchings and their quantum numbers are shown in Figure
\ref{fig:ex_matchings}.
They combine into 25 loop configurations, which are, sorted into their
co--chain groups according to total winding number, depicted in Table
\ref{table:chaingroups}.

The partition function can easily be found by inspection and reads:
\begin{equation}
  P (z) = z \left(- \frac{1}{z} + 3 - z \right) \, .
\end{equation}
The generating function for this example reads
\begin{equation}
  G( \mathbf{q}, \mathbf{z}, y ) = 1 + 3 y^2 z + y^4 z^2 + \left( 1 + y \right)  \left[ 3 y^2 z + 3 q \left( y^2 z + y^4 z^2 \right) + q^2 y^4 z^2 \right] \, .
\end{equation}

\begin{table}[h!]
  \begin{center}
    \begin{tabular}{ccccccc}
      \toprule
      \boldmath $k - l$& $C^0$ & $C^1$ & $C^2$ & $C^3$ & $C^4$ & $C^5$ \unboldmath \cr \midrule
      \rowcolor[gray]{.95}  2 & - & -& - & -& \pb{$\ket{2,1;0,1}_+$  \includegraphics[width=13mm]{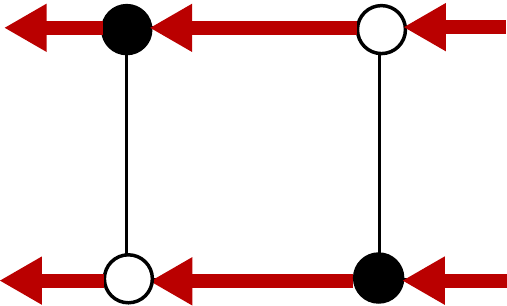}} &\pb{$\ket{2,1;0,1}_{-}$ \includegraphics[width=13mm]{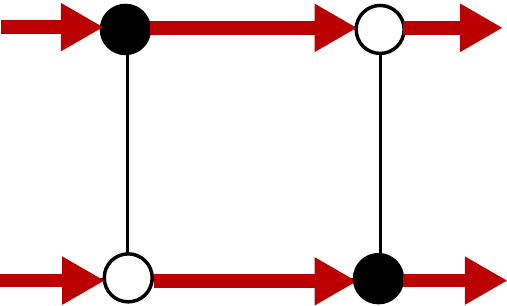}} \cr 
      \midrule
      1 & - & - & \pb{$\ket{1,3;0,1}_+$ \includegraphics[width=13mm]{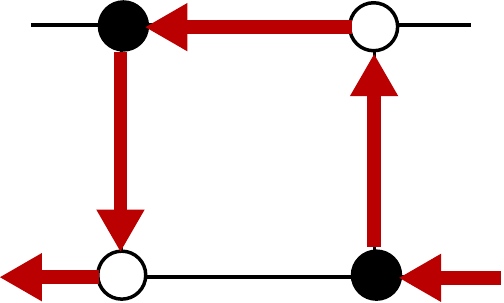}} & \pb{$\ket{1,3;0,1}_-$ \includegraphics[width=13mm]{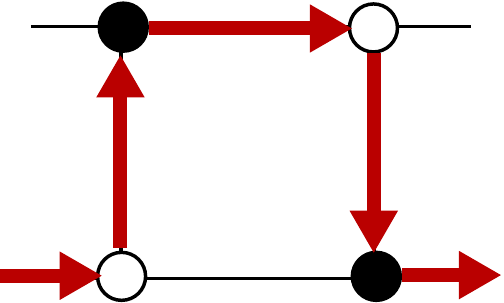}}& \pb{$\ket{2,1;1,3}_+$ \includegraphics[width=13mm]{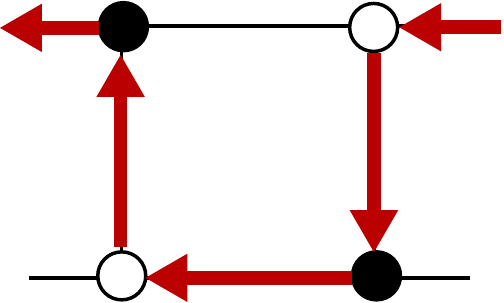}}&\pb{$\ket{2,1;1,3}_{-}$ \includegraphics[width=13mm]{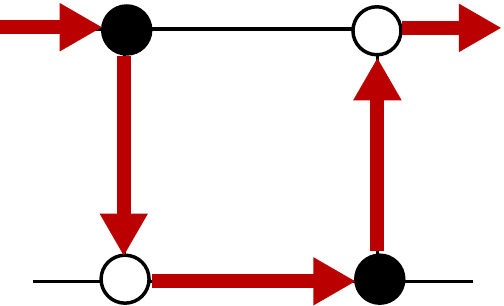}} \cr
      \rowcolor[gray]{.95}  1 & -  & -& \pb{$\ket{1,2;0,1}_+$ \includegraphics[width=13mm]{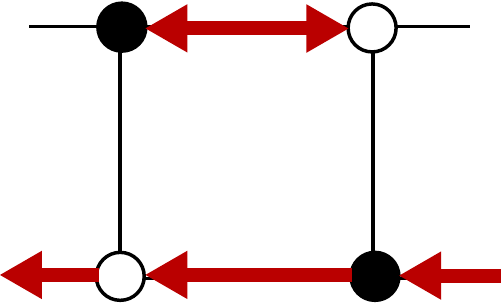} }& \pb{$\ket{1,2;0,1}_-$ \includegraphics[width=13mm]{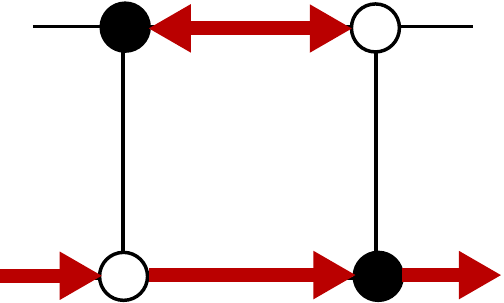}}& \pb{$\ket{2,1;1,2}_+$ \includegraphics[width=13mm]{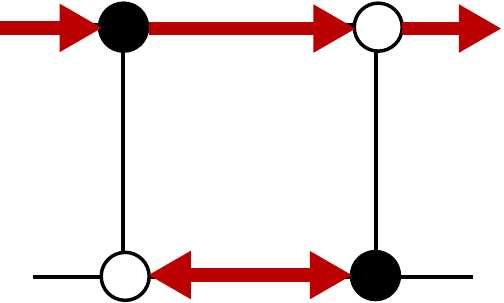}}&\pb{$\ket{2,1;1,2}_{-}$ \includegraphics[width=13mm]{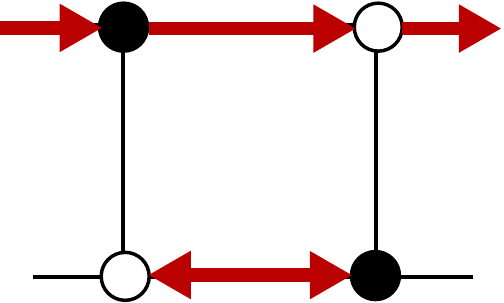}}\cr
      1 & - & - & \pb{$\ket{1,1;0,1}_+$ \includegraphics[width=13mm]{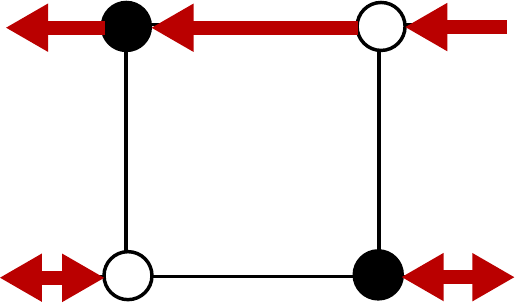}} & \pb{$\ket{1,1;0,1}_-$ \includegraphics[width=13mm]{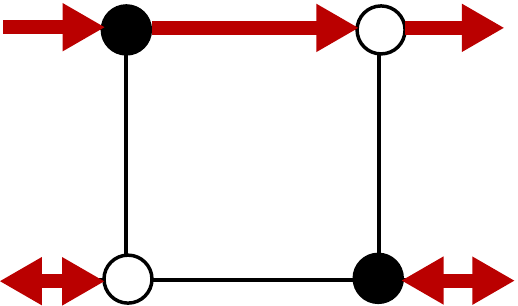}}& \pb{$\ket{2,1;1,1}_+$ \includegraphics[width=13mm]{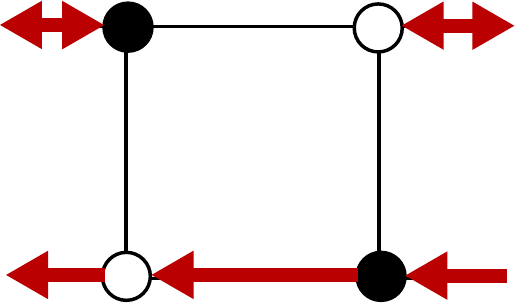}}&\pb{$\ket{2,1;1,1}_{-}$ \includegraphics[width=13mm]{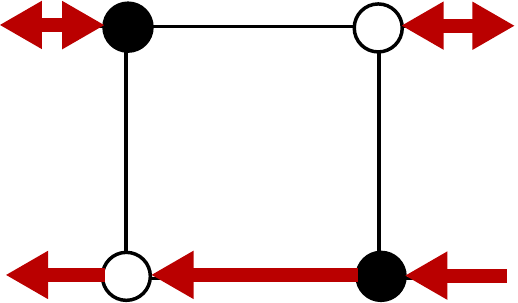}} \cr
      \midrule
      \rowcolor[gray]{.95}  0 & -  & - & \pb{$\ket{1,3;1,2}_+$ \includegraphics[width=13mm]{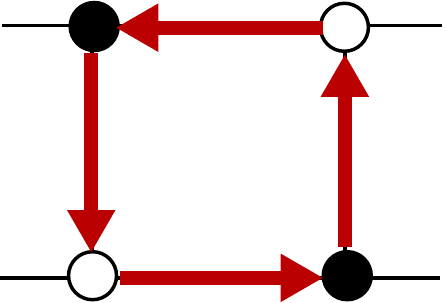}} & \pb{$\ket{1,3;1,2}_-$ \includegraphics[width=13mm]{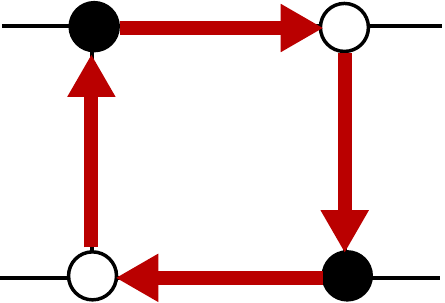}} &- &- \cr
      0 & -  & - & \pb{$\ket{1,3;1,1}_+$ \includegraphics[width=13mm]{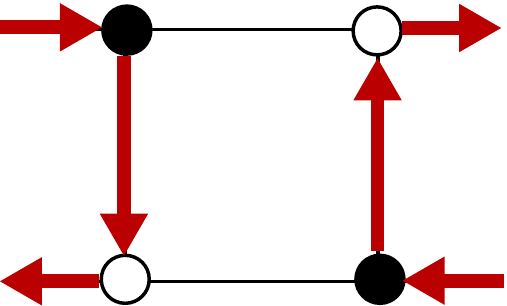}} & \pb{$\ket{1,3;1,1}_-$ \includegraphics[width=13mm]{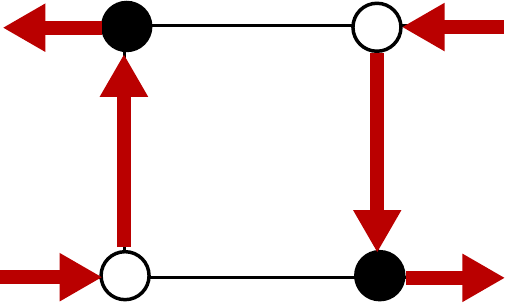}} &- &- \cr
      \rowcolor[gray]{.95}  0 & -   & -& \pb{$\ket{1,2;1,1}_+$ \includegraphics[width=13mm]{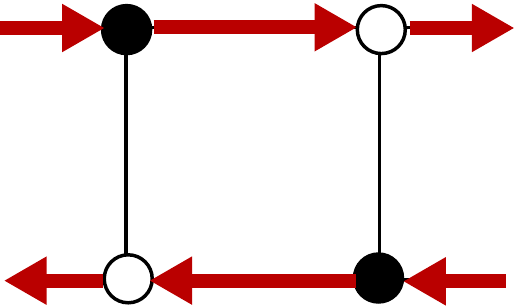}} & \pb{$\ket{1,2;1,1}_-$ \includegraphics[width=13mm]{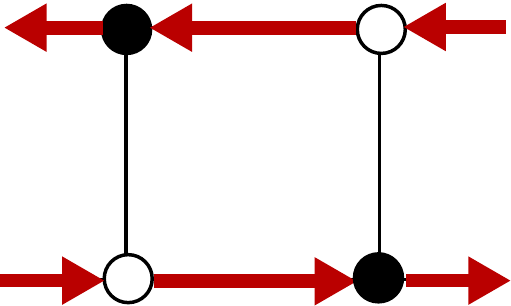}} &- &- \cr
      0 & -  & - & \pb{$\ket{1,3;1,3}$ \includegraphics[width=13mm]{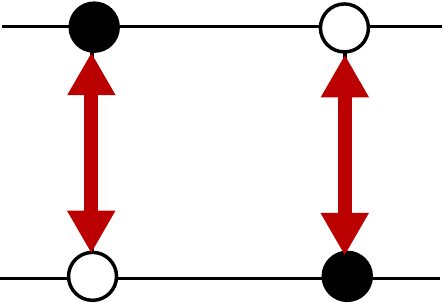}} & - &- &- \cr
      \rowcolor[gray]{.95}  0 & -  & - & \pb{$\ket{1,2;1,2}$ \includegraphics[width=13mm]{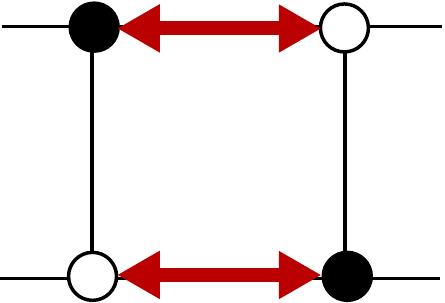}} & - &- &- \cr
      0 & \pb{$\ket{0,1;0,1}$ \includegraphics[width=13mm]{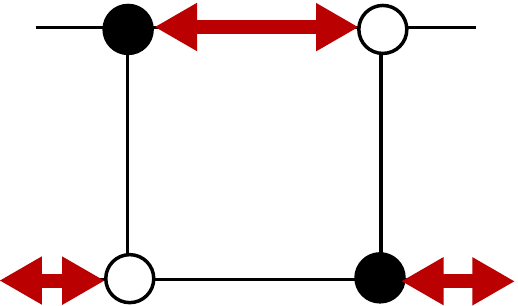}}  & - & \pb{$\ket{1,1;1,1}$  \includegraphics[width=13mm]{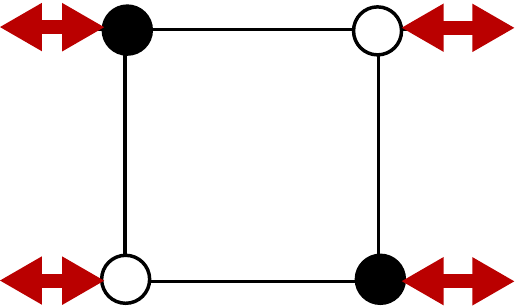}} & -& {\pb{$\ket{2,1;2,1}$ \includegraphics[width=13mm]{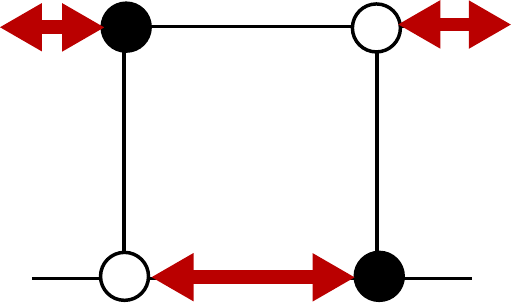}} } &- \cr \bottomrule
    \end{tabular}
    \caption{Co--chain groups for the square on the cylinder}
    \label{table:chaingroups}
  \end{center}
\end{table}

\subsection{Loops as operators}
\label{sec:loops-as-operators}

In this section, we introduce a different notation for the loops that
will allow us to make contact with the geometric and gauge theory descriptions in
Sec.~\ref{sec:geometry} and \ref{sec:results}. Each perfect matching can be seen as a pure
state in a quantum system obeying a normalization condition
$\braket{k,a|l,b} = \delta_{kl} \delta_{ab}$. The general (quantum)
configuration for the system is obtained as a linear combination
\begin{equation}
  \ket{\Psi} = \sum_k \sum_a \lambda_{k,a} \ket{k,a} \, ,
\end{equation}
where $\lambda_{k,a}$ are complex parameters normalized as usual
$\sum_{k,a} \abs{\lambda_{k,a}}^2 =1 $.

\begin{figure}
  \centering
  \subfigure[Complete state graph $C^\ast$]{\includegraphics[width=.4\textwidth]{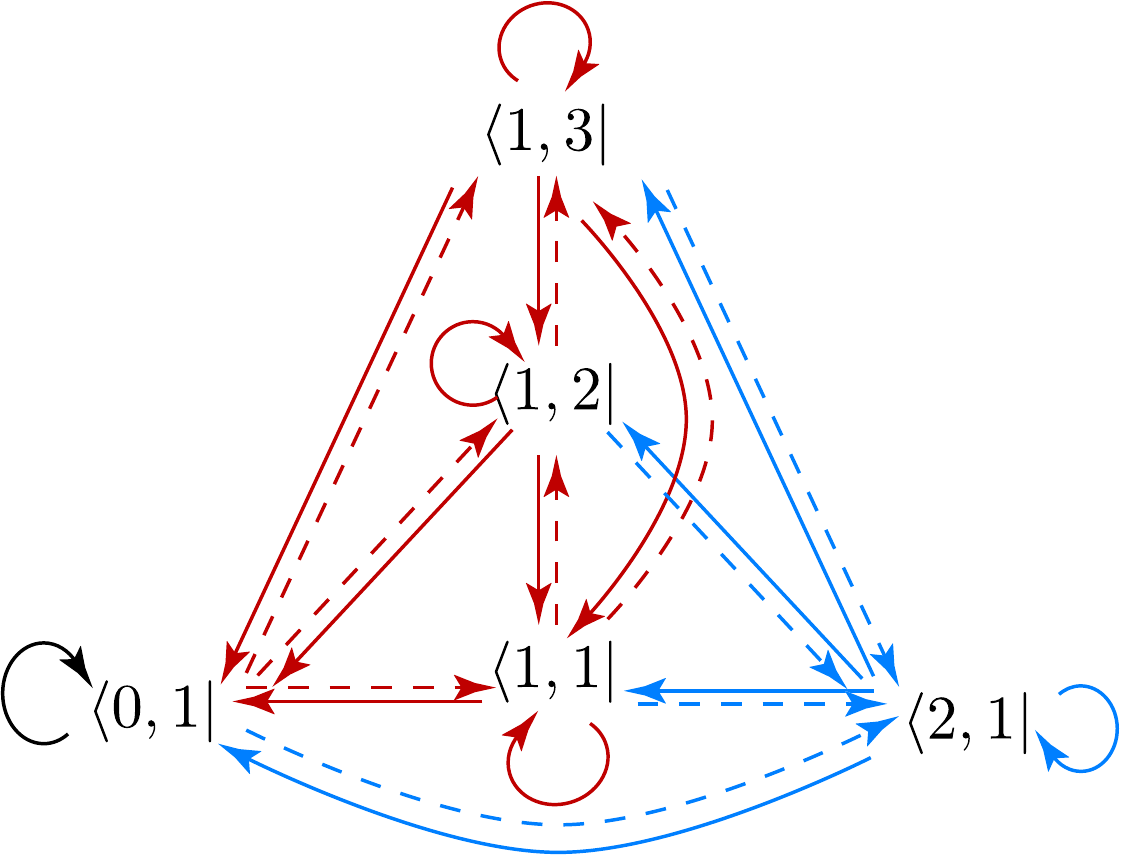}} \hfill
  \subfigure[$C^+ = \bigoplus_{p} C^{2p}$]{\includegraphics[width=.4\textwidth]{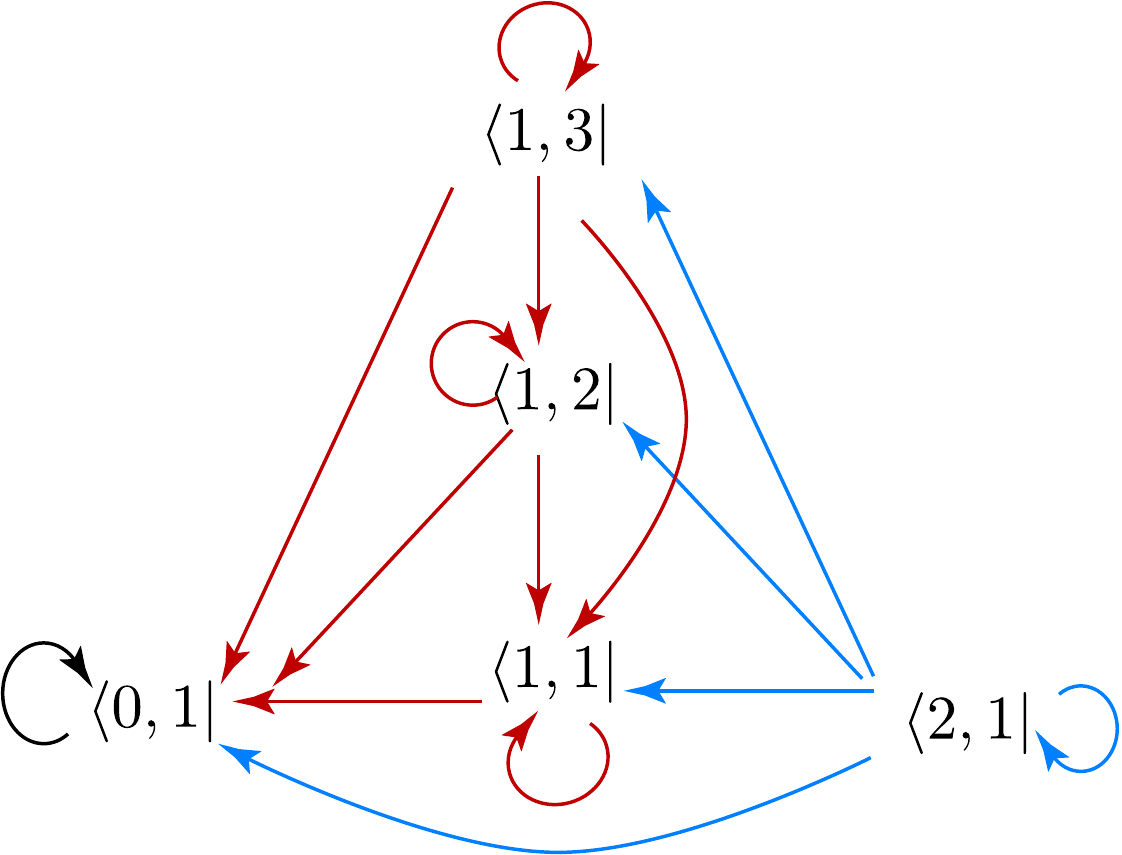}}  \vfill
  \subfigure[$C^- = \bigoplus_{p} C^{2p+1}$]{\includegraphics[width=.35\textwidth]{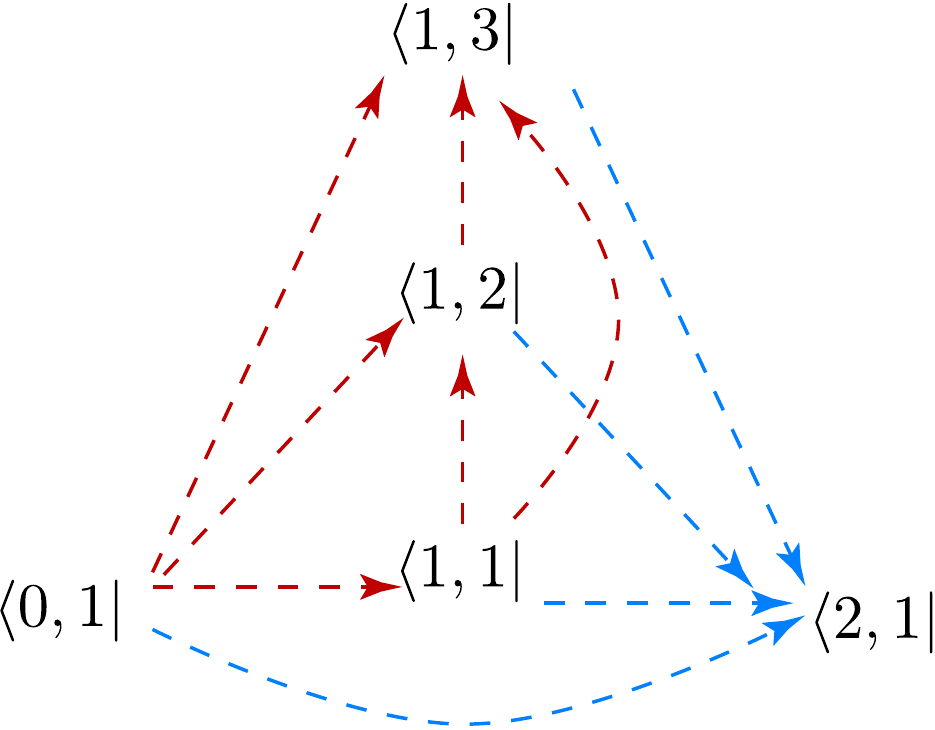}} \hfill
  \subfigure[$H^+ = \bigoplus_{p} H^{2p}$]{\includegraphics[width=.4\textwidth]{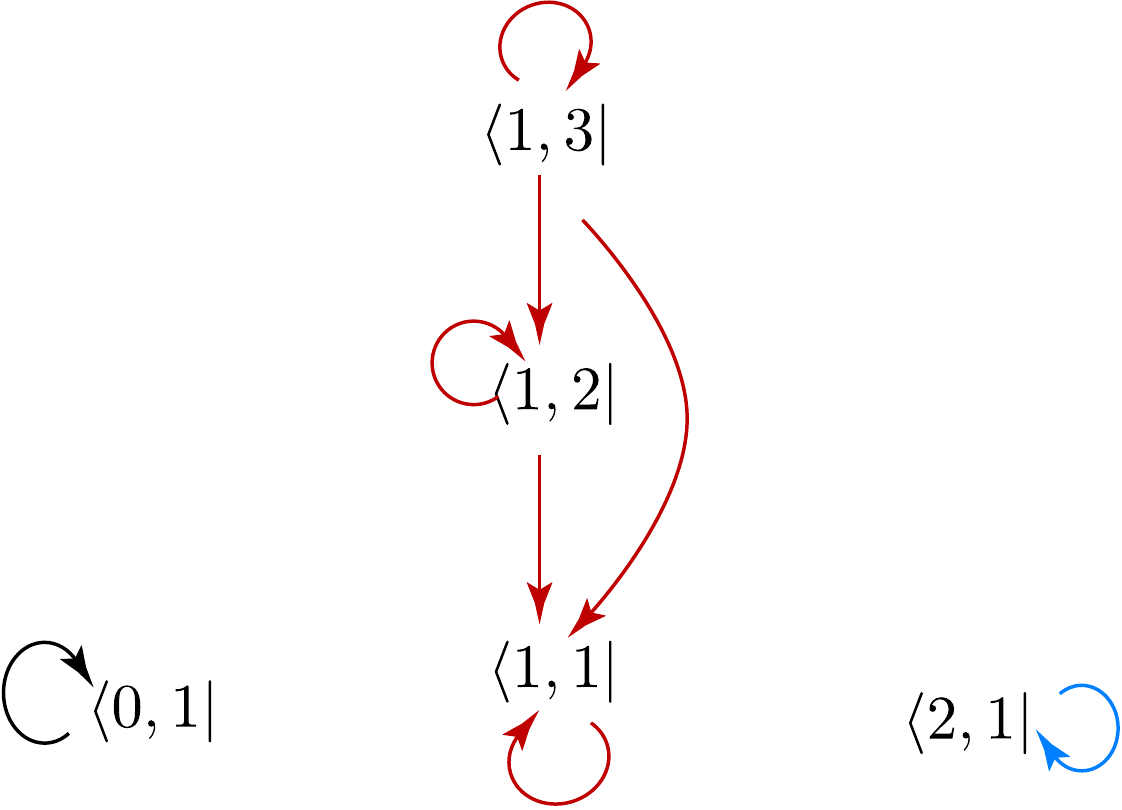}}  \vfill
  \subfigure[$H^- = \bigoplus_{p} H^{2p+1}$]{\includegraphics[width=.35\textwidth]{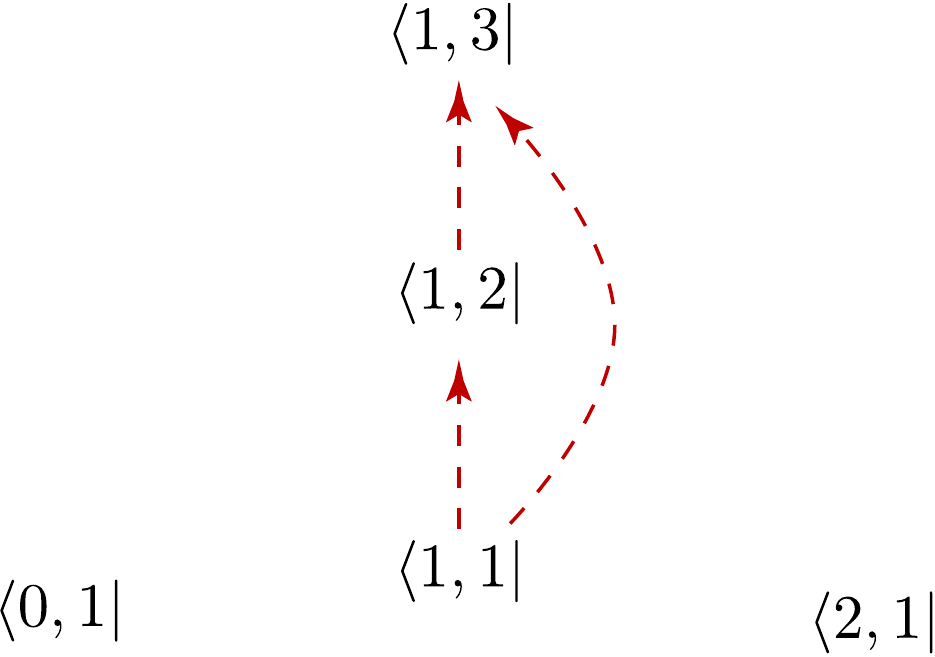}} \hfill
  \subfigure[Loops contributing to the Euler number]{\includegraphics[width=.4\textwidth]{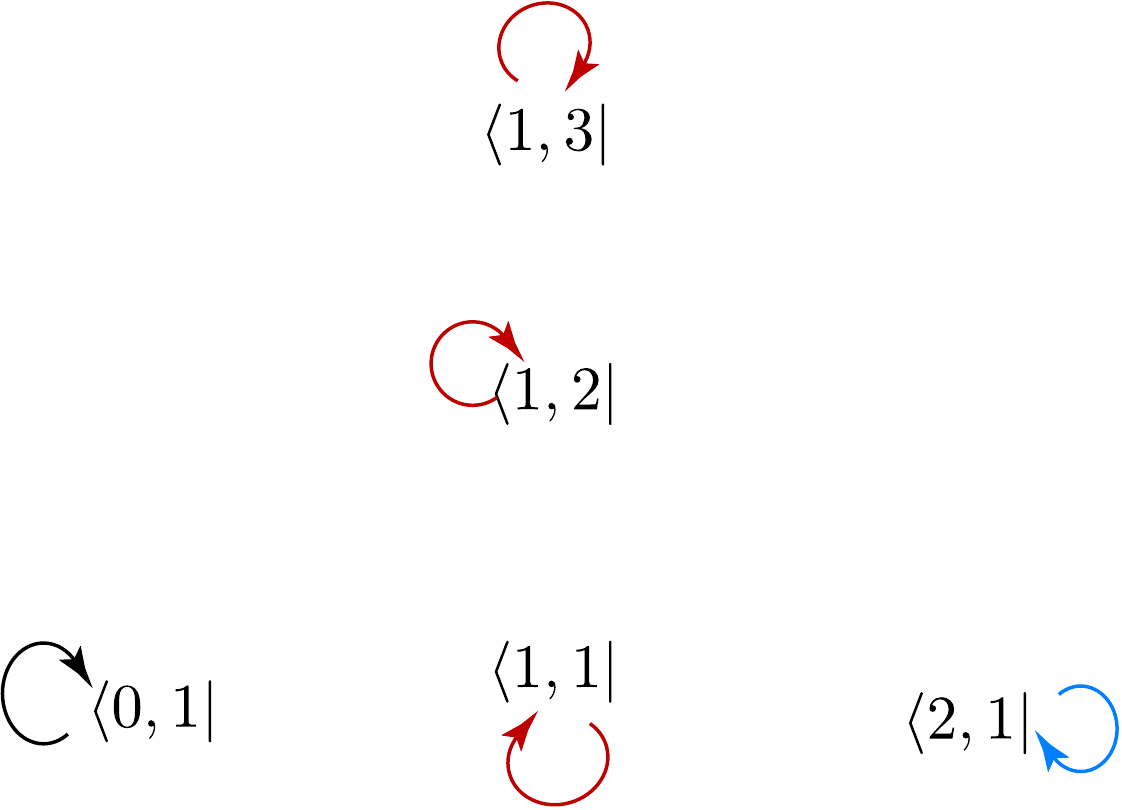}}
  \caption{State graph for the dimer model in
    Sec.~\ref{sec:example:-one-square}}
  \label{fig:state-graph-cylinder}
\end{figure}

Using this point of view, a loop can be identified with a map going
from a state $\ket{k,a}$ to a state $\ket{l,b}$ and as such is
represented in bracket notation by $\ketbra{l,b}{k,a}$ acting as
\begin{equation}
  \left(\ketbra{l,b}{k,a}  \right) \ket{k',a'} = \delta_{k k'} \delta_{a a'} \ket{l,b} \, .
\end{equation}
The linear combination of two loops is still well defined, but a new
operation is naturally defined by the product of two loops:
\begin{equation}
  \left( \ketbra{l,b}{k,a} \right) \times \left( \ketbra{l',b'}{k',a'} \right) \mapsto \ketbra{l,b}{k',a'} \delta_{k l'} \delta_{ab'} \, .
\end{equation}
Note that this is the usual path algebra on a complete graph whose
nodes are labelled by $\ket{k,a}$. It is useful to introduce a new
representation for our states. Consider a graph with $N$ nodes
arranged on the lattice $\setN^2$ at coordinates $\left( k, a \right)$,
and represent the loop $\ketbra{l,b}{k,a}$ as the arrow from
$\left(k,a \right)$ to $\left(l,b \right)$. Our system of loops
corresponds to a complete graph with double lines pointing in
opposite directions plus a loop on top of each vertex (this is a state graph,
see Fig.~\ref{fig:state-graph-cylinder}(a)).  The co--chain groups
(Abelian, freely generated by the loops) receive the following
interpretation:
\begin{itemize}
\item $C^{2p}$ is the group generated by the loops mapping from a
  state with winding $p$ to a state with winding smaller than $p$
  (arrows pointing left), from a state with winding $p$ to another one
  with a smaller degeneracy number (arrows pointing down), or from a
  state to itself (loops). Fig.~\ref{fig:state-graph-cylinder}(b);
\item $C^{2p+1}$ is the group generated by the loops mapping from a
  state with winding smaller than $p$ to one with winding $p$ (arrows
  pointing right), or from a state with winding $p$ to another with
  the same winding and higher degeneracy number (arrows pointing
  up). Fig.~\ref{fig:state-graph-cylinder}(c).
\end{itemize}
In terms of the state graph, this means that the differential map
$Q_{2p} : C^{2p} \to C^{2p+1}$ maps every arrow with a non--vanishing
left component to its opposite and annihilates all the others:
\begin{gather}
  Q_{2p}: \ketbra{l,b}{p,a} \mapsto \ketbra{p,a}{l,b} \, , \\
  Q_{2p}: \ketbra{p,b}{p,a} \mapsto 0 \, .
\end{gather}
On the other hand, $Q_{2p+1} : C^{2p+1} \to C^{2p+2}$ annihilates every map
\begin{equation}
    Q_{2p + 1}: \ketbra{p,a}{l,b} \mapsto 0 \, .  
\end{equation}
It is immediate to see that when considering  the cohomology one finds:
\begin{itemize}
\item $H^{2p} \subset C^{2p}$ is generated by downwards pointing arrows and
  loops. Fig.~\ref{fig:state-graph-cylinder}(d);
\item $H^{2p+1} \subset C^{2p+1}$ is generated by upwards pointing
  arrows. Fig.~\ref{fig:state-graph-cylinder}(e).
\end{itemize}
This implies in turn that taking the Euler character, only the loops
survive since upward and downward pointing arrows are paired and
counted with opposite signs 
(Fig.~\ref{fig:state-graph-cylinder}(f)). They are in one--to--one
correspondence with the nodes and all counted with $\left(+ \right)$
sign, such that
\begin{equation}
  \chi = \sum_p \left( - 1\right)^p \dim (H^p) = \sum_p N_p = N \, .  
\end{equation}

\subsection{The dimer model on the torus and supersymmetric QM}
\label{sec:sqm_torus}

We are now ready to pass to the actual case of interest, \emph{i.e.}
to extend the categorification construction of Sec.~\ref{sec:sqm} to
graphs embedded on a torus. Now we have two non--trivial cycles, so
each matching has two weights, $k_z$ and $k_w$, as explained in
Sec.~\ref{sec:preliminaries}. These weights will be treated on a
different footing, but the final result will restore the expected
symmetry.

The notation is modified as follows:
\begin{itemize}
\item a perfect matching is labelled as $\kket{\mathbf{k},a}$, where
  $\mathbf{k}$ is the vector $\mathbf{k} = \left( k_z, k_w \right) $
  and $a = 1, \dots, N_{\mathbf{k}}$;
\item a loop is the combination of two matchings and is labelled by
  $\ket{\mathbf{k},a;\mathbf{l},b}_\epsilon$, where $k_z \geq l_z$,
  $\epsilon = \pm 1$.
  \begin{equation}
    \begin{array}{ccll}
      \ket{\mathbf{k},a; \mathbf{l},b}_{+} &=& \kket{\mathbf{k},a} \otimes \kket{\mathbf{l},b} & \text{if $\mathbf{k} \succ \mathbf{l}$,} \\ 
      \ket{\mathbf{k},a; \mathbf{l},b}_{-} &=& \kket{\mathbf{l},b} \otimes \kket{\mathbf{k},a} & \text{if $\mathbf{k} \succ \mathbf{l}$,} \\ 
      \ket{\mathbf{k},a; \mathbf{k},b}_{+} &=& \kket{\mathbf{k},a} \otimes \kket{\mathbf{k},b} & \text{if $a>b$,} \\ 
      \ket{\mathbf{k},a; \mathbf{k},b}_{-} &=& \kket{\mathbf{k},b} \otimes \kket{\mathbf{k},a} & \text{if $a>b$,} \\ 
      \ket{\mathbf{k},a; \mathbf{k},a}_{+} &=& \kket{\mathbf{k},a} \otimes \kket{\mathbf{k},a} \, ,\\ 
      \ket{\mathbf{k},a; \mathbf{k},a}_{-} &=& 0 \, ,
    \end{array}
  \end{equation}
  where $\mathbf{k} \succ \mathbf{l} $ if $k_z > l_z $ or if $\set{ k_z = l_z, \,k_w>l_w}$. This corresponds to lexicographic order.
\end{itemize}
Again, the loops can be seen as eigenstates for the following operators:
\begin{align}
  H \ket{\mathbf{k};\mathbf{l}}_\epsilon &= \left( k_z - l_z \right)\ket{\mathbf{k}; \mathbf{l}}_\epsilon \, ,\\
  K_z \ket{\mathbf{k};\mathbf{l}}_\epsilon &= k_z \ket{\mathbf{k};\mathbf{l}}_\epsilon \, ,\\
  K_w \ket{\mathbf{k};\mathbf{l}}_\epsilon &= \max (k_w, l_w) \ket{\mathbf{k};\mathbf{l}}_\epsilon \, ,\\
  \Pi \ket{\mathbf{k}; \mathbf{l}}_\epsilon &= \epsilon \ket{\mathbf{k};\mathbf{l}}_\epsilon \, .
\end{align}
As opposed to the example on the cylinder, a loop state now comes with
two natural signs, $\epsilon_z=\sign{(k_z - l_z)}$, and
$\epsilon_w=\sign{(k_w- l_w)}$. For the grading by the fermion number,
we assign only one sign $\epsilon$ to each loop state according to
Table \ref{table:signs}.
\begin{table}[h!]
  \begin{center}
    \begin{tabular}{ccc}
      \toprule
      \boldmath $\epsilon_z$&$\epsilon_w$&$\epsilon$ \unboldmath \cr \midrule
      \rowcolor[gray]{.95} $+$&$+$&$+$\\
      $-$&$-$&$-$\\
      \rowcolor[gray]{.95} $+$&$-$&$+$\\
      $-$&$+$&$-$\\
      \bottomrule
    \end{tabular}
    \caption{Sign table for torus loop states}
    \label{table:signs}
  \end{center}
\end{table}
This is achieved by 
\begin{equation}
  \label{eq:signdef}
  \epsilon=\epsilon_z+(1-\epsilon_z^2)\,\epsilon_w.
\end{equation}
By choosing the right order on the perfect matchings of the same
weight and choosing a sign for some loop configurations of winding
number 0, the signs of the overall winding numbers around the $z$ and
$w$ cycles can be brought to correspondence with the
signs $\epsilon_z,\,\epsilon_w$ defined above.  We assign a sign to an oriented loop
by choosing a right handed (counterclockwise) system to have positive
sign and a left handed (clockwise) to have negative sign. The signs of
a loop configuration are determined by the overall winding number
around the cycles. Only in cases with an equal number of loops winding
a cycle in opposite directions, a sign has to be chosen arbitrarily.

The fermion number operator only depends on $K_z$ and is defined by
\begin{equation}
  F = 2\, K_z - \frac{1}{2} \left( \Pi - 1 \right) \, .
\end{equation}

States can be again collected into subspaces according to the
eigenvalues of $F$, but in this case, we have an extra grading given by $k_w$. Since $\comm{F, K_w} = 0$, we obtain a direct sum structure, and define:
\begin{gather}
  C^p = \bigoplus_{q=0}^M C^{p,\,q}  \, , \\
  C^{p,\,q} = \Sp \set{\ket{\mathbf{k};\mathbf{l}}_\epsilon | F \ket{\mathbf{k};\mathbf{l}}_\epsilon = p, \, K_w \ket{\mathbf{k};\mathbf{l}}_\epsilon = q} \, .
\end{gather}
The differential operators (or supercharges) are defined precisely as
in Eq.~(\ref{eq:Q-cylinder})
\begin{align}
  \begin{cases}
    Q \ket{\mathbf{k}; \mathbf{l}}_+ = \sqrt{2 \left( k_z - l_z \right)} \ket{\mathbf{k}; \mathbf{l}}_- ,\\
    Q \ket{\mathbf{k}; \mathbf{l}}_- = 0 .
  \end{cases} &&
  \begin{cases}
    Q^\dag \ket{\mathbf{k}; \mathbf{l}}_+ = 0 ,\\
    Q^\dag \ket{\mathbf{k}; \mathbf{l}}_- = \sqrt{2 \left( k_z - l_z \right)} \ket{\mathbf{k}; \mathbf{l}}_+ ,
  \end{cases}
\end{align}
and it is immediate to verify that they respect the grading since
$\comm{K_w, Q} = \comm{K_w, Q^\dag} = 0$. Again we concentrate on $Q$. It can be decomposed into the sum
\begin{align}
  Q_p &= \sum_{q=0}^M Q_{p,\,q} \, , \\
  \intertext{where}
  Q_p &: C^{p,\,q} \to C^{p+1,\,q} \, ,\hspace{2em} q = 0, \dots, M \, . 
\end{align}
As a result, the $C^\ast$ complex is decomposed into $C_{\ast,q}$
complexes:
\begin{equation}
  \label{eq:torus-complex}
  \begin{array}{rccccc}
    0 &\hookrightarrow  C^{0,M} &\xrightarrow{Q_{0,M}} C^{1,M} &\xrightarrow{Q_{1,M}} \dots &\xrightarrow{Q_{N-1,M}} C^{N,M} &\xrightarrow{Q_{N,M}} 0 \, , \\
    & & & \dots \\
    0 &\hookrightarrow  C^{0,q} &\xrightarrow{Q_{0,q}} C^{1,q} &\xrightarrow{Q_{1,q}} \dots &\xrightarrow{Q_{N-1,q}} C^{N,q} &\xrightarrow{Q_{N,q}} 0 \, , \\
    & & &\dots \\
    0 &\hookrightarrow  C^{0,0} &\xrightarrow{Q_{0,0}} C^{1,0} &\xrightarrow{Q_{1,0}} \dots &\xrightarrow{Q_{N-1,0}} C^{N,0} &\xrightarrow{Q_{N,0}} 0 \, .
  \end{array}
\end{equation}

The analysis of the cohomology is the same as in the case of the cylinder and one can easily convince oneself that it consists of states of zero energy:
\begin{equation}
  \bigoplus_{p,q} H^{p,q} = \Sp \set{ \ket{\mathbf{k},a; \mathbf{l},b}_\epsilon | k_z = l_z} .  
\end{equation}
Once more, only the states that are in one--to--one correspondence with the perfect
matchings (\emph{i.e.} the ones that do not contain non--trivial closed loops) are not paired
and thus survive the projection by the $(-1)^F$ operator:
\begin{equation}
  \chi(w)=\Tr \left[ (-1)^F w^{K_w} \right] = \sum_{p,q}  w^q N_{p,q} \,,
\end{equation}
where we introduced the variable $w$ to keep track of the internal
grading. The partition function described in
Eq.~(\ref{eq:torus-partition}) is obtained as the trace of the
operator $(-1)^{F + K_z K_w} \left( - z \right)^{K_z} \left( - w \right)^{K_w}$:
\begin{equation}
\label{eq:Witten-torus}
  P (z,w) = \Tr \left[ (-1)^{F + K_z  K_w} \left( - z \right)^{K_z} \left( - w \right)^{K_w} \right] = \sum_{p,q} (-1)^{p + q + pq}\, z^p w^q\, N_{p,q} \, .
\end{equation}
Note the presence of the factor $(-1)^{p+q+p\,q}$, which coincides with
the signature of the spin structure. Because of this, the expression
differs from the Poincar\'e polynomial for the graded complex
\begin{equation}
  \chi (z,w) =   \sum_{p,q}\, z^p\, w^q\, N_{p,q} \, .
\end{equation}
The two polynomials, on the other hand, contain the same information
and one can pass from one to the other using the identity
\begin{equation}
  2 \,\chi (z,w) = - P(z,w) + P(-z,w) + P(z,-w) + P(-z,-w) \, .  
\end{equation}
In particular, $\chi(w) = \chi(-1,w) $ and $\chi (1,1)$ is the overall
number of perfect matchings on the graph.  As already remarked at the
beginning of this section, it is worth emphasizing that the choice of
the cycles is arbitrary. This means in particular that one could have
used $K_w$ to construct the Fermion number operator and $K_z$ as an
internal grading without affecting the final result.

The generalization to graphs embedded on higher--genus Riemann surfaces
is straightforward and leads to a $\left( 2g -1 \right)$--graded complex.

A simple example to illustrate the above is given in
Appendix~\ref{sec:ex_torus}.


%% file: Gauge.tex
\section{Gauge theory interpretation: Preliminaries}
\label{sec:geometry}

As mentioned in the introduction, there exists a correspondence
between quiver gauge theories describing $D$--branes probing singular
toric surfaces and the dimer
model~\cite{Franco:2005rj,Franco:2005sm}. We will make use of it to
re--interpret the loops in terms of the quiver gauge theory and the
toric geometry.  We will quickly summarize the necessary knowledge of
quiver gauge theories and this correspondence in the following.

\subsection{The quiver gauge theory}

Consider $D3$--branes in type IIB string theory probing a Calabi--Yau
$X$. They correspond to BPS $B$--type branes given by points on
$X$. If the $D$--branes are placed at a singularity, they are expected
to decay into a collection of so--called \emph{fractional} branes.
The resulting world--volume gauge theory can be summarized in a
\emph{quiver} graph $Q_X$ as follows. Each constituent brane appearing
with multiplicity $N_i$ corresponds to a $U(N_i)$ gauge group which is
represented as a node in the quiver graph. The massless open strings
stretching between the fractional branes correspond to chiral fields
in the $(\overline{N}_i, N_j)$ representation and are depicted as
arrows pointing from the $U(N_i)$--node to the $U(N_j)$--node in the
diagram. For the theory to be anomaly free, the number of arrows
$a_{ik}$ going from node $N_i$ to node $N_k$ (in our convention
$a_{ki} = - a_{ik}$) must fulfill
\begin{equation}
  \label{eq:anomaly}
  \sum_i N_i a_{ik} = 0 \, , \hspace{2em} \forall k \, .
\end{equation}
If the ranks of all gauge groups are equal, this reduces to the number
of incoming and outgoing arrows being equal at each node.

We will be concentrating on the case of the Calabi--Yau $X$ being the cone over a singular toric surface $S$.

To fully capture the physics of the $D$--branes, they must be described in terms of the (bounded) derived category of coherent sheaves $D^\flat(X)$. A full explanation of the machinery of the derived category is beyond the scope of this paper, the reader is therefore referred to~\cite{Aspinwall:2004jr}. Luckily, we can avoid working directly in $D^\flat(X)$ by using so--called \emph{exceptional collections} of sheaves supported on a partial resolution of the singular surface $S$. In our case, these sheaves can be mostly thought of as line bundles. The exceptional collections form a convenient basis for the fractional branes and it is possible to construct the quiver from a given exceptional collection and vice versa~\cite{Aspinwall:2004vm}. The whole treatment is based on the fact that the derived category of coherent sheaves is equivalent to the derived category of quiver representations.

The open strings in the topological sector between two $B$--branes $A$
and $B$ are parametrized by $\oplus_p \mathrm{Ext}^p(A,B)$. The mass
of such an open string between two branes with the same grade is given
in string units by
\begin{equation}\label{eq:mass}
m^2=\tfrac{1}{2}(p-1).
\end{equation}
From this, we learn that the massless open strings (\emph{i.e.} the arrows in
the quiver) are counted by $\mathrm{Ext}^1$, while $\mathrm{Ext}^0$
counts the tachyons. For $p>1$, the strings are very massive and are
not seen by the quiver gauge theory.

If oriented loops appear in a quiver (which is always the case for our
quiver gauge theories because of the anomaly--freedom condition in
Eq.~\eqref{eq:anomaly}), the quiver representations become infinite
dimensional, causing the methods employed in~\cite{Aspinwall:2004vm}
to fail. Fortunately, this problem can be evaded by considering
acyclic subquivers. Some of the arrows of the gauge theory quiver
$Q_X$ are linked to the properties of the surface $S$ itself, while
the others parametrize the embedding of $S$ in the Calabi--Yau $X$.
Deleting these latter arrows removes all oriented loops. The resulting
quiver, $Q_S$, allows an ordering relation and is also called the
Beilinson or Bondal quiver in the literature.

The derived category of coherent sheaves on $S$, $D^\flat(S)$, is
equivalent to the derived category of quiver representations for
$Q_S$, denoted by $D^\flat(A-{\mathrm{mod}})$ (where $A$ is the path
algebra of $Q_S$). The basic representations $L_i$ of the path algebra
$A$ are associated to the fractional branes at the nodes of the
quiver. The arrows in the quiver are thus associated with the
$\mathrm{Ext}^1(L_i, L_j)$. The $\mathrm{Ext}^2(L_i, L_j)$ correspond
to relations in the quiver which the paths must obey. There is another
quiver representation given by the projective objects $P_i$, which are
the subspaces of $A$ generated by all paths starting at node $i$. The
sheaves in the exceptional collections we consider are the projective
objects $P_i$, which are in some sense dual to the fractional branes
$L_i$.

Let us illuminate a bit the connection between the quivers $Q_S$ and
$Q_X$ and its geometrical meaning. Consider the embedding of the
surface $S$ into $X$ via $i:\,S\hookrightarrow X$. A $D$--brane in $S$
can obviously also be regarded as a $D$--brane in $X$, \emph{i.e.} we can map
objects in $D^\flat(S)$ injectively to objects in $D^\flat(X)$ using
$i_*$.  
Since there might be more morphisms in $D^\flat(X)$ than $D^\flat(S)$,
some of the open strings between two branes in $S$ may live in $X$ but
outside of $S$. This fact is captured by the relation
\begin{equation}\label{eq:Ext_decomposition}
\mathrm{Ext}^1_X(i_*L_i,i_*L_j)=\mathrm{Ext}^1_S(L_i,L_j)\oplus\mathrm{Ext}^2_S(L_j,L_i).
\end{equation}
Thus, the decomposition of the arrows of the full quiver $Q_X$ into $\mathrm{Ext}^1$ and $\mathrm{Ext}^2$ on $S$ accounts for the embedding of $S$ in $X$.

\subsection{Exceptional collections and helices}

In this section, we collect the definitions on exceptional collections
necessary for later chapters.  An (ordered) collection of sheaves
$\{\mathcal{F}_0, \dots , \mathcal{F}_{n-1}\}$ on $S$ is called
\emph{exceptional} if 
\begin{subequations}
  \label{eq:exceptional}
  \begin{align}
    \mathrm{Ext}^0_S(\mathcal{F}_i, \mathcal{F}_i)&=\setC \\
    \mathrm{Ext}^p_S(\mathcal{F}_i, \mathcal{F}_i)&=0, \quad p\geq1 \\
    \mathrm{Ext}^p_S(\mathcal{F}_i, \mathcal{F}_j)&=0, \quad i>j.
  \end{align}
\end{subequations}
The collection is called \emph{strongly} exceptional, if
$\mathrm{Ext}^p_S(\mathcal{F}_i, \mathcal{F}_j)=0$ for $p\neq0$ and
\emph{complete} if it generates $D^\flat(S)$. Since the $D$--branes on
$S$ can wrap $0-$, $2-$ and $4-$cycles, a collection must contain $n$
sheaves to be complete, where $n=\,$sum over all Betti numbers
($=\chi(S)$). For physics, the collection being strong means that the
basis of fractional branes it generates does not contain ghost matter.

Exceptional collections can be transformed into new exceptional
collections by left and right \emph{mutations}, which represent an
action of the braid group on the set of possible collections.  On a
neighboring pair of sheaves in an exceptional collection, the
mutations act as
\begin{align}
  \label{eq:mutation}
  L_i &: (E_i, E_{i+1})\mapsto(L_{E_i}E_{i+1}, E_i),\\
  R_i &: (E_i, E_{i+1})\mapsto(E_{i+1}, R_{E_{i+1}}E_i).
\end{align}
The sheaves $L_{E_i}E_{i+1}$ and $R_{E_{i+1}}E_i$ are defined via
short exact sequences, see~\cite{Rudakov,Zaslow:1994nk}. Left and
right mutations can be seen as braiding and unbraiding.  A
\emph{helix} $\mathcal{H}$ is an infinite collection of coherent
sheaves $\{\mathcal{F}_i\}_{i\in\setZ}$ such that
\begin{itemize}
\item[(a)] for any $i\in\setZ$,
  $\{\mathcal{F}_{i+1},\dots,\mathcal{F}_{i+n}\}$ is an exceptional
  collection.
\item[(b)] $R^{n-1}\mathcal{F}_i=\mathcal{F}_{i+n}$.
\end{itemize}
The name stems from the fact that after moving $n$ steps to the right,
one is back at the original place up to a translation. $n$ is called
the \emph{period} $n$ of $\mathcal{H}$.  Each collection of the form
$\{\mathcal{F}_{i+1}, \dots,\mathcal{F}_{i+n}\}$ is called a
\emph{foundation} of the helix and $\mathcal{H}$ is uniquely determined by
each of its foundations.

\subsection{The correspondence}

Here we quickly summarize the correspondence worked out in
\cite{Franco:2005rj, Franco:2005sm}.

\begin{enumerate}
\item Start with a toric quiver gauge theory, given by a quiver graph
  and its tree--level superpotential.
\item Draw the quiver diagram on a torus, such that each of the terms
  of the superpotential corresponds to a plaquette. The resulting
  graph is called the \emph{periodic quiver}.
\item Take the graph dual of this graph (i.e. vertices become faces,
  faces become vertices and edges remain edges). Colour the vertices
  coming from the faces associated to a superpotential term with
  negative sign black; the ones coming from a positive term colour
  white. The resulting graph is bipartite and also lives on a torus.
\item Solve the dimer model on this graph by taking the Pfaffian of
  the Kasteleyn matrix. This yields the Newton polynomial. The
  associated \emph{Newton polygon} is given by taking the exponents of
  the monomials in $zw$ to be coordinates in the $(z,w)$--plane. The
  resulting polygon is exactly the toric diagram of the surface the
  quiver gauge theory comes from!
\end{enumerate}

The singular surface $S$ is specified by the corners of the toric diagram. These corner vertices each correspond to Weil divisors and therefore to line bundles. The fully resolved surface is given by adding all the points lying in the intersection of the convex hull of the polygon and the integer $\setZ^2$--lattice to the toric diagram. Each such vertex corresponds to an exceptional divisor which is blown up to resolve the singularity. The surface $S$ can also be resolved partially by blowing up only part of the exceptional divisors. Passing from one partial resolution $S'$ corresponding to the exceptional divisor $E'$ being blown up to another partial resolution $S''$ with $E''$ being blown up corresponds to a sequence of birational transformations (blow down $E'$ and blow up $E''$ instead).
 
The vertices of the toric diagram can be labeled with the multiplicities of the perfect matchings with the corresponding weight. There are certain patterns in the multiplicities which are worth pointing out:
\begin{itemize}
\item The vertices at the corners of the toric diagram always have multiplicity one, since they correspond to the highest or lowest weight states in either $z$ or $w$, which are unique.
\item The multiplicities on the edges of the graph follow the rule of Pascal's triangle (i.e. $1,\,2,\,1$ or $1,\,3,\,3,\,1$, etc.). Indeed, the polynomial consisting only of the monomials associated to the vertices of an edge has the form $(z+w)^n$, or can be brought to this form by multiplication with a prefactor $z^{n_0}w^{m_0}$ or by redefinition of the variables.
\end{itemize}
For the multiplicities of the vertices in the interior of the toric diagram, a pattern is much less obvious.

\begin{figure}
  \centering
  \begin{diagram}
    \begin{minipage}{29mm}
      \includegraphics[width=28mm]{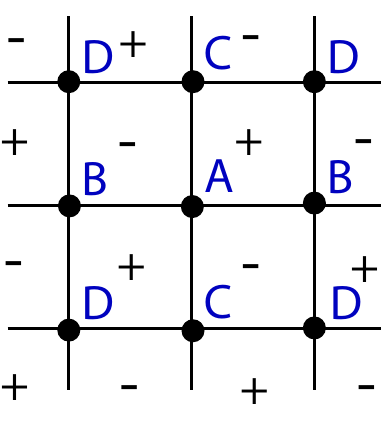}
    \end{minipage} & \rCorresponds^{\text{\hspace{1em}graph dual}} &
    \begin{minipage}{26mm}
      \includegraphics[width=25mm]{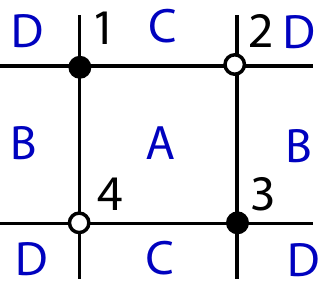}
    \end{minipage} \\
    \\
    \uTo^{\text{on the torus}} & & \dTo^{\det(K)} \\
    \\
    \begin{minipage}{23mm}
      \includegraphics[width=22mm]{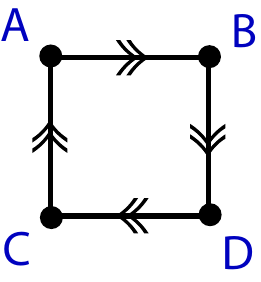}
    \end{minipage} & \rCorresponds &
    \begin{minipage}{33mm}
      \includegraphics[width=32mm]{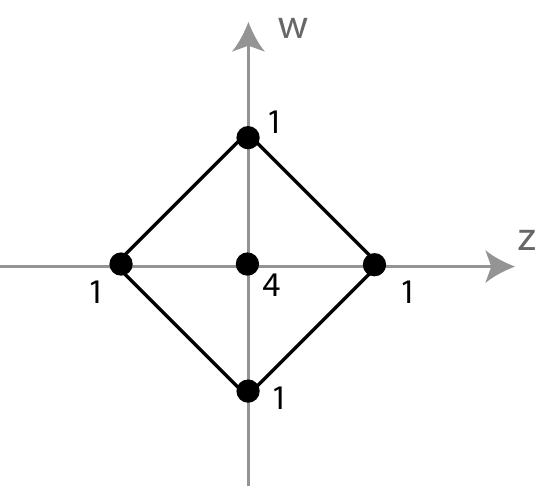}
    \end{minipage}
  \end{diagram}
\caption{Web of correspondences for one square on the torus/ quiver gauge theory on $\mathbb{F}_0$}
    \label{table:corr}
\end{figure}

We illustrate the above by the example of one square on the torus, which corresponds to the quiver gauge theory on $\mathbb{F}_0$ in Figure \ref{table:corr}.

\newarrow{Corresponds}<--->
The tree-level superpotential is given by
\begin{equation}\label{eq:supo}
W=-2\,X_{DC}X_{CA}X_{AB}X_{BD}+2\,X_{CA}X_{AB}X_{BD}X_{DC},
\end{equation}
the Newton polynomial is 
\begin{equation}
  \label{eq:partition_square}
  P (z,w) = zw \left(- \frac{1}{z}-\frac{1}{w} + 4 - z -w \right) \, .
\end{equation}


%% file: Results.tex
\section{Gauge theory interpretation: Results}
\label{sec:results}

In Section~\ref{sec:loops-as-operators}, we have seen that our fermion loop configurations can be interpreted as maps from one perfect matching to another.
Geometrically, a marked difference exists between the inner vertices of a toric diagram and those on the boundary. The inner vertices correspond to exceptional divisors and are compact. The vertices on the boundary must be subdivided into the ones at the corners of the diagram and the others. The latter are also exceptional divisors and are of the form $\mathbb{P}^1\times \mathbb{C}$ with possible blow--ups, \emph{i.e.} have a compact and a non--compact factor. The corner vertices are non--compact and have the form $\mathbb{C}\times\mathbb{C}$ plus possible blow--ups. It is therefore not surprising that it makes a difference whether we only map between internal matchings or if boundary matchings appear on one side of the map. When the size of the graph is very large, \emph{i.e.} $N,\,M\to\infty$, maps between internal matchings indeed are the generic case.

It was shown in~\cite{Hanany:2006nm} that for each internal matching
$\mathrm{PM}_i\in\{\mathrm{PM}^{int}$\}, an acyclic quiver
$Q_{\mathrm{PM}_i}$ is obtained by deleting all edges which are part
of $\mathrm{PM}_{i}$ from the dual quiver $Q_X$. We can construct an
exceptional collection of sheaves for each such
$Q_{\mathrm{PM}_i}$. We associate $Q_{\mathrm{PM}_i}$ to the quiver
$Q_{S_i}$ corresponding to the partial resolution $S_i$ of $S$
obtained by blowing up the divisor $E_i$ to which $\mathrm{PM}_{i}$
belongs. The acyclic quiver $Q_S$ represents the system as seen by an
observer living only on $S$, while from the point of view of the whole
Calabi--Yau $X$, the low--energy theory remains unchanged, regardless
of the resolution. The different $Q_{S_i}$ associated to perfect
matchings belonging to the same internal point can be seen as
different ways the gauge theory is embedded in the surface $S_i\subset
X$. The acyclic quivers associated to different resolutions $S_i$ and
$S_j$ also correspond to different decompositions of the arrows in
$Q_X$ into $\mathrm{Ext}^1_S$ and ${Ext}^2_S$, see
Eq.~(\ref{eq:Ext_decomposition}), but this difference should rather be
seen as due to the different topologies of the partially resolved
surfaces $S_i,\,S_j$ the exceptional collections are supported on.

On the quiver side, a map between internal matchings corresponds thus
to a map $f_{ji}:Q_{S_i} \mapsto Q_{S_j}$, while on the geometry side,
it corresponds to mapping from one exceptional collection to another.
These maps are \emph{always graph isomorphisms of the full quiver}
$Q_X$, in the sense that the adjacency remains unchanged and only the
labels of the nodes are permuted. This means that the physical gauge
quiver $Q_X$ remains unchanged by the loop maps, \emph{i.e.} the
quiver gauge theory is too coarse to see their effect.  At the level
of the acyclic quivers $Q_{S_i}$, the maps correspond to gauge
dualities. Geometrically, the decomposition of the arrows of $Q_X$
into $\mathrm{Ext}^1_S$ and $\mathrm{Ext}^2_S$ changes, which
corresponds to a change of embedding.

In the case of a boundary perfect matching, cycles will remain in the
reduced quiver $Q_{\mathrm{PM}_{bd}}$ which causes the quiver
representation to be infinite dimensional and the construction of the
exceptional collection to fail. This failure of the known methods to
extract useful information for these cases prevents us from giving a
satisfactory interpretation of maps involving boundary matchings.

There is one special case in which the loop states admit a direct
interpretation in the quiver gauge theory. It is the case of loops
arising from the composition of two adjacent boundary matchings. It
was noticed in~\cite{Hanany:2005ss} that this results in a so--called
\emph{zig--zag path}.  A zig--zag path is a path which turns
alternatingly maximally left and maximally right at the vertices. This
maximal turning condition results in also the edges in the dual graph
forming a closed path, which is not the case for a general path. A
closed zig--zag path on the dimer model graph $\gra$ therefore
results in a closed zig--zag path in the dual quiver, which
corresponds to a gauge invariant trace operator.

\subsection{Matchings and acyclic quivers}
\label{sec:acyclic}

The problem of finding a minimal set of arcs in a directed graph upon
the deletion of which the graph becomes acyclic is well studied in
mathematics and computer science~\cite{Karp:1972}. Such a set is
called a minimal \emph{feedback arc set} (\textsc{fas}). There exists
a precise relation between the problem of finding all \textsc{fas} of
a digraph and the problem of identifying all the perfect matchings in
its dual
graph.\\
Consider a bipartite plane graph $\gra$ with $N$ nodes and its graph
dual $\gra^\prime$. The one--cycles in $\gra^\prime$ are generated by
the plaquettes $\set{p_j}_{j=1}^N$ which correspond to the vertices of
$\gra$. Removing the edge $e_{ij}$ shared by the cycles $p_i $ and
$p_j$ breaks both cycles. A minimal \textsc{fas} is obtained as the
collection of $N/2$ edges $\set{e_{i_k j_k}}_{k=1}^{N/2}$ shared by
disjoint pairs of plaquettes $p_{i_j}$ and $p_{i_k}$. In the dual
graph $\gra$, this corresponds to selecting a set of edges joining all
the nodes and touching all of them only once. In other words, the
\textsc{fas} in $\gra^\prime $ is a perfect matching in $\gra$. The
situation is different when $\gra $ is embedded on a Riemann surface
of genus $g>0$, because in addition to the one--cycles generated by
the plaquettes, there are $2g$ equivalence classes of cycles of
non--trivial holonomy. In this case, the winding cycles in
$\gra^\prime$ are generated by the zig--zag paths. It follows that
being a perfect matching in $\gra$ is only a necessary condition for a
set of edges to be a \textsc{fas} in $\gra^\prime$. It was shown
in~\cite{Hanany:2006nm} that a perfect matching corresponding to an
internal point in the toric diagram is always a \textsc{fas}. Removing
a boundary matching, on the other hand, always preserves at least one
zig--zag path.

\bigskip

As mentioned before, the gauge quiver $Q_X$ is the graph dual of the
graph $\gra$ on which the dimer model lives.  The acyclic quiver
$Q_{S_n}$ is obtained from $Q_X$ by deleting all the edges contained
in the $n$th perfect matching
$\mathrm{PM}_n\in\{\mathrm{PM}^{int}\}$. The deleted arrows are
represented by dashed lines and correspond to relations in the quiver,
\emph{i.e.} $\mathrm{Ext}^2$s. Such an acyclic quiver allows an
ordering of its nodes such that there are no arrows between the nodes
$i$ and $j$ if $i<j$. If $Q_{S}$ has more than one starting or end
point, an ordering ambiguity arises. Indeed, $Q_{S}$ is in general a
partially ordered set which requires two ordering relations, say left
to right and bottom to top.

The maps $f^{(0,0)}_{m,n}$ corresponding to $(0,0)$--loops turn out to
have a particularly simple structure: \emph{they always correspond to
  a cyclic permutation of the nodes in one of the two ordering
  relations} (\emph{i.e.} up to re--ordering in the vertical
direction). Maps of non--trivial weight correspond to a general
permutation of the nodes. In particular, they do not preserve either
of the ordering relations. In the following, we will attempt to give
these maps an interpretation as gauge dualities of the acyclic quivers
$Q_{S_i}$.

\begin{figure}
  \centering
  \subfigure[Perfect matching $\mathrm{PM}_1$ (weight $(0,0)$) and quiver $Q_{{\mathrm{PM}}_1}$]{\includegraphics[height=40mm]{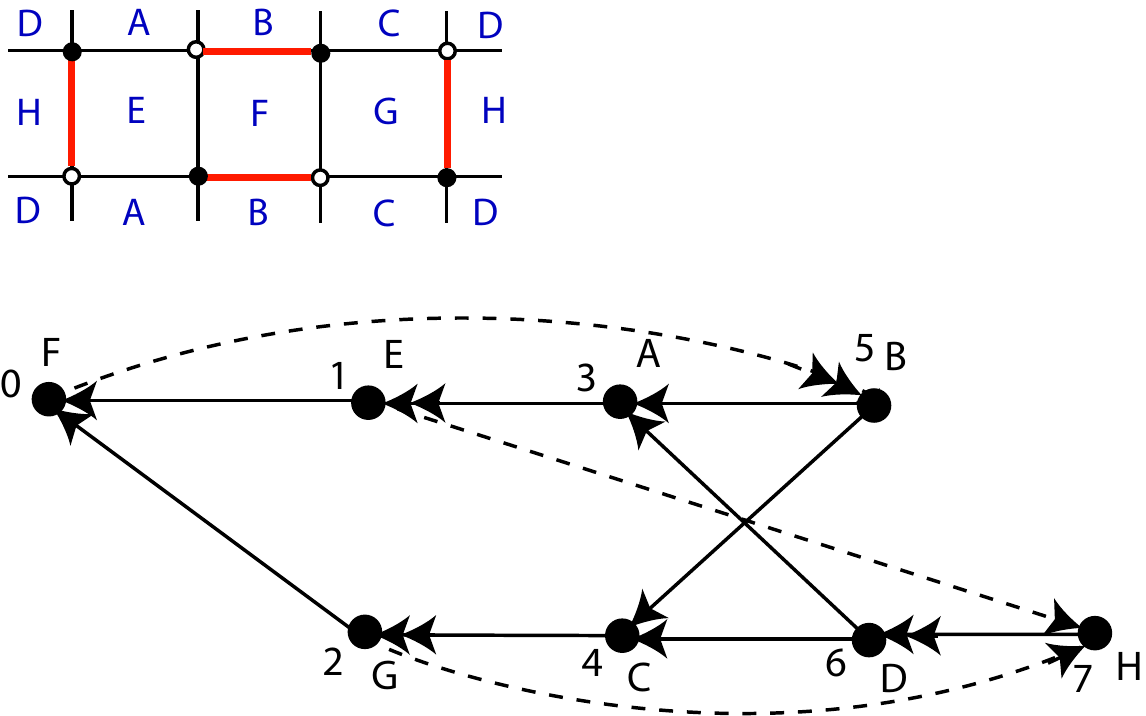}} \hfill
  \subfigure[Perfect matching $\mathrm{PM}_2$ (weight $(0,0)$) and quiver $Q_{{\mathrm{PM}}_2}$]{\includegraphics[height=40mm]{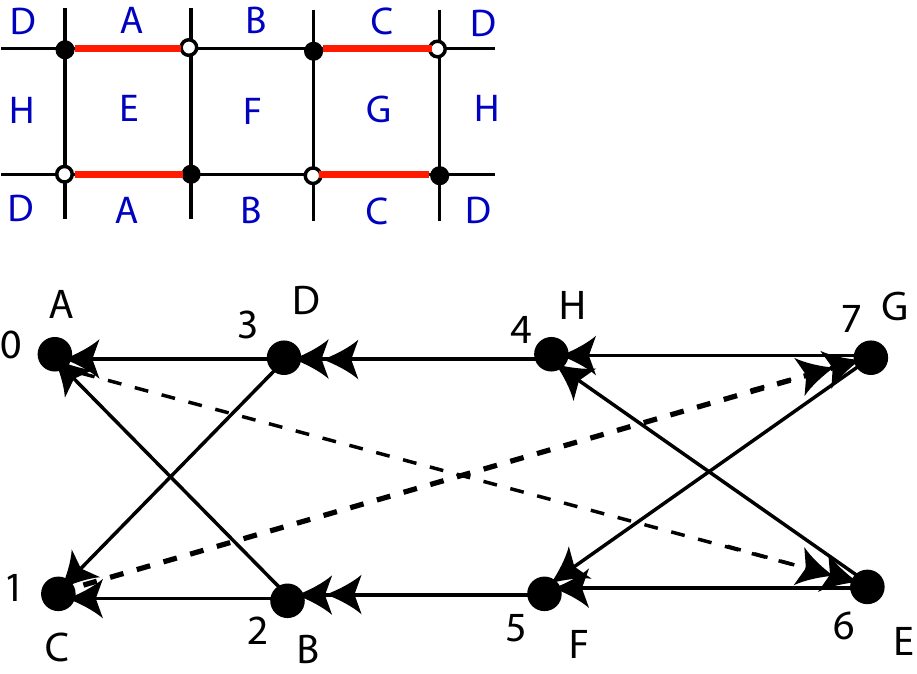}}  \vfill
  \subfigure[Perfect matching $\mathrm{PM}_3$ (weight $(1,0)$) and quiver $Q_{{\mathrm{PM}}_3}$]{\includegraphics[height=34mm]{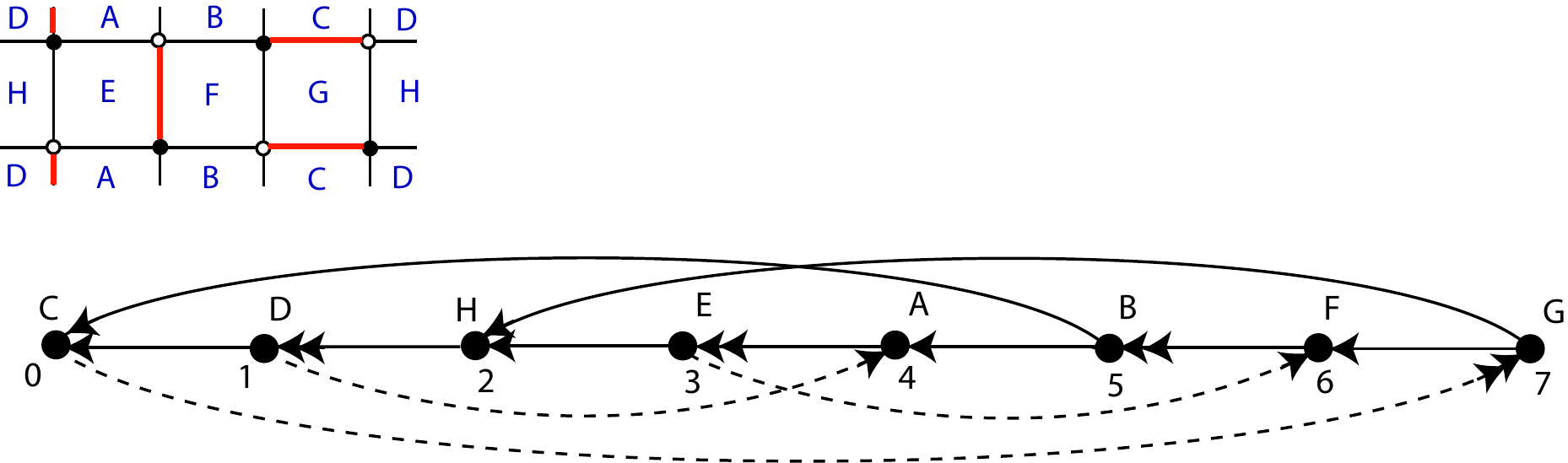}} \hfill  \caption{Example: Three inner perfect matchings and associated acyclic quivers}
  \label{fig:iso}
\end{figure}

An example of the above maps is given in Figure \ref{fig:iso}. This
example is big enough to have more than one inner point for the toric
diagram. Fig.  \ref{fig:iso}.(a) shows a perfect matching of weight
$(0,0)$ and the corresponding dual quiver with the deleted edges
depicted by dashed arrows. The map from this quiver to the one shown
in Fig. \ref{fig:iso}.(b) obviously corresponds to the cyclic
permutation $f_{2,1}^{(0,0)}:\, (F,E,G,A,C,B,D,H)\mapsto
(A,C,B,D,H,F,E,G)$. Graphically, $\mathrm{PM}_1$ can be recovered from
$\mathrm{PM}_2$ by first moving $G$ in front of all nodes its arrows
are attached to, then $E$, and finally $F$. In the process, dashed
arrows which change their direction from pointing left to pointing
right become solid and solid arrows which change from pointing left to
right become dashed.  Notice the ordering ambiguity in the vertical
direction. We could have ordered $Q_{{\mathrm{PM}}_2}$ also as
\emph{e.g.} $(C,A,B,D,F,H,E,G)$. Now consider the map from
$Q_{{\mathrm{PM}}_1}$ to $Q_{{\mathrm{PM}}_3}$, shown in
Fig. \ref{fig:iso}.(c). It corresponds to the permutation
$f_{3,1}^{(1,0)}:\, (F,E,G,A,C,B,D,H)\mapsto (C,D,H,E,A,B,F,G)$ which
has an obviously more complicated structure and does not preserve
either ordering relation.

\subsection{Seiberg dualities and graph isomorphisms}
\label{sec:doubleseiberg}

In the following, we would like to connect the graph isomorphisms of
$Q_X$ arising in the maps $f_{m,n}$ to known gauge dualities of the
acyclic quivers. Seiberg dualities~\cite{Seiberg:1994pq}, which are IR
dualities between gauge theories, are obvious candidates. It was first
remarked in~\cite{Cachazo:2001sg} and later elaborated upon
in~\cite{Aspinwall:2004vm, Herzog:2004qw} that certain Seiberg
dualities are realized as a sequence of exceptional mutations.  Here, we are
only interested in producing quivers which are subquivers of $Q_X$. In
general, this will not be the case for Seiberg dualities because the
rank of the gauge groups can change, as in
Eq.~\eqref{eq:Seiberg-ranks}.

We want to show here that the effect of performing two consecutive
Seiberg dualities on the same node is a graph isomorphism on the
physical quiver $Q_X$ and a cyclic permutation on the nodes of the
quiver $Q_S$ with respect to one of the ordering relations. Let $i_0$
be a starting node in the quiver, $A_{\text{out}}$ the set of
endpoints of the arrows starting in $i_0$, and $A_{\text{in}}$ the set
of starting points for the arrows ending in $i_0$. The effect of a
Seiberg duality on $i_0$ is the following~\cite{He:2004rn}: $i_0$ is
moved to the left of the nodes in $A_{\text{out}}$, the directions of
the arrows coming to and going from $i_0$ are reversed, and the arrows
between $A_{\text{out}}$ and $A_{\text{in}}$ are changed to preserve
consistency. More precisely, for $A \in A_{\text{out}}$ and $B \in
A_{\text{in}}$,
\begin{equation}
  \begin{cases}
    a_{A i_0}^\prime = a_{i_0 A} \, ,\\
    a_{i_0 B}^\prime = a_{B i_0} \, ,\\
    a_{A B}^\prime = a_{AB} - a_{i_0 A} a_{B i_0} \, ,
  \end{cases}
\end{equation}
where $a_{ij}$ is the number of arrows going from node $i$ to node $j$
(with the convention $a_{ji} = - a_{ij}$) in the initial quiver $Q_S$,
and $a^\prime_{ij}$ the corresponding in the quiver $Q_S^\prime$
obtained after the duality. The ranks of the gauge groups are
changed as follows:
\begin{equation}
  \label{eq:Seiberg-ranks}
  \begin{cases}
    N^\prime_j = N_j & \text{if $j \neq i_0$} \, \\
    N^\prime_{i_0} = \displaystyle{\sum_{A \in A_{\text{out}}}} a_{i_0 A} N_A - N_{i_0} \, . 
  \end{cases}
\end{equation}
In general, $N^\prime_{i_0} \neq N_{i_0}$ and the transformation
is not an isomorphism of $Q_X$. Let us now apply the 
transformation again on the same node $i_0$. The arrows joining $i_0$ with
the other nodes have been reversed, so the role of $A_{\text{in}}$ and
$A_{\text{out}}$ is interchanged:
\begin{align}
  A^\prime_{\text{out}} = A_{\text{in}} \, ,&&
  A^\prime_{\text{in}} = A_{\text{out}} \, .
\end{align}
The node $i_0$ is now moved to the left of $A^\prime_{\text{out}} =
A_{\text{in}}$ and the arrows are changed as follows:
\begin{equation}
  \begin{cases}
    a^{\prime\prime}_{B i_0} = a^\prime_{i_0 B} = a_{B i_0} \, \\
    a^{\prime \prime}_{i_0 A} = a^\prime_{A i_0} = a_{i_0 A} \, \\
    a^{\prime\prime}_{BA} = a^\prime_{BA} - a^\prime_{i_0 B} a^\prime{A i_0} = - \left( a_{AB} - a_{i_0 A} a_{B i_0} \right) -  a_{i_0 A} a_{B i_0}  = a_{BA} \, .
  \end{cases}
\end{equation}
The ranks now read:
\begin{equation}
  \begin{cases}
    N^{\prime \prime}_j = N^\prime_j = N_j & \text{for $j \neq i_0$}\\
    N^{\prime \prime}_{i_0} = \displaystyle{ \sum_{B \in A^\prime_{\text{out}}} a^\prime_{i_0 B} N^\prime_B - N^\prime_{i_0} = \sum_{B \in A_{\text{in}}} a_{B i_0} N_B  - \sum_{A \in A_{\text{out}}} a_{i_0 A} N_A + N_{i_0} = N_{i_0}},
  \end{cases}
\end{equation}
where we used the anomaly--cancellation condition around $i_0$ in
$Q_X$.

\begin{figure}
  \centering
  \subfigure[$Q_S$]{\includegraphics[height=2.7cm]{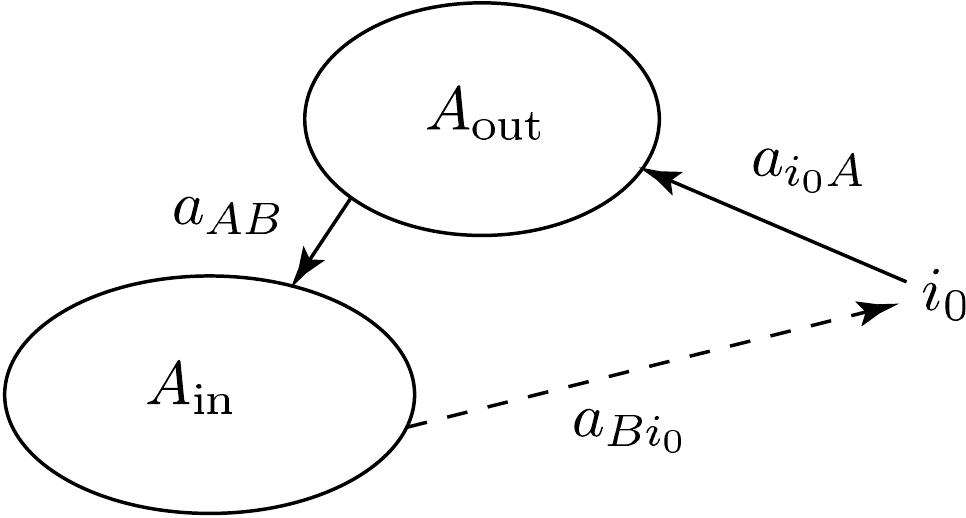}} \hfill
  \subfigure[$Q_S^\prime$]{\includegraphics[height=2.7cm]{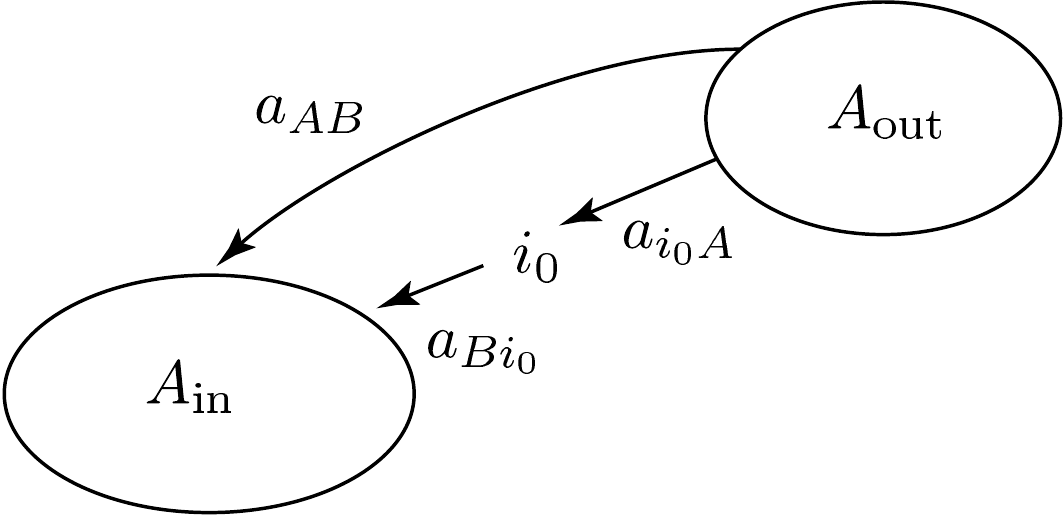}} \hfill
  \subfigure[$Q_S^{\prime\prime}$]{\includegraphics[height=2.7cm]{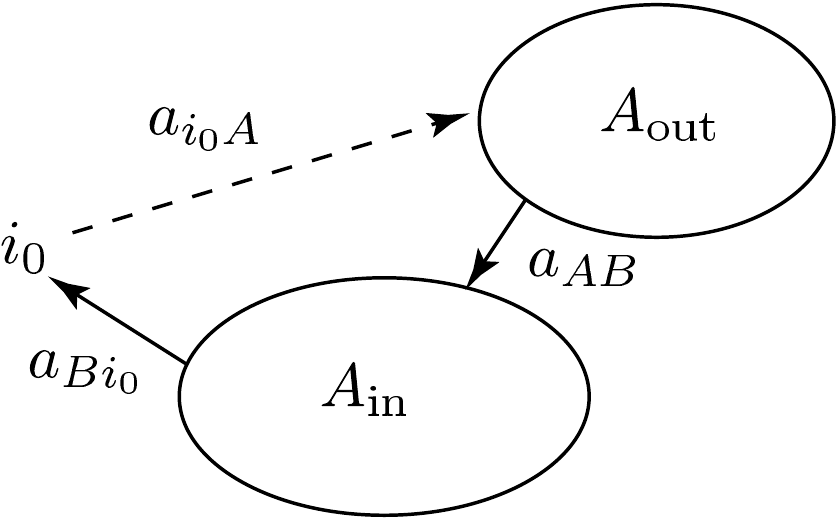}} 
  \caption{Double Seiberg duality around the node $i_0$.}
  \label{fig:Double-Seiberg}
\end{figure}

As expected, applying the Seiberg duality twice just amounts to an
isomorphism of the gauge quiver $Q^{\prime \prime}_X \sim Q_X$. On
the other hand, the node $i_0$ has been changed from a starting point
in $Q_S$ to an end point in $Q_{S^{\prime \prime}}$, and continuous
and dashed arrows from/to $i_0$ have been interchanged (see
Fig.~\ref{fig:Double-Seiberg}).

\subsection{Interpretation of the $\left(0,0\right)$--loops}
\label{sec:0-0-loops}

\subsubsection{Plaquette moves}
\label{sec:plaquette-moves}

With the results of the last section, we can now attempt an
interpretation of the $(0,0)$--loops, which form the diagonal
component of the cohomology $H^{(k,k)}$ for the complex in
Eq.~\eqref{eq:torus-complex}.

\begin{figure}
 \begin{center}
  \includegraphics[width=50mm]{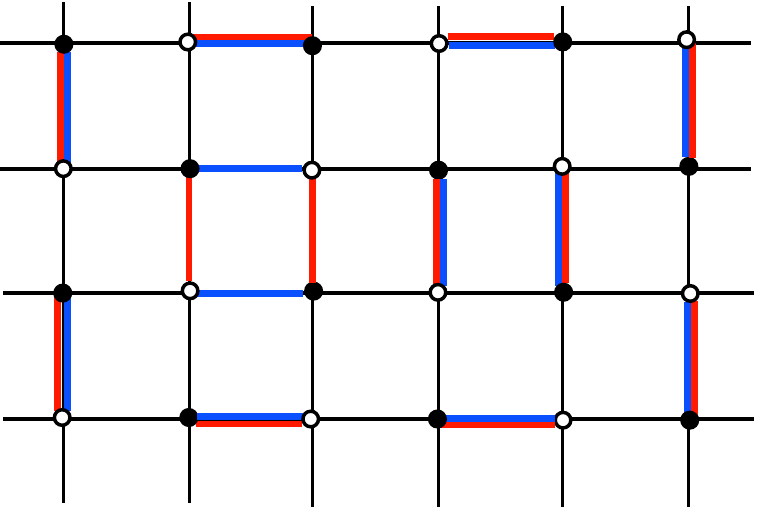} 
  \caption{Plaquette move in the square graph}
  \label{fig:plaquette}
   \end{center}
\end{figure}

Consider a loop configuration which contains the boundary of a basic plaquette and otherwise only consists of double--line perfect matchings.
We define a \emph{plaquette move} to be a map that switches between the two perfect matchings making up the above plaquette loop while leaving the rest of the configuration invariant, see Figure \ref{fig:plaquette}. This is obviously a $(0,0)$ map and corresponds to the minimal loop. It can be shown that the set of loops with windings $\left(\bar p, \bar q \right)$ fixed is connected, with the arcs
given by the plaquette moves. It follows that all maps of weight $(0,0)$ are generated by plaquette moves, \emph{i.e.} we can move between all perfect matchings of fixed weight $\left(\bar p, \bar q \right)$ by a sequence of plaquette moves.

The incoming and outgoing arrows of a node $i_0$ in $Q_X$ correspond to the edges
of the plaquette $i_0$ in the dual dimer model graph \gra. From this dual point of
view, the double Seiberg duality described in the last section is precisely a
plaquette move since the deleted edges around $i_0$ are interchanged while all
other edges are left untouched~\cite{Heckman:2006sk}. 
Since any $\left( 0, 0 \right)$--loop can be decomposed 
into a product of plaquette moves, it can be interpreted as a sequence of double Seiberg dualities on the acyclic quiver.

It was shown in~\cite{Herzog:2003zc} that a cyclic permutation of the nodes of $Q_X$ corresponds to choosing a different foundation of the same helix. So whenever we map between perfect matchings belonging to the same internal point of the toric diagram, we remain in the same helix.

Moreover, as we already pointed out, a Beilinson quiver
$Q_S \subset Q_X$ corresponds to the embedding of a (partially
resolved) singular surface $S$ in the Calabi--Yau three--fold $X$. We
can therefore read a $\left( 0,0 \right)$--loop as a map between two
possible embeddings $i : S \hookrightarrow X$.

\subsubsection{Relation to crystal melting and black holes}
\label{sec:melting}

The dimer configurations of the infinite hexagon graph are in one--to--one correspondence to the configurations of a melting crystal corner~\cite{Okounkov:2003sp}. It was shown in~\cite{Heckman:2006sk} that they also correspond to certain \textsc{bps} black holes given by $D$--branes wrapping collapsed cycles in $\setC^3/\IZ_n\times\IZ_n$ in the large $n$ limit. The black holes are generated by non--geometry changing flop transitions, which in the dimer picture are identified with so--called \emph{local dimer moves}. The local dimer moves of~\cite{Heckman:2006sk} correspond precisely to the plaquette moves of the hexagon graph. The $S_R^2$ transformations on the fractional branes to which their local dimer moves are related, correspond to the double Seiberg dualities described in Section 
\ref{sec:doubleseiberg}. Thus, when specializing to the hexagon graph, the $(0,0)$--loops are the non--geometry changing flop transitions in $\setC^3/\IZ_n\times\IZ_n$ which generate those \textsc{bps} black hole charge configurations that are parametrized by the 3d crystal melting configurations. On a plane graph, everything is captured by the $(0,0)$--loops since no non--trivial winding states exist.

\subsection{Interpretation of the loops with non--trivial winding}
\label{sec:winding_loops}

As mentioned before, the maps corresponding to loops of non--trivial
winding result in a general permutation of the nodes of the gauge
quiver $Q_X$. Geometrically, we map from one partial resolution of $S$
to another, which is a birational map. The associated exceptional
collections produce equivalent derived categories.

Unlike the $(0,0)$--maps, loops with non--trivial winding can in
general not be written as a sequence of Seiberg dualities. We have
seen that plaquette moves cannot generate winding loops, therefore we
cannot write the latter as a sequence of double Seiberg
dualities. Single Seiberg dualities in general change the ranks of the
gauge groups, see Eq.~\eqref{eq:Seiberg-ranks}, and thus do not lead to
a graph isomorphism of the gauge quiver $Q_X$.

Since the full quiver gauge theory remains unchanged, the action of
the weight changing maps $f^{(p,q)}_{m,n}$ on the acyclic quivers can
still be seen as a generalized gauge duality on the theory on the
(partially resolved) surface $S$.

What is the geometrical meaning of this? Morally speaking, an
exceptional collection forms a basis for the physical branes. All
partial resolutions lead to the same quiver gauge theory in $X$, so at
the level of the derived category, the associated exceptional
collections are equivalent. Passing from one to the other corresponds
to a change of basis. We have seen that with the $(0,0)$--maps, we
change between foundations of a helix. This is a very particular
change of basis, which preserves the order of the sheaves in the
collection and only amounts to tensoring the canonical bundle a
certain number of times to all sheaves in the collection. We have seen
on the other hand, that the general permutation of nodes which takes
place in a weight--changing map does not preserve the order of the
sheaves in the collection. This means that we no longer remain in the
same helix. At the level of the sheaves, this map is realized by a
general sequence of mutations.


%% file: Conclusions.tex
\section{Conclusions}
\label{sec:conclusions} 

In this note, we relate a statistical mechanical system, the dimer
model on a torus, to a system of lattice quantum field theory,
\emph{i.e.} the massless free fermion. The loop configurations
obtained after a diagrammatic expansion of the fermion determinant
(Sec.~\ref{sec:loop-expans-ferm}) can be interpreted as states in
\textsc{sqm} after employing categorification techniques inspired by
Khovanov's work (Sec.~\ref{sec:categorification}). The Newton
polynomial of the dimer model becomes the generalized Euler
characteristic of a co--chain complex (Sec.~\ref{sec:sqm_torus}).  The
states with vanishing overall winding become the supersymmetric ground
states of the system. Since all loop states except the double line
perfect matchings are paired, taking the Witten index of this system
again counts the number of perfect matchings
(Eq.~\eqref{eq:Witten-torus}).\\
In the following, we make use of the dimer model--quiver gauge theory
correspondence to give the loop configurations an interpretation in
yet another picture. Interpreting the loops as maps from one perfect
matching to another (Sec.~\ref{sec:loops-as-operators}), maps between
internal perfect matchings associated to the same point in the toric
diagram become changes of foundation of the associated helix and
sequences of double Seiberg dualities in the acyclic quiver
(Sec.~\ref{sec:doubleseiberg}). When specializing to the honeycomb
graph, the plaquette moves which generate all $(0,0)$--maps correspond
directly to the \textsc{bps} black hole configurations which are parametrized
by crystal melting configurations (Sec.~\ref{sec:melting}).

Mapping between matchings associated to different internal points is a
more severe change and cannot be achieved as a sequence of Seiberg
dualities. In the language of exceptional collections, we map from one
helix to another.

In Table~\ref{table:dictionary}, we summarize the results of this
paper by giving a dictionary between the loop states, the fermion
states, the gauge theory and the geometry.

\bigskip

From this point, one can go to several different directions for
further research.
\begin{itemize}
\item The dimer model admits a description as a free fermion model
  based on a transfer matrix approach
  (see~\cite{lieb:2339,Sutherland:1968,alet:041124} and
  also~\cite{Okounkov:2003sp}) which, at first sight, is different
  from the one we introduced in this note. It would be very
  interesting to clarify the link between these two descriptions.
\item Furthermore, it would be interesting to use a more general
  definition for the Dirac operator on a graph than the one in
  Eq.~\eqref{eq:Lattice-Dirac}, to prove the equivalence to the dimer
  model completely independently of the type of graph and its
  representation. Indeed, this could provide an elegant solution to
  the problem of choosing a Dirac operator on a random graph that
  depends only on the the adjacency.
\item The algebraic structure we introduce in
  Sec.~\ref{sec:categorification} also deserves further
  investigation. A better understanding of the complex and the mapping
  to \textsc{sqm} might indeed cast a new light on the dual quiver
  gauge theory.
\item A final word must be spent on the geometric interpretation in
  Sec.~\ref{sec:geometry}. To our knowledge, the meaning of the
  boundary perfect matchings (and, correspondingly, of the
  infinite-dimensional representations of the path algebras) in terms
  of derived category is not yet clear. This is an obvious
  prerequisite to a complete understanding of the geometrical meaning
  of all the loop states that appear in the theory.
\end{itemize}

The study of the dimer model, which due to its combinatorial nature provides a relatively easy playground, has proven to be very fruitful thanks to its manifold connections to other physical systems.
We believe that these inter--relations have not yet been fully exploited and deserve further attention.

\begin{landscape}
  \begin{table}[]
    \begin{center}
      \begin{tabular}{cccp{4.5cm}p{4cm}}
        \toprule
        & \textbf{Loop state} & \textbf{Gauge theory on $S$}&
        \begin{minipage}{4.4cm}
          \center \textbf{Geometry}
        \end{minipage} &
        \begin{minipage}{3.9cm}
          \center \textbf{Free fermion}
        \end{minipage}
        \\ \midrule
        \rowcolor[gray]{.95}
        \cellcolor[gray]{1} &  double line $\mathrm{PM}$ & Identity &
        \begin{minipage}{4.4cm}
          \center Identity
        \end{minipage}
        & time evolution with no local charge creation\\[.3cm]
        \cellcolor[gray]{1} $\{\mathrm{PM}^{int}\} \to \{\mathrm{PM}^{int}\}$ &$(0,0)$&Seiberg duality  & change of foundation in helix & time evolution in which no net charge is created\\[0.3cm]
        \rowcolor[gray]{.95}
        \cellcolor[gray]{1} & $(p,q)\neq (0,0)$ & general gauge duality &
        general sequence of mutation, change of helix, birational
        transformation
        &
        initial conditions with gradient $q$ and creation of net charge $p$
        \\   \midrule
        $\{\mathrm{PM}^{int}\} \leftrightarrow  \{\mathrm{PM}^{bd}\}$ &   no corner $ \mathrm{PM}$ & ? & birational transformation & \hspace{1.8cm}" \\  \midrule
        \rowcolor[gray]{.95}
        \cellcolor[gray]{1}$\set{\mathrm{PM}^{bd}} \to \set{\mathrm{PM}^{bd}}$ &
        \begin{minipage}{3.3cm}
          \tabtop adjacent boundary matchings. Zig-zag path
        \end{minipage}
        & \begin{minipage}{3.5cm}
          gauge-invariant multi--trace operator
        \end{minipage}
        &  & \hspace{1.8cm}" \\[.7cm]
        & no corner PM  & 
        & birational transformation & \hspace{1.8cm}" \\ \bottomrule
      \end{tabular}
      \caption{Dictionary}
      \label{table:dictionary}
    \end{center}
  \end{table}
\end{landscape}


%% file: Appendix.tex
\section{Perfect matchings in the operator formalism}
\label{sec:perf-match-oper}

In Section~\ref{sec:loop-expans-ferm}, we found the loop states to be
combinations of two perfect matchings. It would be interesting to
recover the same result in terms of the operator formalism of
Section~\ref{sec:loops-as-fermionic}. To do so, we start by
distinguishing between even and odd plaquettes. We will say that a
plaquette with coordinates $\mathbf{z} = (z, w) $ is \emph{even}
(resp. \emph{odd}) if $z + w $ is even (resp. odd). We use the
definition of the height function from the superposition of two
perfect matchings $\mathrm{PM}_1-\mathrm{PM}_2$ given in Section
\ref{sec:preliminaries}. If an odd plaquette contains either an upward
arrow on the left side, a downward arrow on the right side, a right
pointing arrow on the bottom side or a left pointing arrow on the
upper side, it belongs to the first perfect matching $\mathrm{PM}_1$,
in the converse case, it belongs to $\mathrm{PM}_2$. For an even
plaquette, the arrows are reversed. The strategy to construct an
operator which only preserves the operators corresponding to one of
the perfect matchings, say $\mathrm{PM}_1$, when acting on a state
$\ket{\Psi}$ is simple. Consider \emph{e.g.} all the black nodes. For each of
them add all the annihilators corresponding to the arrows belonging to
$\mathrm{PM}_2$. If $\ket{\Psi}$ is an allowed state, only one
operator is paired for each node which gives a $1$ as a result of the
anticommutation, while all the others will just annihilate the
state. This results in only the operators belonging to $\mathrm{PM}_1$
surviving and so yielding the result we are looking for. We can define
the operators ${PM}_1$ and ${PM}_2$ explicitly as follows:
\begin{subequations}
  \label{eq:loop-splitting}
  \begin{align}
    {PM}_1 &= \prod_{z = 1}^{N} \prod_{w = 1}^{M} \frac{1 + \left( -1 \right)^{z + w}}{2} \left[ a(z,w) + d(z, w) + b(z, w - 1) + c(z-1,w) \right] \, ,\\
    {PM}_2 &= \prod_{z = 1}^{z} \prod_{w = 1}^{w} \frac{1 +
      \left( -1 \right)^{z + w}}{2} \left[ b(z,w) + c(z,
      w) + a(z, w - 1) + d(z-1,w) \right] \, .
  \end{align}
\end{subequations}
A state $\ket{\Psi} = \kket{m_1} \otimes \kket{m_2}$ is decomposed
into its two constituent perfect matchings by their action:
\begin{equation}
  PM_i \ket{\Psi} = \kket{m_i} \, , \hspace{1em} i = 1,2 \, .
\end{equation}
Now we are able to give a consistency condition for a state to
correspond to a loop appearing in the expansion of the fermionic
action. The idea is that the state should be decomposable into two
matchings, which both touch each node once. Let us therefore define
\begin{subequations}
  \begin{align}
    P_1 (z,w) &= \textstyle{\frac{1 + \left( -1 \right)^{z + w}}{2}}
    \left[ N_a(z,w) + N_d(z, w) + N_b(z, w - 1) +
      N_c(z-1,w)
      - 1\right] \, , \\
    P_2 (z,w) &= \textstyle{\frac{1 + \left( -1 \right)^{z + w}}{2}}
    \left[ N_b(z,w) + N_c(z, w) + N_a(z, w - 1) +
      N_d(z-1,w) - 1\right] \, .
  \end{align}
\end{subequations}
An allowed state $\ket{\Psi}$ must satisfy the local conditions
\begin{equation}
  \label{eq:state-consistency}
  P_1 (\mathbf{z}) \ket{\Psi} = P_2 (\mathbf{z}) \ket{\Psi} = 0 \, , \hspace{1em} \forall \mathbf{z} \, .  
\end{equation}

The action of the height operators in Eq.~\eqref{eq:global-height} is
still defined on the perfect matching states $\kket{\Psi}$. To be
consistent with the constructions in Section \ref{sec:preliminaries},
we introduce a sign to distinguish between $\kket{m_1}$ and
$\kket{m_2}$:
\begin{subequations}
\label{eq:matching-charges}
  \begin{align}
    H_z \kket{m_i} &= \left(-1 \right)^{i+1} \left[ \sum_{\zeta =
        1}^{N} N_a ( \zeta, \bar w) - N_b ( \zeta, \bar w)
    \right] \kket{m_i} \, , \\
    H_w \kket{m_i} &= \left(-1 \right)^{i+1} \left[ \sum_{\omega =
        1}^{M} N_c ( \bar z, \omega) - N_d ( \bar z, \omega)
    \right] \kket{m_i} \, .
  \end{align}
\end{subequations}
We find that each loop can be identified by four charges,
corresponding to the eigenvalues of the winding operators acting on
the two perfect matchings. The relation between the winding of a loop
and the weights of the constituent perfect matchings is
\begin{subequations}
  \begin{align}
    H_z \ket{\Psi} &= H_z \kket{m_1} - H_z \kket{m_2} \, ,\\
    H_w \ket{\Psi} &= H_w \kket{m_1} - H_w \kket{m_2} \, .
  \end{align}
\end{subequations}
Extending the definition of the height function to perfect matchings
requires some care, since the equality in
Eq.~\eqref{eq:height-equality} does not hold for perfect matchings. A
possible way out consists in symmetrizing the sum of the two
expressions and define
\begin{equation}
  h (z, w ) \kket{m_i} = \frac{\left(-1 \right)^{i+1}}{2} \left[ \sum_{\zeta = 0}^{z} N_a ( \zeta, w) - N_b ( \zeta, w) + \sum_{\omega = 0}^{w} N_c ( z, \omega) - N_d ( z, \omega) \right] \kket{m_i} \, . 
\end{equation}
In this way, one recovers once more
\begin{equation}
  h (\mathbf{z}) \ket{\Psi} = h (\mathbf{z}) \kket{m_1} - h (\mathbf{z}) \kket{m_2} \, .
\end{equation}

\section{Generating functions}
\label{sec:generating-functions}

In this section, we give the generating function which contains the full
information of the fermion loop gas, \emph{i.e.} all the
loop states. The dimer model partition function and the Witten index
can be recovered from it.

\subsection{Generating function for the fermion loops on the cylinder}
\label{sec:generating-function}

We compute the generating functions
for the loops defined as
\begin{equation}
  G (q, z, y) = \Tr \left[ q^H z^K y^F \right] \, ,  
\end{equation}
which keeps track of all the information. Expanding the trace one finds
an explicit expression depending on the matching degeneracies $N_k$:
\begin{equation}
  \label{eq:generating-cylinder}
  G( q, z, y ) = \sum_{k=0}^N N_k\, y^{2k} z^k + \left( 1 + y \right) \left\{ \frac{1}{2}\sum_{k=0}^N N_k \left( N_k - 1 \right) y^{2k} z^k + \sum_{l=1}^N q^l \sum_{k=l}^N N_k\, N_{k-l}\, y^{2k} z^k \right\} \, . 
\end{equation}

All the physical information about the system is contained in the
generating function $G$. For example:
\begin{itemize}
\item the Euler number for the complex in Eq.~\eqref{eq:SQM-complex}
  is $\,\chi = \Tr \left[ \left( -1 \right)^F \right] = G(1,1,-1)$ ;
\item the Poincar\'e polynomial for the complex is $\,\chi (z) = \Tr \left[ \left( -1
    \right)^F z^k \right] = G(1,z,-1)$\,;
\item the total number of perfect matchings is $N = \Tr \left[ \left(
      -1 \right)^{K+F} \right] = G(1,-1,-1)$~;
\item the multiplicity of weight-$k$ matchings is $N_k = \frac{1}{k!}
  \left. \frac{\del^k}{\del z^k } G(q,\frac{z}{y},y)\right|_{q=1,z=0,y=-1}$~;
\item the dimension of the $k$-th chain group is $\,\dim C_k =
  \frac{1}{k!}  \left. \frac{\del^k}{\del y^k}
    G(q,z,y)\right|_{q=1,z=1,y=0}$~;
\item the number of loops with winding number $h$ is $L_h =
  \frac{1}{h!}  \left. \frac{\del^h}{\del q^h}
    G(q,z,y)\right|_{q=0,z=1,y=1}$~.
\end{itemize}

\subsection{Generating function for the fermion loops on the torus}
\label{sec:generating-function-1}

Let us now extend the definition of generating function for the loops
to graphs embedded on the torus. Following the formula for the
cylinder in Eq.~\eqref{eq:generating-cylinder} we define:
\begin{equation}
  G (\mathbf{q}, \mathbf{z}, y) = \Tr \left[ \mathbf{q}^{\mathbf{H}} \mathbf{z}^{\mathbf{K}} y^F \right] \, ,
\end{equation}
where the notation $\mathbf{q}^{\mathbf{H}} $ is a shortcut for
$\mathbf{q}^{\mathbf{H}} \doteq q_z^{H_z} q_w^{H_w}$ and
$\mathbf{z}^{\mathbf{k}} \doteq z^{k_z} w^{k_w}$. Expanding in the loop basis we find:
\begin{multline}
  G ( \mathbf{q}, \mathbf{z}, y ) %
  = \sum_{\mathbf{k}=(0,0)}^{\mathbf{N}} N_{\mathbf{k}}
  y^{2 k_z} \mathbf{z}^\mathbf{k} + \frac{ 1 + y}{2} \left\{
     \sum_{\mathbf{k} = (0,0)}^{\mathbf{N}}
     N_{\mathbf{k}} \left( N_{\mathbf{k}} - 1 \right) y^{2k_z} \mathbf{z}^{\mathbf{k}} +  \right. \\
   \left. + \sideset{}{'}\sum_{\mathbf{l} = (0,0)}^{\mathbf{N}} \mathbf{q}^{\mathbf{l}} \sum_{\mathbf{k}
       = \mathbf{l}}^{\mathbf{N}} \left( N_{\mathbf{k}}
       N_{\mathbf{k}-\mathbf{l}} + N_{k_z (k_w-l_w)} N_{(k_z-l_z)k_w}
     \right)  y^{2 k_z} \mathbf{z}^{\mathbf{k}} \right\} \, ,
\end{multline}
where the primed sum $\displaystyle{\sideset{}{'}\sum_{\mathbf{l}}}$
means that the value $\mathbf{l}=(0,0)$ is to be omitted. All the
physics of the system is encoded into $G ( \mathbf{q}, \mathbf{z}, y
)$. In particular:
\begin{itemize}
\item the Euler number for the complex in Eq.~\eqref{eq:torus-complex}
  is $\,\chi(w)=\Tr \left[ (-1)^F w^{K_w} \right]$ \\ $= G((1,1),(1,w),-1)$~;
\item the Poincar\'e characteristic for the complex is $ \chi (z,w) = \Tr \left[ (-1)^F
    \mathbf{z}^{\mathbf{k}} \right]$\\  $= G((1,1),\mathbf{z},-1)$~;
\item the number of perfect matchings with height $\mathbf{k}$ is
  $N_{\mathbf{k}} = \frac{1}{\mathbf{k}!}
  \left. \frac{\del^{\abs{\mathbf{k}}}}{\del \mathbf{z}^{\mathbf{k}}}
    G \right|_{\mathbf{q}= (1,1), \mathbf{z} = (0,0), y = -1}$~, where
  we introduced the notation $\mathbf{k}! \doteq k_z! k_w !$~,
  $\abs{\mathbf{k}} \doteq k_z + k_w $~, and
  $\frac{\del^{\abs{\mathbf{k}}}}{\del \mathbf{z}^{\mathbf{k}}} \doteq
  \frac{\del^{k_z + k_w}}{\del z^{k_w} \del w^{k_w}}$~;
\item the dimension of the $C_{\mathbf{k}}$ chain group is $\dim
  C_{\mathbf{k}} = \frac{1}{\mathbf{k!}}
  \left. \frac{\del^{\abs{\mathbf{k}}}}{\del y^{k_z} \del w^{k_w}} G
  \right|_{\mathbf{q}= (1,1), \mathbf{z} = (0,1), y = 1}$~;
\item the number of loops with winding $\mathbf{h}$ is $L_{\mathbf{h}}
  = \frac{1}{\mathbf{h}!} \left. \frac{\del^{\abs{\mathbf{h}}}}{\del
      \mathbf{q}^{\mathbf{h}}} G \right|_{\mathbf{q}= (0,0),
    \mathbf{z} = (1,1), y = 1}$~.
\end{itemize}

\section{Example: One square on the torus}
\label{sec:ex_torus}

To illustrate the construction given in Section~\ref{sec:sqm_torus}, we now present the smallest possible example
on the torus, consisting only of one square.  The eight possible
matchings and their quantum numbers are shown in Figure
\ref{fig:ex_matchings}.
\begin{figure}[h!]
  \begin{center}
    \includegraphics[width=110mm]{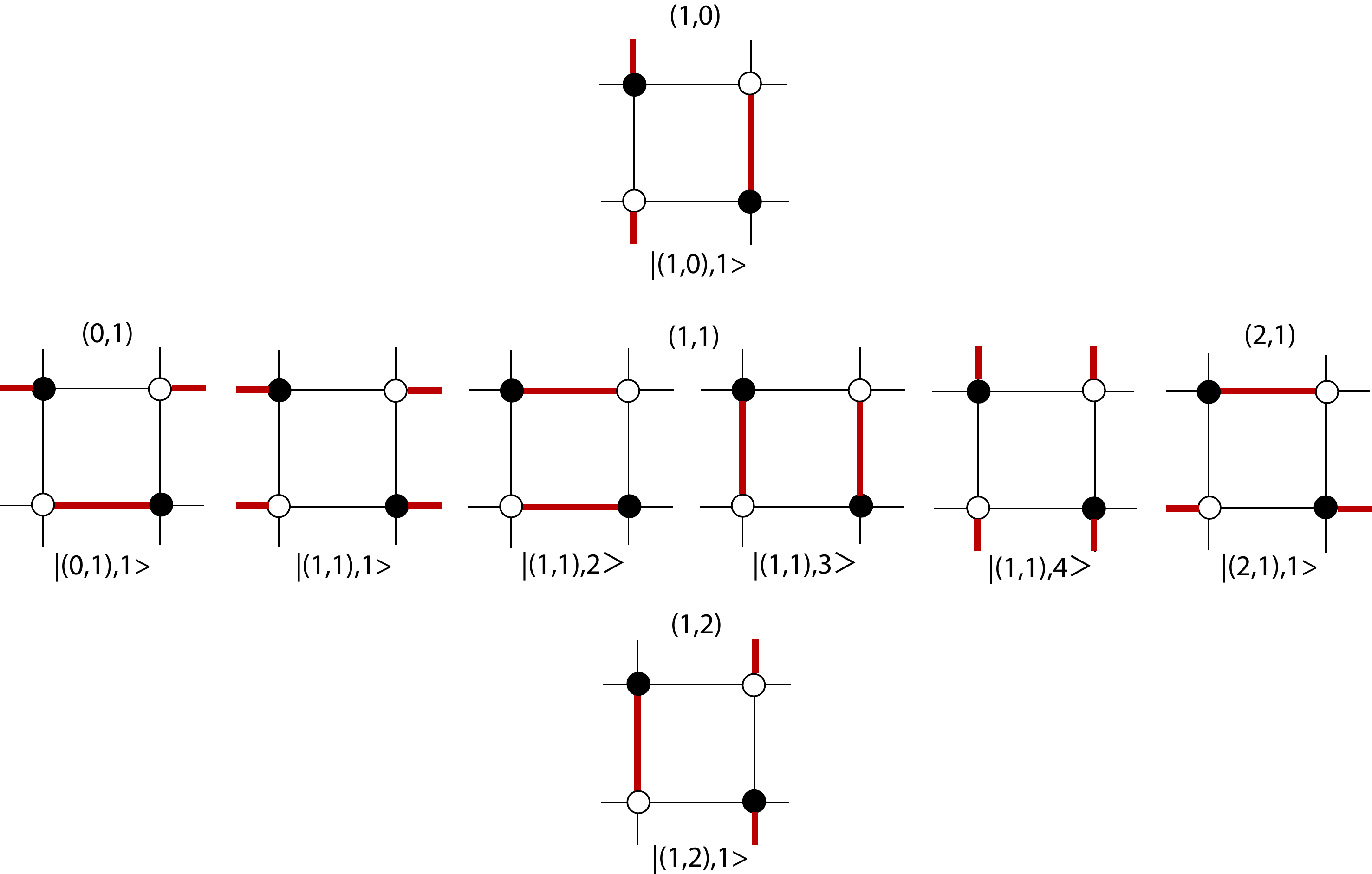}
    \caption{Matchings for the square on the torus}
    \label{fig:ex_matchings-T}
  \end{center}
\end{figure}
They combine into 64 loop configurations, which are, sorted into their
co--chain groups according to total winding number, depicted in Table~\ref{chaingroupsT}.

The partition function of the dimer model is the one given in (\ref{eq:partition_square}),
the generating function for this example reads
\begin{eqnarray}
  G( \mathbf{q}, \mathbf{z}, y ) &=& w + y^2 z + 4 w y^2 z + w^2 y^2 z + w y^4 z^2 +\nonumber \\
 && +\left( 1 + y \right) \left[ \left( 6 w y^2 z + 4 q_z + 4 q_w +  q_zq_w \right) w y^2 z + \left( 4 q_w + q_w^2 + q_zq_w \right) w^2 y^2 z + \right. \nonumber\\
 && \left. + \left( 4 q_z + q_z^2 + q_zq_w \right) w y^4 z^2 + q_zq_w w^2 y^4 z^2 \right].
\end{eqnarray}

\afterpage{%
  \begin{center}
    \begin{longtable}{ccccccc}
      \caption[Co--chain groups for the square on the torus]{Co--chain groups
        for the square on the torus}
      \label{chaingroupsT}\\
      
      \toprule {$\,K_w\,$}& $C^0$ &{$C^1$} 
      &{$C^2$}& {$C^3$} & {$C^4$} & {$C^5$}  \cr \midrule
      \endfirsthead

      \multicolumn{7}{l}
      {\tablename\ \thetable{} -- continued from previous page} \\
      \toprule
      $\,K_w\,$& $C^0$ &{$C^1$} & {$C^2$}& {$C^3$} & {$C^4$} & {$C^5$} \cr
      \midrule
      \endhead
      
      \midrule \multicolumn{7}{r}{{Continued on next page}} \\ \bottomrule
      \endfoot
      
      \bottomrule
      \endlastfoot
      
      \rowcolor[gray]{.95} 2& -& - & \pb{$\ket{1,2,1;0,1,1}_+$  \includegraphics[height=10mm]{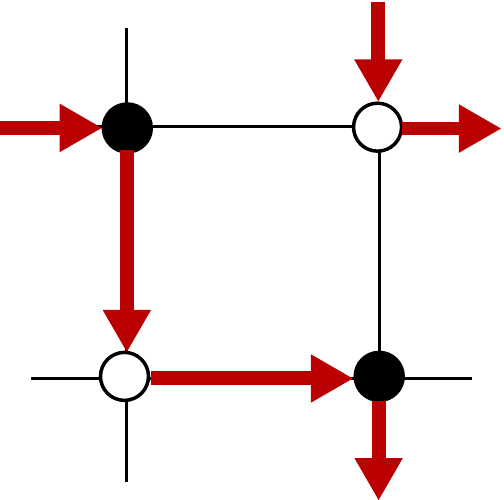}} & \pb{$\ket{1,2,1;0,1,1}_{-}$  \includegraphics[height=10mm]{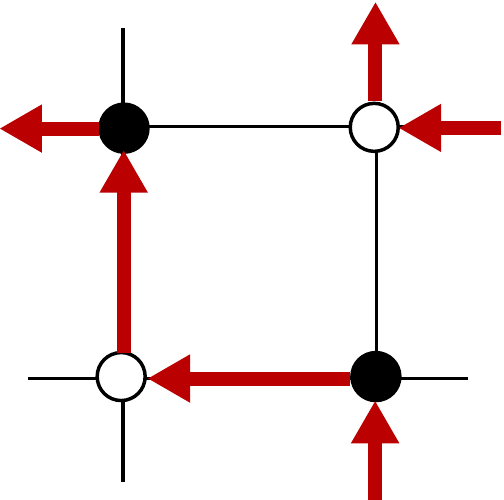}} & - & - \cr
      2 & -& - & \pb{$\ket{1,2,1;1,0,1}_+$ \includegraphics[height=10mm]{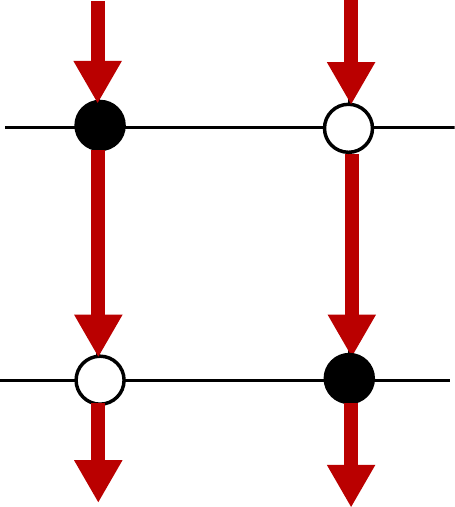}}&\pb{$\ket{1,2,1;1,0,1}_{-}$ \includegraphics[height=10mm]{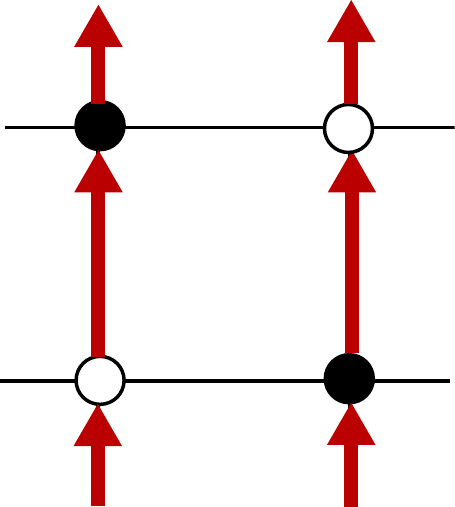}} & - & - \cr
      \rowcolor[gray]{.95} 2 & -& - &  \pb{$\ket{1,2,1;1,1,4}_+$ \includegraphics[height=10mm]{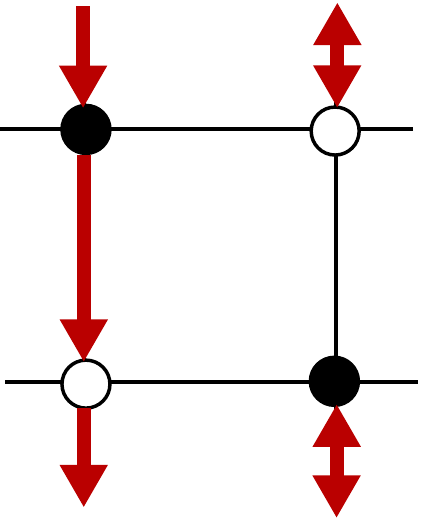}} &\pb{$\ket{1,2,1;1,1,4}_{-}$ \includegraphics[height=10mm]{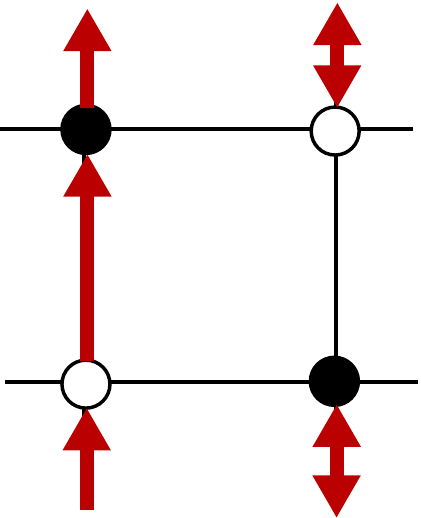}} & - & - \cr
      2 & -& - &  \pb{$\ket{1,2,1;1,1,3}_+$ \includegraphics[height=10mm]{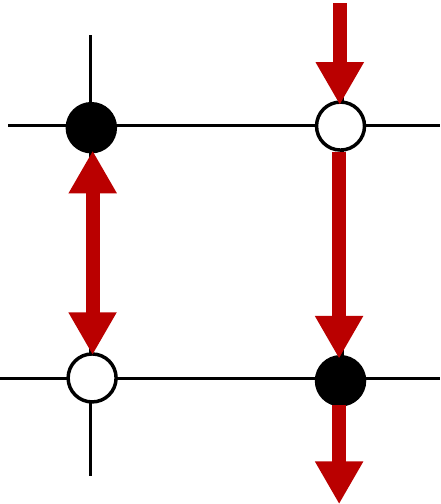}}&\pb{$\ket{1,2,1;1,1,3}_{-}$ \includegraphics[height=10mm]{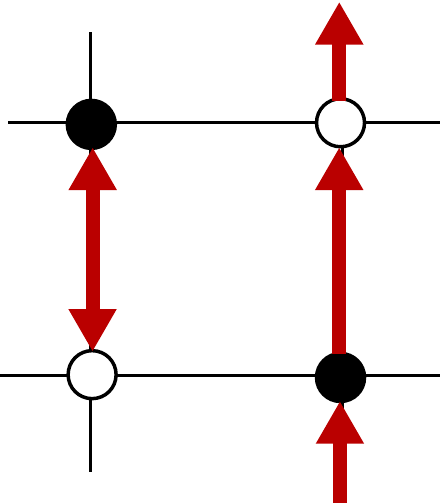}} & - & - \cr
      \rowcolor[gray]{.95} 2 & -& - &  \pb{$\ket{1,2,1;1,1,2}_+$  \includegraphics[height=10mm]{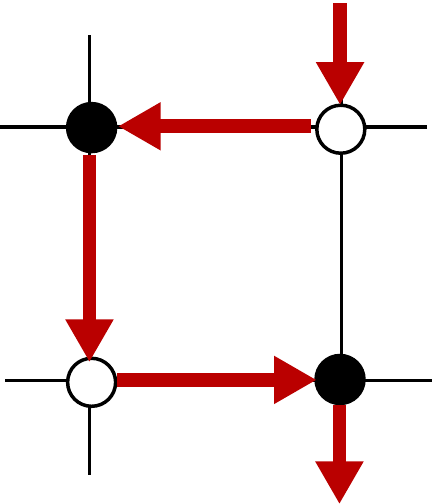}}&\pb{$\ket{1,2,1;1,1,2}_{-}$ \includegraphics[height=10mm]{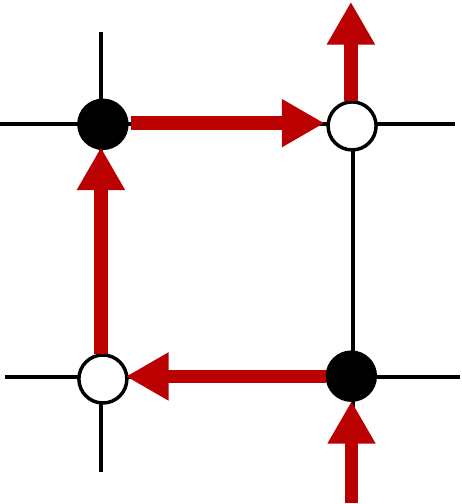}} & - & - \cr
      2 & -& - &  \pb{$\ket{1,2,1;1,1,1}_+$ \includegraphics[height=10mm]{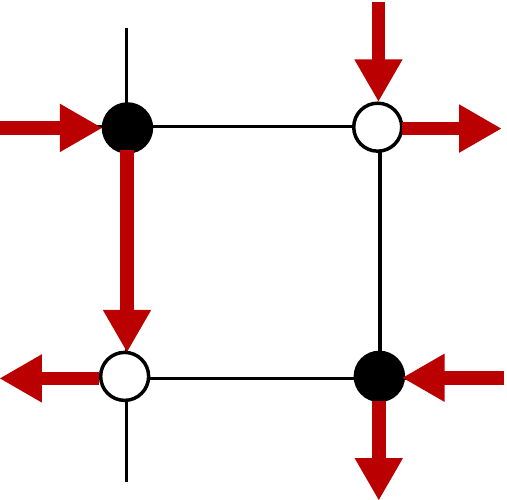}} &\pb{$\ket{1,2,1;1,1,1}_{-}$ \includegraphics[height=10mm]{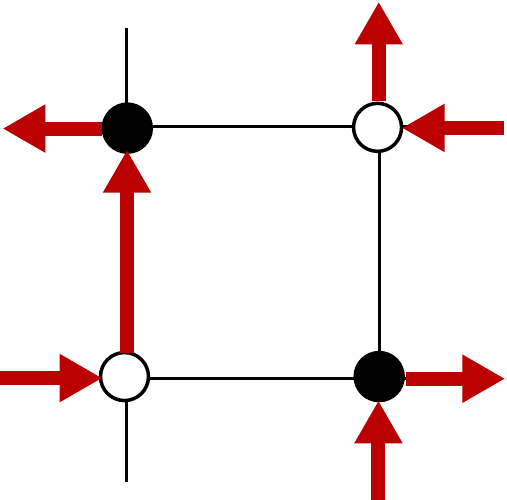}} & - & - \cr
      \rowcolor[gray]{.95} 2 & -& - &  \pb{$\ket{1,2,1;1,2,1}$ \includegraphics[height=9mm]{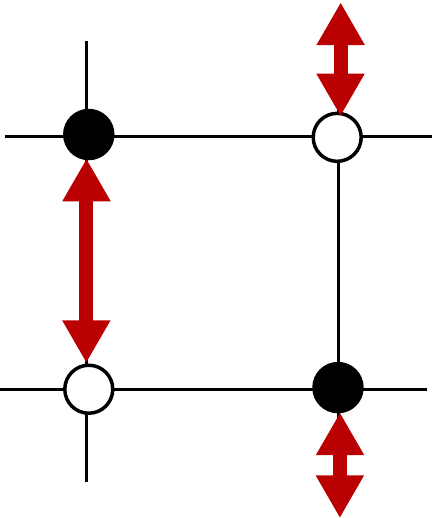}}&-&\pb{$\ket{2,1,1;1,2,1}_+$ \includegraphics[height=10mm]{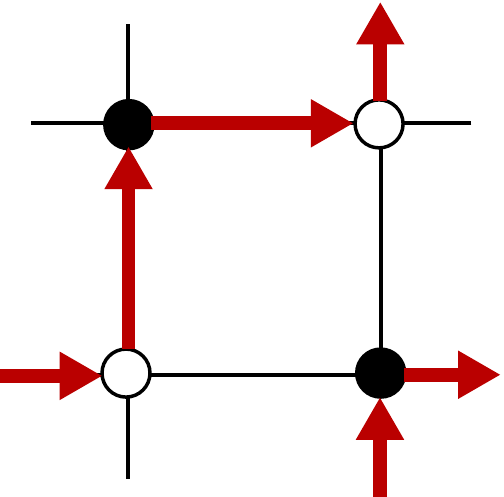}} &\pb{$\ket{2,1,1;1,2,1}_-$ \includegraphics[height=10mm]{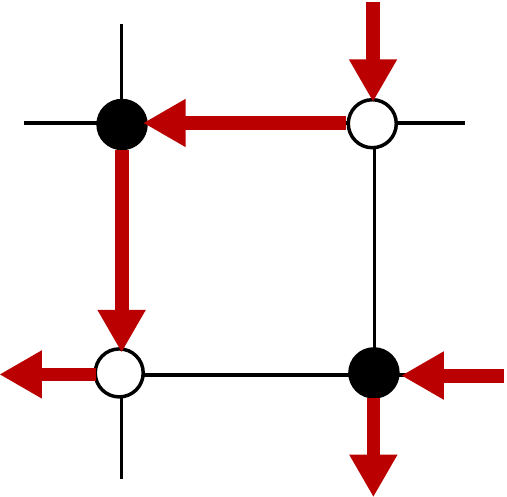}} \cr
      \midrule
      1 & -& - & \pb{$\ket{1,1,4;0,1,1}_+$ \includegraphics[height=10mm]{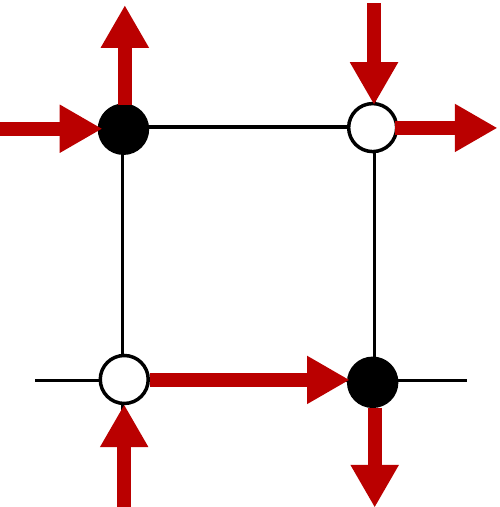}} & \pb{$\ket{1,1,4;0,1,1}_-$ \includegraphics[height=10mm]{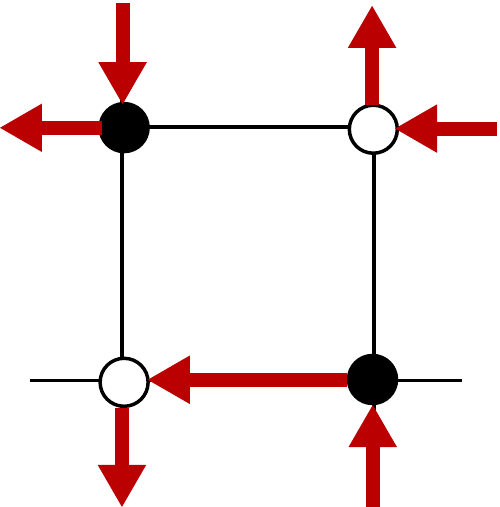}} & - &-\cr
      \rowcolor[gray]{.95} 1 & -& - & \pb{$\ket{1,1,3;0,1,1}_+$ \includegraphics[height=10mm]{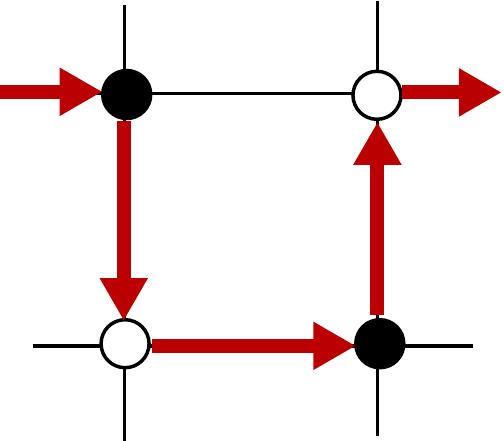}}& \pb{$\ket{1,1,3;0,1,1}_-$ \includegraphics[height=10mm]{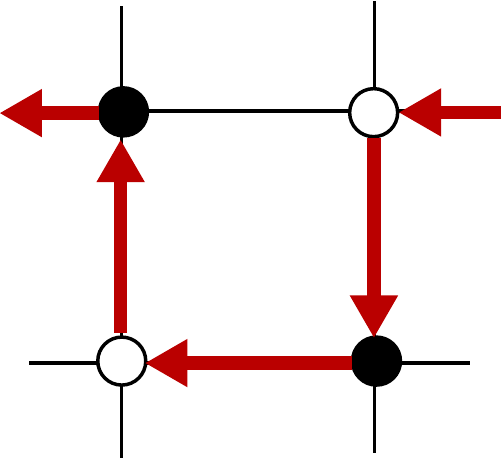}}&  -&-\cr
      1 & -& - & \pb{$\ket{1,1,2;0,1,1}_+$ \includegraphics[height=10mm]{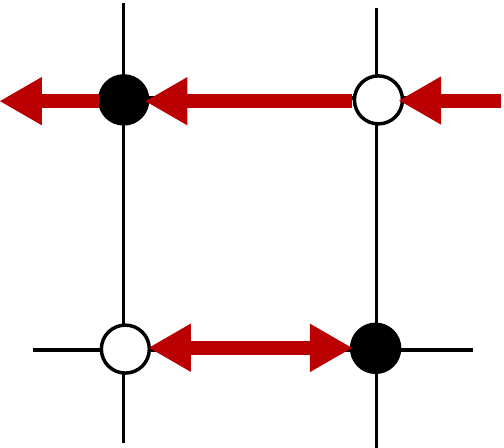}} & \pb{$\ket{1,1,2;0,1,1}_-$ \includegraphics[height=10mm]{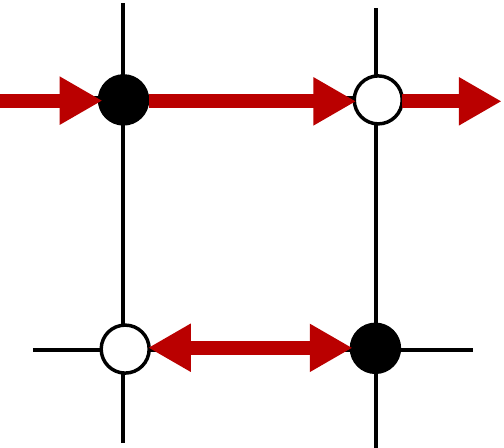}} &  -&-\cr
      \rowcolor[gray]{.95} 1 & -& - & \pb{$\ket{1,1,1;0,1,1}_+$ \includegraphics[height=10mm]{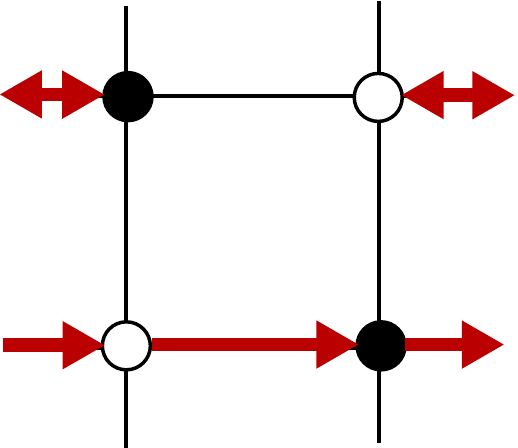}}& \pb{$\ket{1,1,1;0,1,1}_-$ \includegraphics[height=10mm]{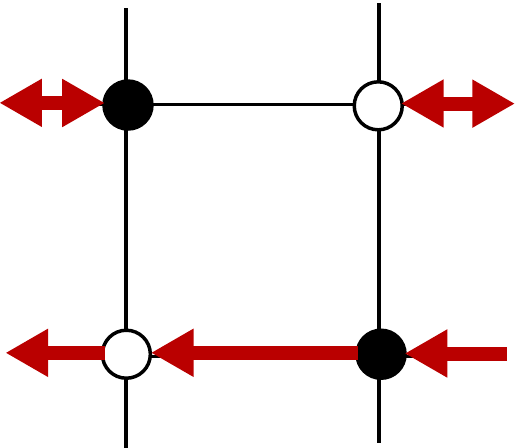}}&  -&-\cr
      1 & -& - & \pb{$\ket{1,0,1;0,1,1}_+$ \includegraphics[height=10mm]{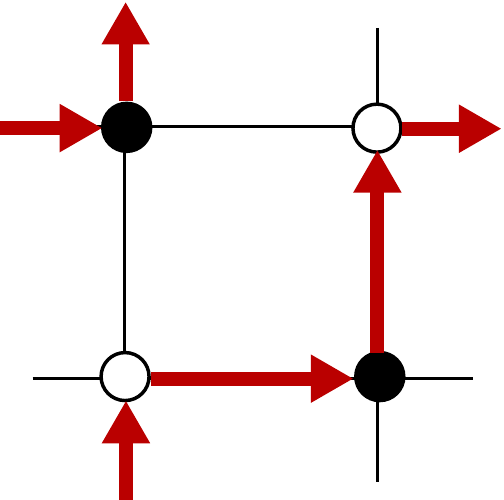}}& \pb{$\ket{1,0,1;0,1,1}_-$ \includegraphics[height=10mm]{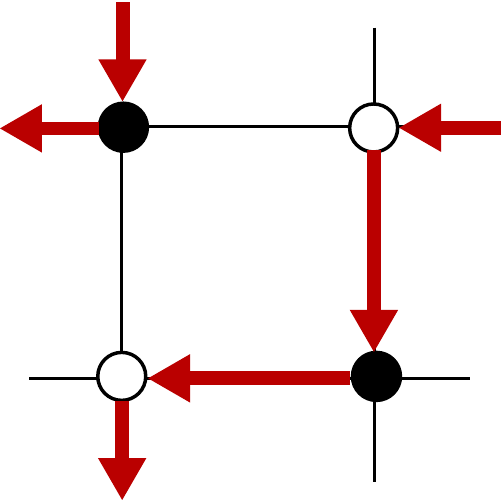}}&  -&-\cr
      \rowcolor[gray]{.95} 1 & -& - & \pb{$\ket{1,1,4;1,0,1}_+$ \includegraphics[height=10mm]{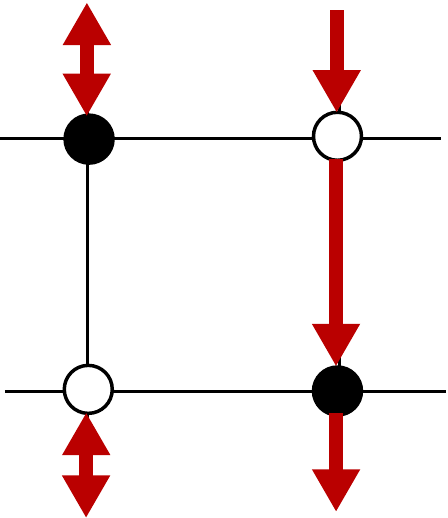}}& \pb{$\ket{1,1,4;1,0,1}_-$ \includegraphics[height=10mm]{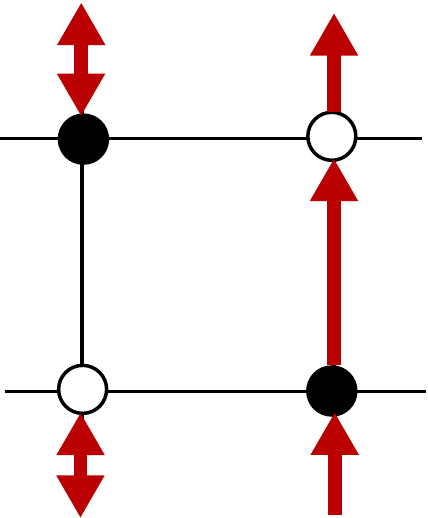}}&  -&- \cr
      1 & -&  - & \pb{$\ket{1,1,3;1,0,1}_+$ \includegraphics[height=10mm]{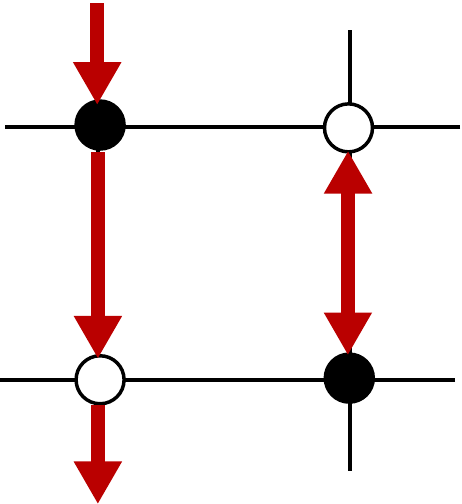}}& \pb{$\ket{1,1,3;1,0,1}_-$ \includegraphics[height=10mm]{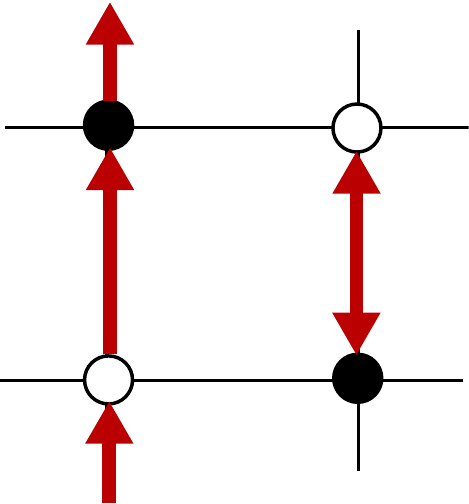}}&  -&-\cr
      \rowcolor[gray]{.95} 1 & -& - & \pb{$\ket{1,1,2;1,0,1}_+$ \includegraphics[height=10mm]{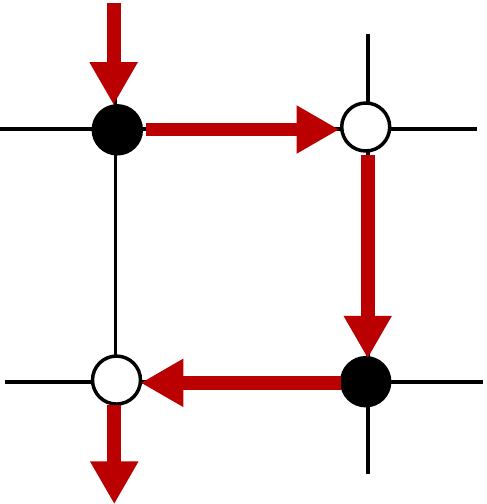}} & \pb{$\ket{1,1,2;1,0,1}_-$ \includegraphics[height=10mm]{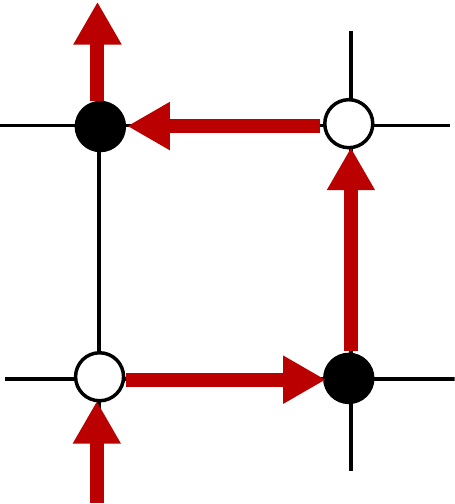}}&  -&-\cr
      \rowcolor[gray]{.95} 1 & -& - & \pb{$\ket{1,1,1;1,0,1}_+$ \includegraphics[height=10mm]{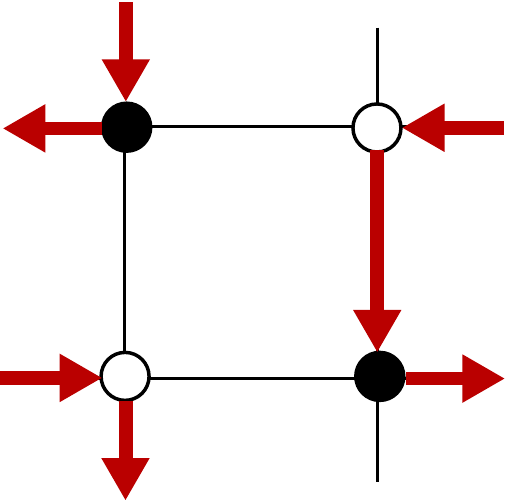}} & \pb{$\ket{1,1,1;1,0,1}_-$ \includegraphics[height=10mm]{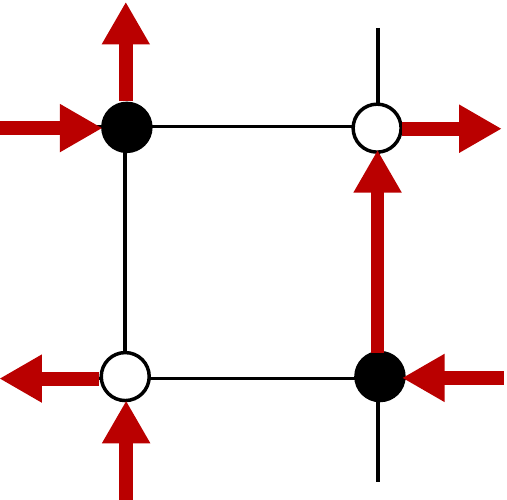}}&  -&-\cr
      1 & -& - & \pb{$\ket{1,1,4;1,1,3}_+$ \includegraphics[height=10mm]{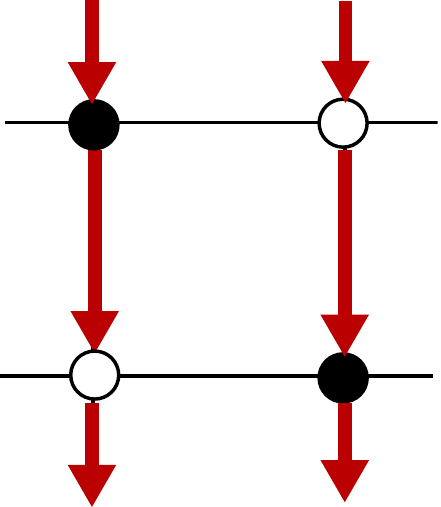}}& \pb{$\ket{1,1,4;1,1,3}_-$ \includegraphics[height=10mm]{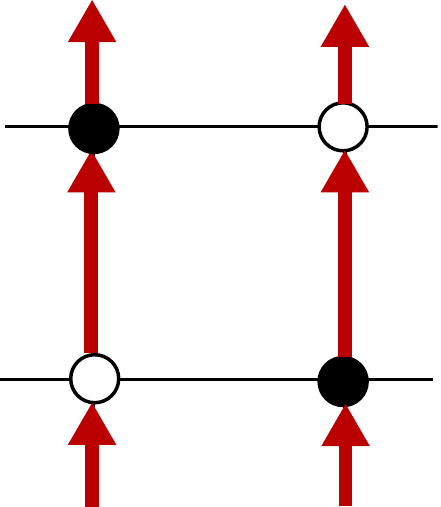}}&  -&-\cr
      \rowcolor[gray]{.95} 1 & -& - & \pb{$\ket{1,1,4;1,1,2}_+$ \includegraphics[height=10mm]{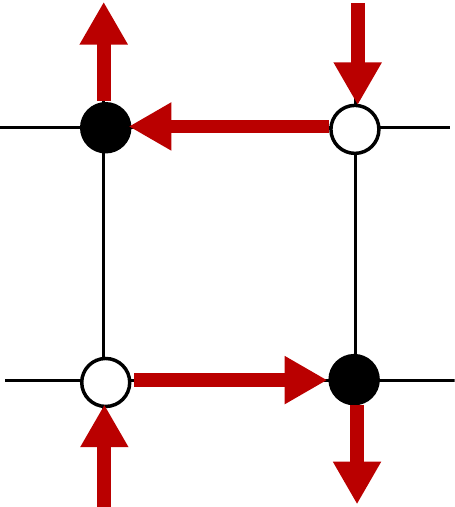}}& \pb{$\ket{1,1,4;1,1,2}_-$ \includegraphics[height=10mm]{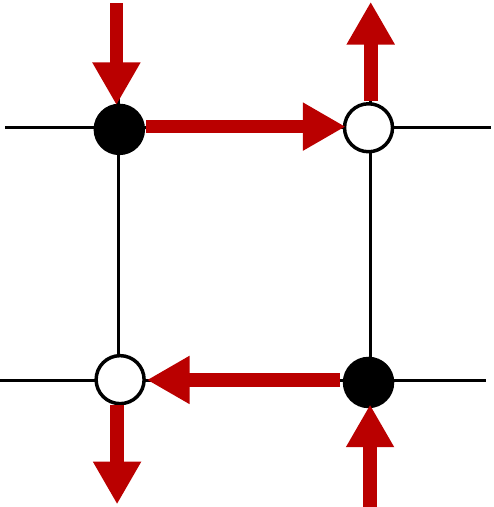}}&  -&-\cr
      1 & -& - & \pb{$\ket{1,1,4;1,1,1}_+$ \includegraphics[height=10mm]{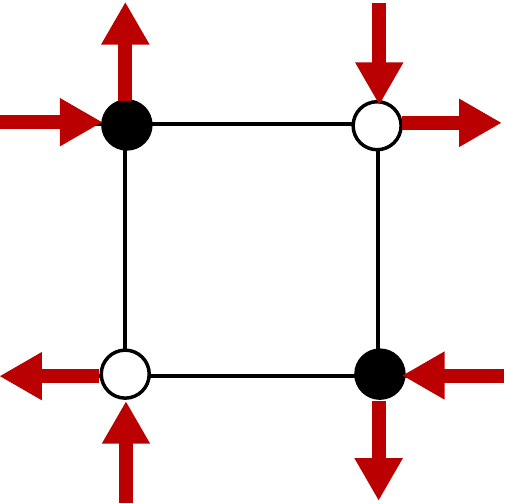}} & \pb{$\ket{1,1,4;1,1,1}_-$ \includegraphics[height=10mm]{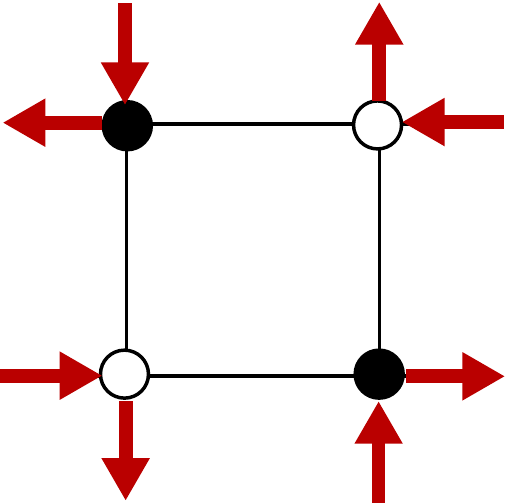}}&  -&-\cr
      \rowcolor[gray]{.95} 1 & -& - & \pb{$\ket{1,1,3;1,1,2}_+$ \includegraphics[height=10mm]{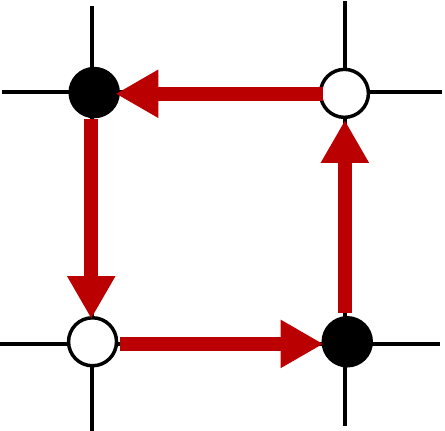}}& \pb{$\ket{1,1,3;1,1,2}_-$ \includegraphics[height=10mm]{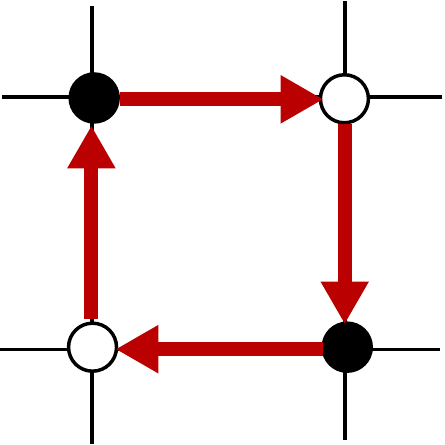}} & \pb{$\ket{2,1,1;0,1,1}_+$ \includegraphics[height=10mm]{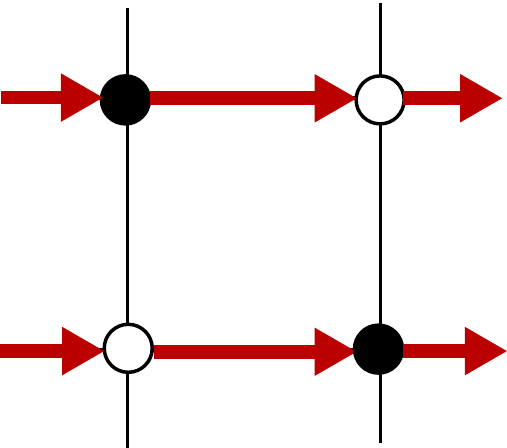}}  &\pb{$\ket{2,1,1;0,1,1}_-$ \includegraphics[height=10mm]{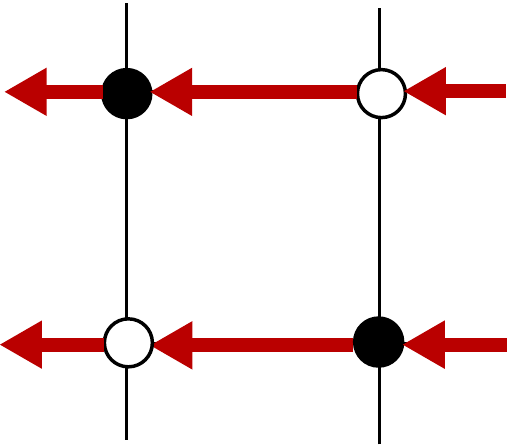}} \cr
      1 & -& - & \pb{$\ket{1,1,3;1,1,1}_+$ \includegraphics[height=10mm]{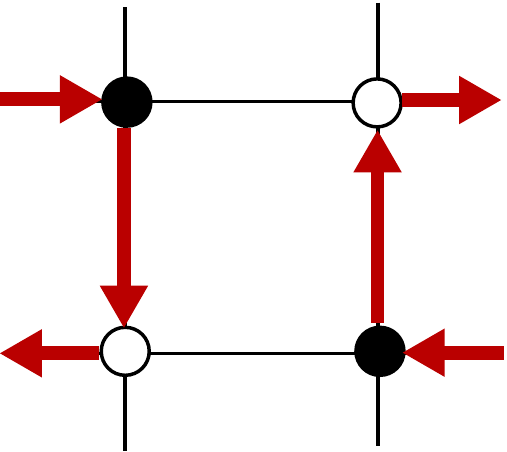}} & \pb{$\ket{1,1,3;1,1,1}_-$ \includegraphics[height=10mm]{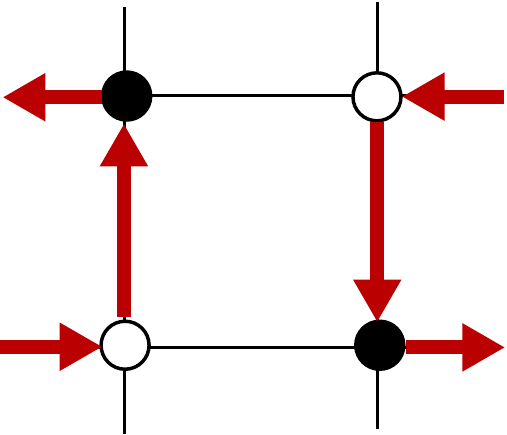}} & \pb{$\ket{2,1,1;1,0,1}_+$ \includegraphics[height=10mm]{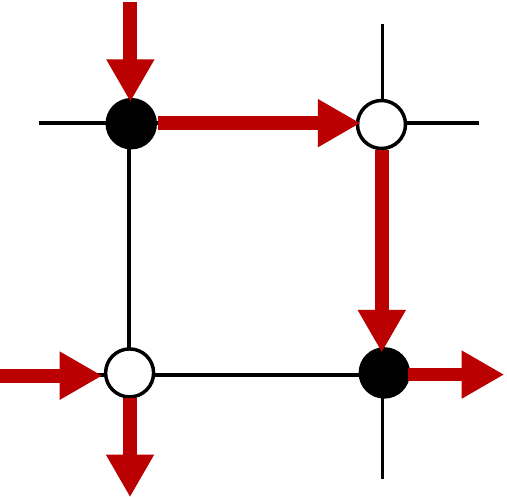}} & \pb{$\ket{2,1,1;1,0,1}_-$ \includegraphics[height=10mm]{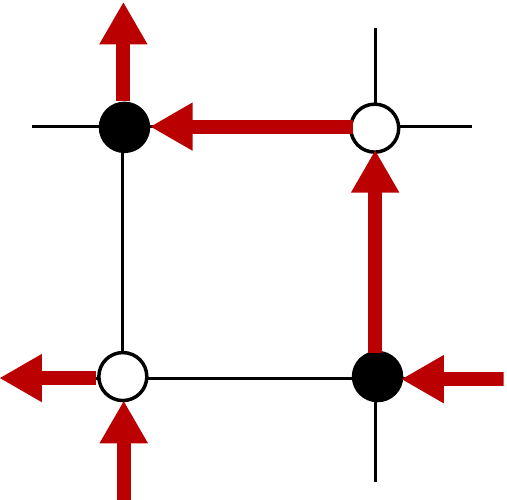}} \cr
      \rowcolor[gray]{.95} 1 & -& - & \pb{$\ket{1,1,2;1,1,1}_+$ \includegraphics[height=10mm]{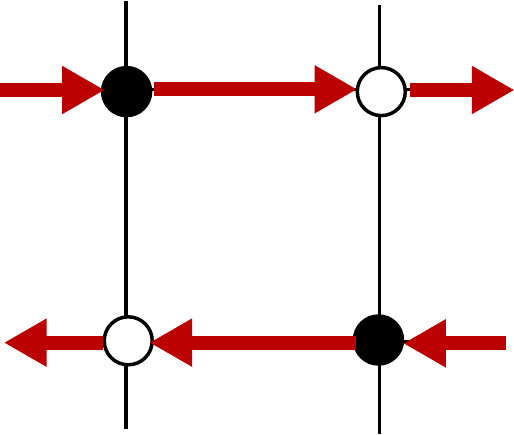}} & \pb{$\ket{1,1,2;1,1,1}_-$ \includegraphics[height=10mm]{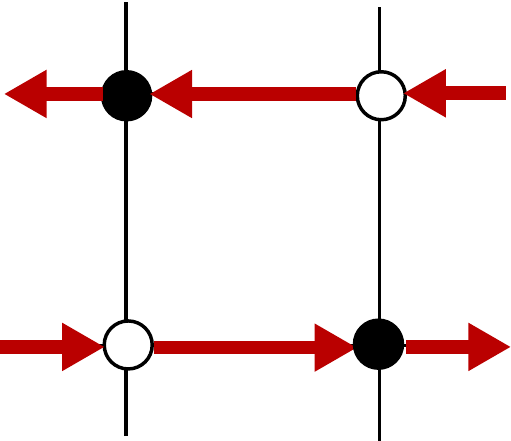}} &\pb{$\ket{2,1,1;1,1,4}_+$ \includegraphics[height=10mm]{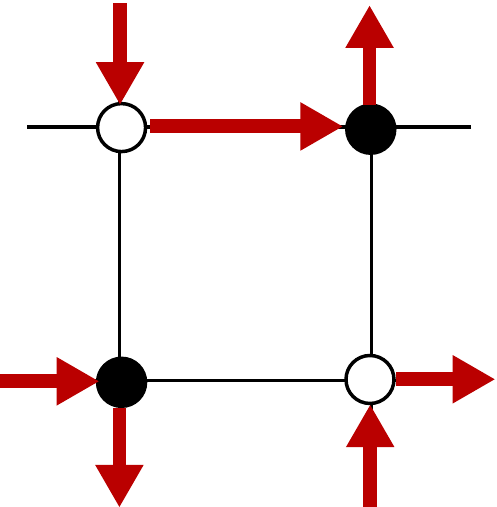}}  &\pb{$\ket{2,1,1;1,1,4}_-$  \includegraphics[height=10mm]{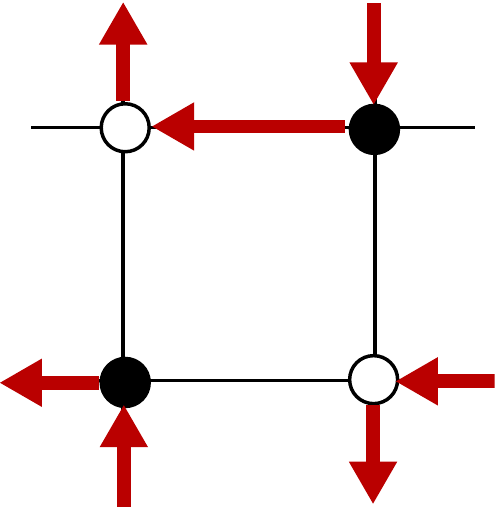}} \cr
      1 & -& - & \pb{$\ket{1,1,4;1,1,4}$ \includegraphics[height=10mm]{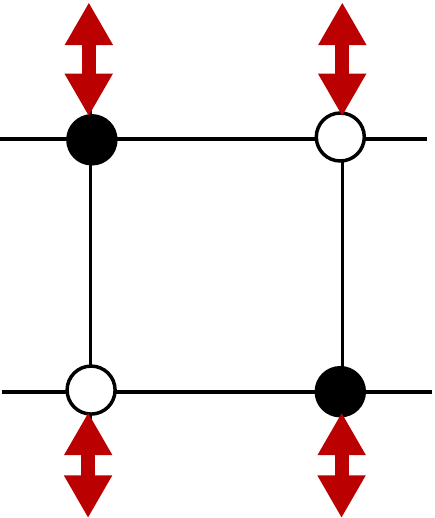}} &- & \pb{$\ket{2,1,1;1,1,3}_+$ \includegraphics[height=10mm]{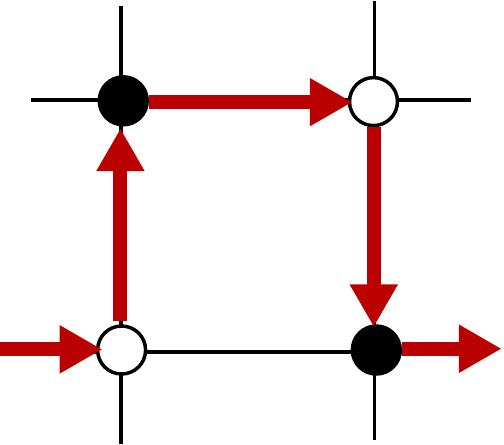}}  &\pb{$\ket{2,1,1;1,1,3}_-$ \includegraphics[height=10mm]{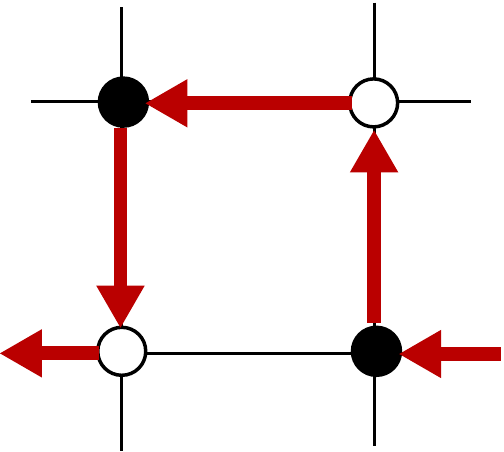}} \cr
      \rowcolor[gray]{.95} 1 & -& - & \pb{$\ket{1,1,3;1,1,3}$ \includegraphics[height=10mm]{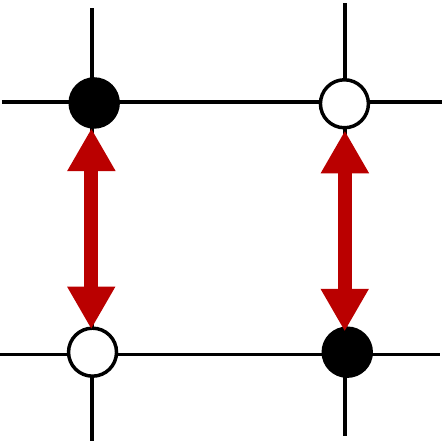}} &- & \pb{$\ket{2,1,1;1,1,2}_+$ \includegraphics[height=10mm]{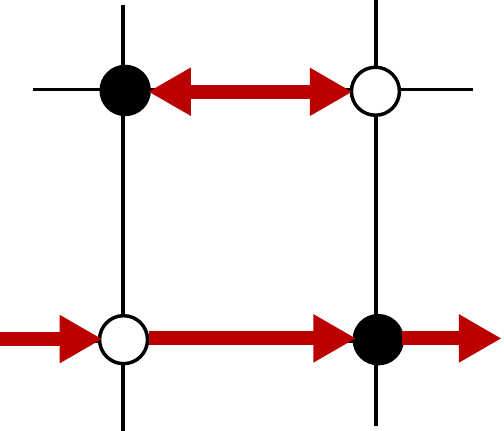}} &\pb{$\ket{2,1,1;1,1,2}_-$ \includegraphics[height=10mm]{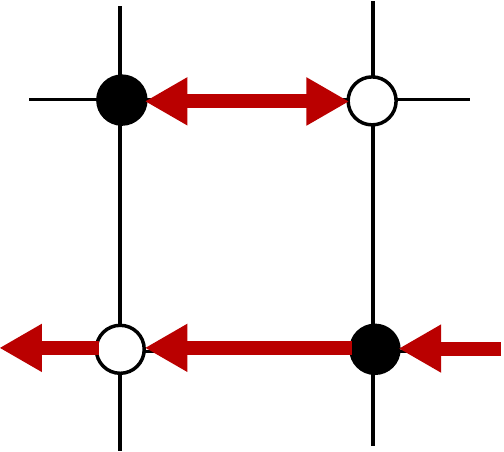}} \cr
      1 & -& - & \pb{$\ket{1,1,2;1,1,2}$ \includegraphics[height=10mm]{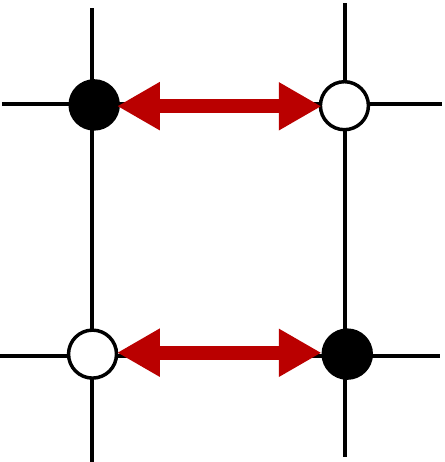}} &-& \pb{$\ket{2,1,1;1,1,1}_+$ \includegraphics[height=10mm]{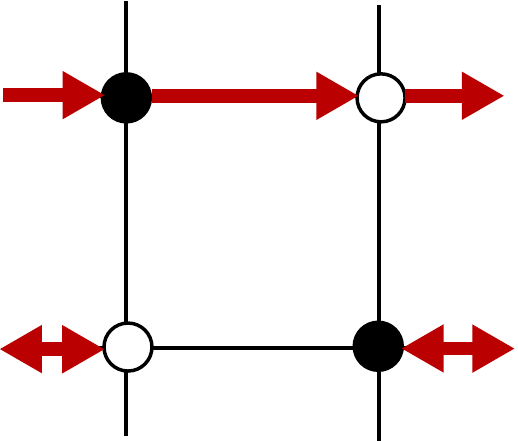}} &\pb{$\ket{2,1,1;1,1,1}_-$ \includegraphics[height=10mm]{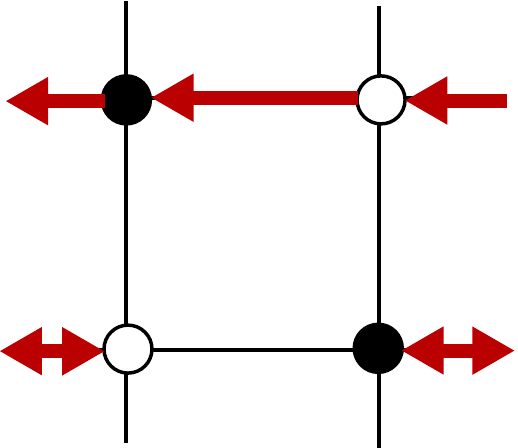}} \cr
      \rowcolor[gray]{.95} 1 & \pb{ $\ket{0,1,1;0,1,1}$ \includegraphics[height=10mm]{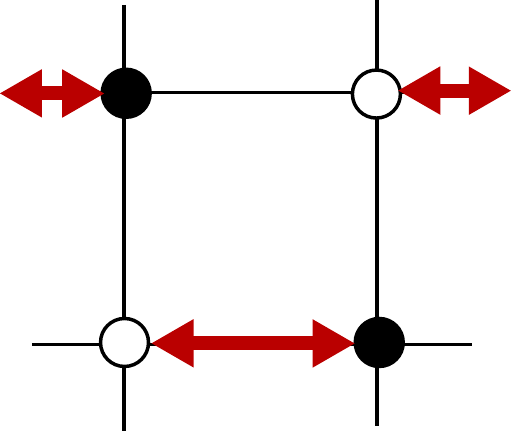}}  & -& \pb{$\ket{1,1,1;1,1,1}$ \includegraphics[height=10mm]{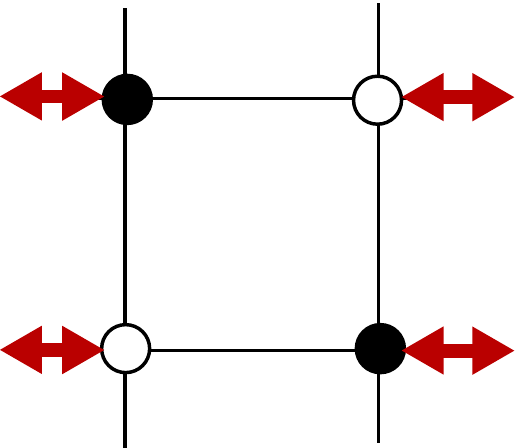}}&-& {\pb{$\ket{2,1,1;2,1,1}$ \includegraphics[height=10mm]{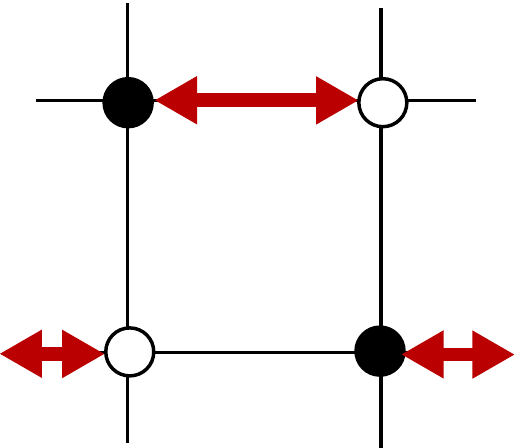}}}&-\cr
      \midrule
      0 & -& -  & \pb{$\ket{1,0,1;1,0,1}$ \includegraphics[height=10mm]{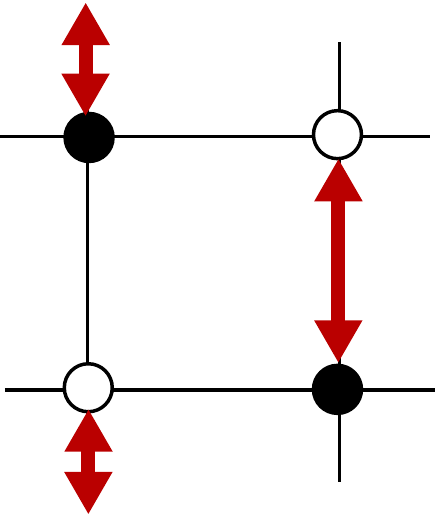}} & - &-  &- \cr
    \end{longtable}
  \end{center}
}
